\documentclass[a4paper,12pt]{article}
\usepackage{jheppub,esint,psfrag,bm}
\usepackage[utf8]{inputenc}

\usepackage{epsfig,amssymb,amsmath,psfrag,subfigure,rotate,color,wasysym,xcolor}
\usepackage[bbgreekl]{mathbbol}
\usepackage{tikz}

\allowdisplaybreaks 
\newcommand{\insertfig}[2]{\includegraphics[width=#1cm]{#2}}

\DeclareSymbolFontAlphabet{\mathbbm}{bbold}
\DeclareSymbolFontAlphabet{\mathbb}{AMSb}%

\newcommand{\dd}{\mathrm d}
\newcommand{\cG}{\mathcal G}
\newcommand{\cO}{\mathcal O}
\newcommand{\cF}{\mathcal F}
\newcommand{\Lu}{L_u}
\newcommand{\Lv}{L_v}
\newcommand{\logeq}{\mathrel{\overset{{\rm null}}{=}}}
\newcommand{\logsim}{\mathrel{\overset{{\rm null}}{\sim}}}

\DeclareMathOperator*{\SumInt}{%
\mathchoice%
  {\ooalign{$\displaystyle\sum$\cr\hidewidth$\displaystyle\int$\hidewidth\cr}}
  {\ooalign{\raisebox{.14\height}{\scalebox{.7}{$\textstyle\sum$}}\cr\hidewidth$\textstyle\int$\hidewidth\cr}}
  {\ooalign{\raisebox{.2\height}{\scalebox{.6}{$\scriptstyle\sum$}}\cr$\scriptstyle\int$\cr}}
  {\ooalign{\raisebox{.2\height}{\scalebox{.6}{$\scriptstyle\sum$}}\cr$\scriptstyle\int$\cr}}
}

\def\XXint#1#2#3{{\setbox0=\hbox{$#1{#2#3}{\int}$ }
\vcenter{\hbox{$#2#3$ }}\kern-.6\wd0}}

\def \sech {{\rm sech}}

\def \be  {\begin{equation}}
\def \ee  {\end{equation}}
\def \ba  {\begin{eqnarray}}
\def \ea  {\end{eqnarray}}
\def \baa {\begin{eqnarray*}}
\def \eaa {\end{eqnarray*}}

\def \lab #1 {\label{#1}}

\newcommand\re[1]{(\ref{#1})}
\def\d{\hbox{{d}\kern-.20em\hbox{l}}}

\def \matrix #1 {\left(\begin{array}{cc} #1 \end{array}\right)}

\def \tr {\mathop{\rm tr}\nolimits}
\def \e  {\mathop{\rm e}\nolimits}
\newcommand\lr[1]{{\left({#1}\right)}}

\newcommand \vev [1] {\langle{#1}\rangle}
\newcommand \VEV [1] {\left\langle{#1}\right\rangle}
\newcommand \ket [1] {|{#1}\rangle}

\newcommand{\ft}[2]{{\textstyle\frac{#1}{#2}}}

\def\1{\mathbbm{1}}

\title{Flux tube and crossing}

\author{A.V.~Belitsky}
\affiliation{Department of Physics, Arizona State University,  Tempe, AZ 85287-1504, USA}  

\abstract{We develop a formalism for studying four-point correlation functions of weight-$k$ half-BPS operators in the null limit,
i.e., when all nearest-neighbor distances become light-like. We use crossing relations to constrain the form of these correlators.
This currently restricts our analysis to cases involving at least one dimension-two operator. The restriction ensures that only one
R-symmetry representation propagates in each of the two partial-wave channels of the OPE. Despite this single internal-symmetry
representation, the sum receives contributions from a plethora of higher-spin towers that are degenerate at the level of bare
dimensions. Interactions lift their degeneracy. In the planar limit, the null frame of the correlator sources a stable gauge-field flux.
To tackle the resulting sums, we recast them as averages over the GKP vacuum. Although the spectrum of excitations on the GKP
background is well known, we calculate perturbative expansions for the measures and form factors governing the creation and
absorption of flux-tube excitations by operator sources. Once these ingredients are fixed, we calculate all remaining correlators
through three loops and predict the four-loop contribution for one particular infinite family.}

\begin{document}

\maketitle
\flushbottom
\setcounter{footnote} 0

\section{Introduction}

Correlation functions are physical observables in conformal field theories, but they are, of course, ubiquitous in all field theories.
Unprotected correlators are hard to calculate perturbatively and even harder to constrain nonperturbatively. In 
{\sl planar} maximally supersymmetric Yang--Mills theory (a.k.a. $\mathcal{N} = 4$ sYM), which we choose as our playground, 
an integrability-based formalism exists for computing them at finite 't Hooft coupling \cite{Basso:2015zoa,Fleury:2016ykk}. 
However, the sheer complexity of summing over elementary magnon excitations propagating on the BMN vacuum has thrown a 
wrench in the works. We know only one correlator exactly: the octagon, for which the aforementioned sums can 
be performed \cite{Coronado:2018ypq,Kostov:2019auq,Belitsky:2019fan}. The octagon arises when all R-charges of the half-BPS
operators making up a specially polarized four-point correlator are taken to infinity
\cite{Coronado:2018ypq}. 

At the opposite extreme, that of weight-two half-BPS operators, quite a lot is known in $\mathcal{N} = 4$ sYM. For instance, 
the basis of loop integrals in their perturbative expansion is available to an astounding twelve-loop order for four points
\cite{Bourjaily:2025iad}. In the {\sl null limit}, these correlators enter a triality relation with on-shell scattering amplitudes and 
Wilson loops on null polygonal contours \cite{Alday:2007hr,Drummond:2007cf,Brandhuber:2007yx}. The latter connection makes the 
problem amenable to yet another integrability-based technique grounded in the existence of a highly excited state with infinitely 
many derivatives \cite{Basso:2013aha,Belitsky:2014sla}. Via the particle-hole transformation in the underlying integrable model, it 
can be thought of as a ground state, known as the GKP vacuum \cite{GubKlePol02}. From the field-theoretical point of view, it is a 
stable flux of gauge fields that is sourced by the frame of the correlator \cite{Alday:2010zy}. The null limit corresponds to the colored 
fields propagating sequentially from one space-time point of the correlator to another at the speed of light, copiously emitting gluons 
and forming a Wilson line (up to subtle recoil effects). From the dual string point of view, the GKP string is stretched between the 
opposite sides of the perimeter and sweeps a two-dimensional worldsheet as it moves from side to side. For the dimension-two 
operators, one has to consider correlators with more than four insertions to excite the ground state and thereby probe the effective
theory of excitations living on this string. This was formalized in the framework of the pentagon operator product expansion (OPE). 
Alternatively, as we will see in this work, one can consider operators of higher R-weights.

The objects we study are four-point correlation functions of half-BPS operators of various weights,
\begin{align}
\label{GenericCorrelator}
G_{k_1,k_2,k_3,k_4} = \vev{\cO_{k_1} (x_1) \cO_{k_2} (x_2) \cO_{k_3} (x_3) \cO_{k_4} (x_4)}
\, .
\end{align}
Our main motivation is to start unraveling the transition between the two extreme cases alluded to above, i.e., to build a bridge between 
correlators of R-charge two and those of infinite R-charge, for which the null limit is well understood. Another fact that adds to the riddle is 
that the octagon is a massive scattering amplitude in disguise, so one can hope that it can be tackled by a generalization of the pentagon 
OPE for massless amplitudes. Indeed, the leading critical exponent driving its null limit obeys two different equations: one based on the 
BMN vacuum description and the other on the GKP vacuum. We therefore anticipate that a flux-tube theory should emerge from this.
In the present work, we catch another glimpse of it in null correlators of finite R-weight. The excitations are the same as for null Wilson
loops, but they couple differently to the contour produced by half-BPS operators in the correlation function.

The main calculational tools we employ are the light-cone OPE and {\sl crossing relations}. This is the well-known conformal 
bootstrap paradigm \cite{Ferrara:1973vz,Polyakov:1974gs}, which has been at the forefront of conformal field theory for the past fifty years. The idea 
is simple at its core but yields strong constraints on correlators. In a {\sl unitary} theory, the identity operator admits the resolution
\begin{align}
\1 = \sum_{i} | \cO_i \rangle \langle \cO_i |
\, ,
\end{align}
over a complete set of $s$- or $t$-channel states $| \cO_{s,t} \rangle$ that have nontrivial overlap with the external operators. 
These states belong to unprotected (super)multiplets, with $i$ enumerating all of their possible quantum numbers. Associativity of the OPE then yields 
the {\sl crossing relation}
\begin{align}
\label{GenericCrossing}
\sum_s \vev{\cO_{k_1} (x_1) \cO_{k_2} (x_2)  | \cO_s } \vev{ \cO_s | \cO_{k_3} (x_3) \cO_{k_4} (x_4)}
=
\sum_t \vev{\cO_{k_2} (x_2) \cO_{k_3} (x_3)  | \cO_t } \vev{ \cO_t | \cO_{k_4} (x_4) \cO_{k_1} (x_1)}
\, ,
\end{align}
in terms of the three-point functions $\vev{\cO \cO | \cO}$. Of course, for generic operators and generic kinematics of the external points, the interplay 
between the two channels is very hard to unravel. Generally, a state in the $s$-channel is mapped to an infinite tower of states in the $t$-channel, 
and vice versa. There are, however, limits in which this interplay simplifies substantially---in particular, for special assignments of external R-charges and in the null limit
considered here, i.e.,
\begin{align}
x_{i,i+1}^2 \to 0
\, .
\end{align}

After the trivial factors of free propagators forming the frame are stripped off, the null limit remains singular. From the OPE standpoint, the divergence stems from 
exchanging towers of high-spin states that develop anomalous dimensions that grow logarithmically with their spin $S$,
\begin{align}
\label{ADsLogS}
\gamma_i \sim g^2 \log S + \dots
\, .
\end{align}
In the planar limit, these states are created by single-trace Wilson operators with $S$ covariant derivatives. Sudakov scaling is universal.
It is intrinsic to all gauge theories, conformal or otherwise, since it manifests the infrared catastrophe caused by the copious emission of 
gauge fields from color sources. In the present context, these are the fields forming the frame of the correlator that create and absorb the gauge-field 
flux stretching from side to side. This flux tube can be excited by inserting additional field species that can then propagate on the background of 
covariant derivatives. The flux does not get screened in the planar limit. Presently, we know quite a lot about the spectrum of flux-tube excitations from 
early studies of the integrability of dilatation operators in QCD and, later, $\mathcal{N} = 4$ sYM. What we aim to uncover in the present study is how 
these excitations couple to external operators. Our main contribution is the following quantitative resolution of the identity operator in terms
of flux-tube states, the so-called {\sl holes}:
\begin{align}
\label{CompletenessHoles}
\1 = \SumInt | \mbox{holes} \rangle \langle \mbox{holes} |
\, .
\end{align}
It provides a microscopic flux-tube theory for products of three-point functions.

Our presentation is organized as follows. First, we pin down the two correlators that we analyze. We then recall their perturbative 
expansions through three loops and present their null limits in Sect.\ \ref{3LdataSection}. This is the {\sl data} on which the rest of our analysis relies.
Next, we turn to the simpler of the two mixed correlators, distinguished by having only one high-twist tower in one of its OPE channels. We solve
the crossing relations and discuss their ambiguity. This {\sl gauge} freedom will be paramount in reconciling perturbative calculations with the
flux-tube picture. We then introduce the latter and, using the known dispersion relations for holes, completely fix their integration measures and
their creation and absorption form factors from a handful of low-weight correlators. These ingredients allow us to predict all but one of the
four-loop expansion coefficients. We next turn to the mixed correlator that supports high-twist towers in both channels and verify the consistency
of the perturbative and flux-tube data, confirming this microscopic picture of null correlators. We close with conclusions and directions for further
study. Several appendices collect formulas too cumbersome for the body of the paper.

\section{Four-point correlation functions of half-BPS operators}

In this paper, we study four-point correlation functions of single-trace half-BPS operators of weights $k_i\geq 2$ in planar
$\mathcal{N} = 4$ sYM theory,
\begin{align}
G_{k_1,k_2,k_3,k_4} = \vev{\cO_{k_1} (x_1, Y_1) \cO_{k_2} (x_2, Y_2) \cO_{k_3} (x_3, Y_3) \cO_{k_4} (x_4, Y_4)}
\, .
\end{align}
The operators are built from the six real scalars $\phi_I$ of the theory,
\begin{align}
\cO_k (x, Y) = Y_{I_1} \dots Y_{I_k} \tr [\phi_{I_1} \dots \phi_{I_k}] (x)
\, .
\end{align}
The auxiliary complex six-dimensional null vector $Y_I$ projects the product of scalar fields onto the symmetric traceless
representation $[0,k,0]$ of the R-symmetry group $\mathrm{SO}(6)\simeq\mathrm{SU}(4)$. The half-BPS condition fixes the
scaling dimension of $\cO_k (x, Y)$ to be equal to its weight $k$.

For arbitrary 't Hooft coupling
\begin{align}
g^2 = \frac{g_{\rm\scriptscriptstyle YM}^2 N_c}{(4 \pi)^2}
\, ,
\end{align}
the correlator separates into its Born contribution and a coupling-dependent correction,
\begin{align}
\label{CorrSplit}
G_{k_1,k_2,k_3,k_4} = G^{\rm Born}_{k_1,k_2,k_3,k_4} + G^{\rm Loop}_{k_1,k_2,k_3,k_4}
\, .
\end{align}
The Born term is obtained by Wick contracting the scalar fields and is a sum of products of free propagators
$(Y_i\cdot Y_j)/x_{ij}^2$. As a consequence of the partial non-renormalization theorem, the second term in Eq.\ \re{CorrSplit}
takes the factorized form
\begin{align}
\label{LoopG}
G^{\rm Loop}_{k_1,k_2,k_3,k_4}
=
C_{k_1,k_2,k_3,k_4} R(1,2,3,4) \sum_{b_{ij}} \prod_{1 \leq i < j \leq 4} \left( \frac{Y_i \cdot Y_j}{x_{ij}^2} \right)^{b_{ij}}
\frac{\cF_{\{ b_{ij} \}} (u, v; g^2)}{x_{13}^2 x_{24}^2}
\, .
\end{align}
Here $C_{k_1,k_2,k_3,k_4}$ is a normalization factor that accounts for color and R-charge factors, combinatorics, and the
normalization of scalar propagators. The universal prefactor $R(1,2,3,4)$ is a polynomial in $x_{ij}^2=(x_i-x_j)^2$ and
$Y_i\cdot Y_j$, homogeneous of degree two in each polarization and of conformal weight one at each point; it reflects the
superconformal constraint on loop effects. Explicit expressions for both factors can be found in Ref.\ \cite{Chicherin:2015edu}.

The product of scalar propagators in Eq.\ \re{LoopG} defines the {\sl frame} of the contributing Feynman graphs. It is
parametrized by the sextuple of nonnegative integers $\{ b_{ij} \}$ with $1 \leq i < j \leq 4$. Setting $b_{ji}=b_{ij}$, charge
conservation at each insertion requires $\sum_{j\neq i}b_{ij}=k_i-2$ ($i=1,\dots,4$). If the four operators are placed at the
adjacent vertices of a rectangle, the nearest-neighbor bridges contain $b_{i,i+1}+1$ propagators (with cyclic indices), while
the diagonal bridges contain $b_{13}$ and $b_{24}$ propagators. The difference is due to the extra factor $1/(x_{13}^2 x_{24}^2)$
extracted from $\cF$ in Eq.\ \re{LoopG}.

All remaining dynamical information is contained in the conformally invariant functions $\cF_{\{b_{ij}\}}(u,v;g^2)$. They depend
on the two cross ratios
\begin{align}
\label{uv}
u = \frac{x_{12}^2 x_{34}^2}{x_{13}^2 x_{24}^2}
\, , \qquad
v = \frac{x_{23}^2 x_{41}^2}{x_{13}^2 x_{24}^2}
\, ,
\end{align}
and admit an expansion in powers of the 't Hooft coupling. We study the correlators in the null limit
\begin{align}
\label{NullLimitDef}
x_{12}^2, x_{23}^2, x_{34}^2, x_{41}^2 \to 0
\, ,
\end{align}
in which both cross ratios vanish simultaneously.

For the simplest correlator, $G_{2,2,2,2}$, the charge constraints force all $b_{ij}$ to vanish, $b_{ij} = \{0, \dots, 0\}$.
Its loop correction is therefore determined by a single function $\cF$. In the null limit \re{NullLimitDef}, the leading
contribution to this function in both light-cone OPE channels, $x_{12}^2\to 0$ and $x_{23}^2\to 0$, comes from the tower of
unprotected twist-two operators. Crossing symmetry relates their contributions in the two channels, leading to the map
\begin{align}
\mbox{twist-}2\quad \leftrightarrow\quad \mbox{twist-}2
\, .
\end{align}
The resulting crossing equations were solved in the null limit to all orders in perturbation theory in Ref.\ \cite{Alday:2013cwa}.

We extend this analysis to higher-weight half-BPS operators. In general, several R-symmetry representations contribute to the
light-cone OPE in each channel, and crossing couples their contributions. The allowed representations follow from the
R-symmetry selection rules. In the channel $x_{12}^2\to 0$, the tensor product decomposes, for $k_2\geq k_1$, as
\begin{align}
\label{TensorDec}
[0,k_1,0] \otimes [0,k_2,0]
=
\mathop{\oplus}_{n = 0}^{k_1} \mathop{\oplus}_{m = 0}^{k_1 - n} [m, k_1 + k_2 - 2n - 2m, m]
\, .
\end{align}
The representations with $n=0,1$ correspond to protected operators. They contribute only to the first term in Eq.\ \re{CorrSplit}
and are irrelevant to us. The loop correction \re{LoopG} receives contributions from the superconformal primaries of long
multiplets, which correspond to the range $n\geq 2$ in Eq.\ \re{TensorDec}.

When one of the operators has weight two, $k_1=2$, the long-multiplet part of the sum \re{TensorDec} reduces to the single term
$n=2$, $m=0$. Thus, in the OPE $\cO_{2} (x_1) \cO_{k_2} (x_2)$, long superconformal primaries can transform only in
$[0,k_2-2,0]$. An exchanged multiplet must also be allowed in the OPE of the opposite pair $\cO_{k_3} (x_3) \cO_{k_4} (x_4)$.
For the correlators $G_{2,k_2,k_3,k_4}$ considered below, this leaves a single R-symmetry representation of long superconformal
primaries in each channel.

The same simplification is visible directly in Eq.\ \re{LoopG}. For $k_1=2$, nonnegativity forces $b_{12}=b_{13}=b_{14}=0$, and
the remaining three charge constraints uniquely determine
\begin{align}
\label{bijWeight2}
b_{23}=\frac{k_2+k_3-k_4}{2}-1\,,
\qquad
b_{24}=\frac{k_2+k_4-k_3}{2}-1\,,
\qquad
b_{34}=\frac{k_3+k_4-k_2}{2}-1\,,
\end{align}
whenever an allowed structure exists. The loop correction then depends on a single function $\cF_{\{b_{ij}\}}(u,v;g^2)$. This
does not imply a single exchanged operator. At fixed bare twist, several operators can share the same spin and R-symmetry
quantum numbers while acquiring distinct anomalous dimensions. The null limit \re{NullLimitDef} suppresses contributions of
higher bare twist relative to the leading unprotected twist, but it does not resolve the multiplicity within the leading tower
when that twist exceeds two.

\begin{figure}[t]
\begin{center}
\mbox{
\begin{picture}(0,145)(180,0)
\put(0,10){\insertfig{12}{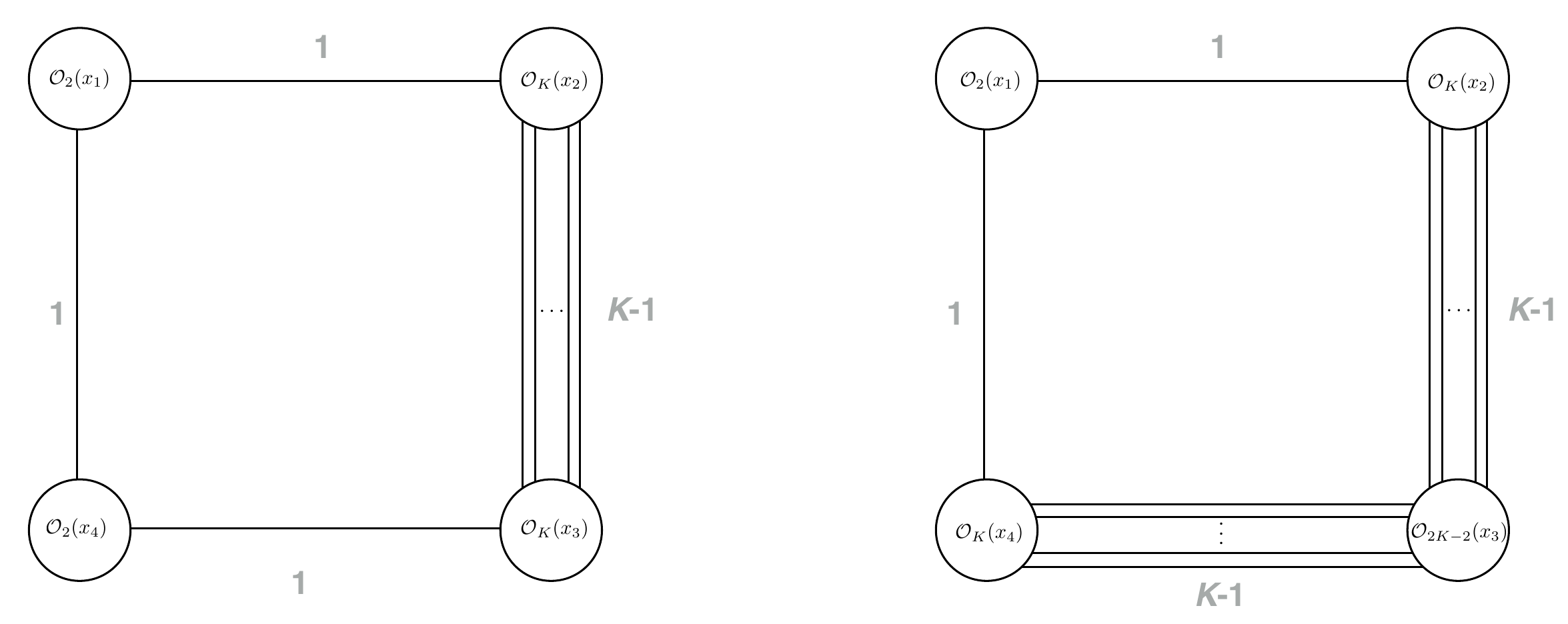}}
\put(58,-7){(a)}
\put(258,-7){(b)}
\end{picture}
}
\end{center}
\caption{\label{CorrelatorSkeletonsFig} Planar null-frame skeletons of the correlators in Eq.\ \re{TwistMap}: (a) $G_{2,K,K,2}$
and (b) $G_{2,K,2K-2,K}$. Each edge represents a bridge of scalar propagators, labeled by its multiplicity $b_{ij}+1$. Thick
edges denote bridges with $K-1$ propagators. Both diagonal bridges are absent, $b_{13}=b_{24}=0$. At each vertex, the incident
multiplicities add up to the weight of the corresponding operator.}
\end{figure}

We consider two families of correlators, $G_{2,K,K,2}$ and $G_{2,K,2K-2,K}$ with $K>2$. Their frames, defined by the products of
scalar propagators in Eq.\ \re{LoopG}, are shown in Fig.\ \ref{CorrelatorSkeletonsFig}. In the null limit, crossing relates the
contributions of the towers of operators with the minimal bare twist in the two channels,
\begin{alignat}{3}
\label{TwistMap}
&G_{2,K,K,2}: \qquad && \mbox{twist-}{K} && \leftrightarrow \mbox{twist-}2 \, ,
\nonumber\\[2mm]
&G_{2,K,2K-2,K}: \qquad && \mbox{twist-}{K} && \leftrightarrow \mbox{twist-}{K} \, ,
\end{alignat}
where the left column refers to the $s$-channel ($x_{12}^2\to 0$) and the right one to the $t$-channel ($x_{23}^2\to 0$).

For the correlators in Eq.\ \re{TwistMap}, we abbreviate the corresponding conformally invariant functions as
\begin{align}
\label{FKdef}
\cF_{[K,2]}
\equiv
\cF_{\{0,0,0,K-2,0,0\}}
\, , \qquad
\cF_{[K,K]}
\equiv
\cF_{\{ 0,0,0,K-2,0,K-2\}}
\, ,
\end{align}
where the bracketed entries on the left-hand sides specify the leading bare twists in the $s$- and $t$-channels, and the
subscripts on the right-hand sides give the bridge labels in the ordering $\{b_{12},b_{13},b_{14},b_{23},b_{24},b_{34}\}$. These
differ from the nomenclature adopted in Ref.\ \cite{Chicherin:2015edu} by a mere permutation of labels. The two functions in
Eq.\ \re{FKdef} coincide at $K=2$.

Crossing, i.e., the exchange of the two cross ratios, acts on the conformal functions as
\begin{align}
\label{CrossingF}
\cF_{\{b_{12},b_{13},b_{14},b_{23},b_{24},b_{34}\}}(u,v;g^2)
=
\cF_{\{b_{14},b_{13},b_{12},b_{34},b_{24},b_{23}\}}(v,u;g^2)
\, .
\end{align}
This permutation leaves the bridge structure of $\cF_{[K,K]}$ invariant, so that this function is symmetric,
$\cF_{[K,K]}(u,v;g^2)=\cF_{[K,K]}(v,u;g^2)$. For $\cF_{[K,2]}$, it exchanges the channel assignments together with the
corresponding bridge labels.

The choice of correlators in Eq.\ \re{TwistMap} reduces the R-symmetry content of the exchanged long superconformal primaries
to a single representation in each channel. It does not, however, remove the multiplicity of twist-$K$ operators. Unlike in the
twist-two case, crossing constrains only sums over their OPE data and generally cannot resolve the individual contributions.
The resulting freedom is a central feature of our analysis. We constrain it using explicit perturbative results and the
flux-tube description developed below.

\section{Data}
\label{3LdataSection}

Since explicit perturbative results provide the main data for our construction of the flux-tube phenomenology, let us recall what is
available in the literature for $G_{2,K,K,2}$ and $G_{2,K,2K-2,K}$. Although bases of integrals defining these correlators are known
through five loops in the planar limit \cite{Chicherin:2018avq}, the integrals have been evaluated in terms of generalized polylogarithms only
through three loops \cite{Drummond:2013nda}. For the reader's convenience, we list these conformal off-shell integrals in Appendix \ref{app:basis}.
Here, we merely give the perturbative expansions in this basis, using our convention for the normalization of the 't Hooft coupling:
\begin{align}
\cF (u, v; g^2) = \sum_{L \geq 1} g^{2L} \cF^{(L)} (u, v)
\, .
\end{align}

\begin{figure}[t]
\begin{center}
\mbox{
\begin{picture}(0,145)(230,0)
\put(0,10){\insertfig{16}{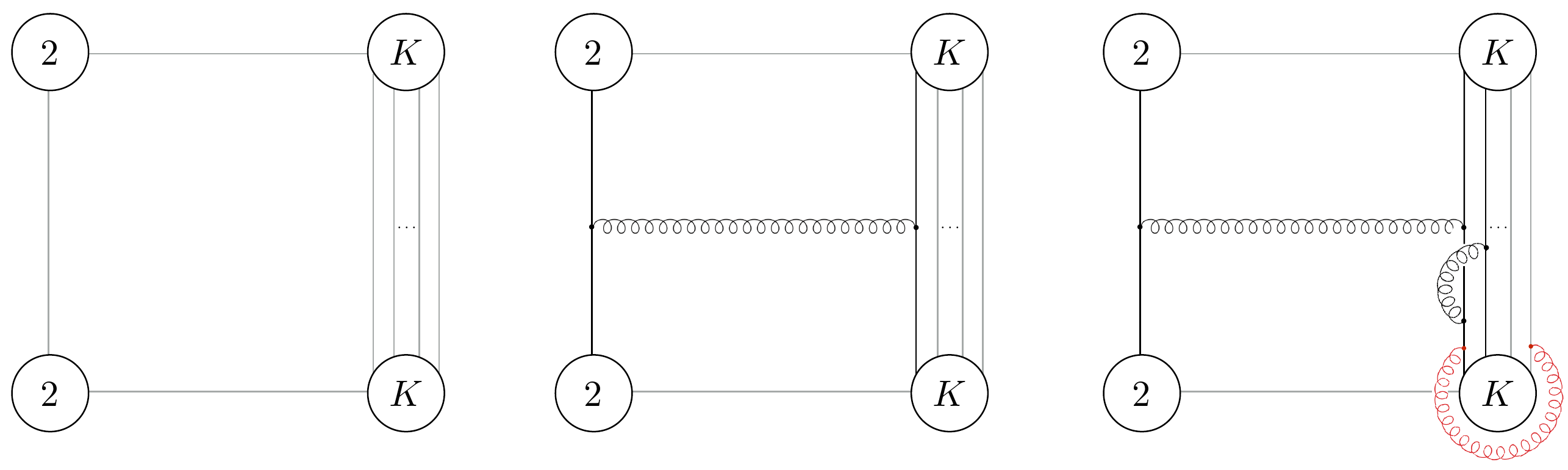}}
\put(56,-5){(a)}
\put(214,-5){(b)}
\put(374,-5){(c)}
\end{picture}
}
\end{center}
\caption{\label{KdepFig} Emergence of the $K$-dependence in the conformal function $\cF_{[K,2]}$ (see the text for details).
The Born, one-loop, and two-loop graphs are shown in (a), (b), and (c), respectively. In panel (c), the red gluon exchange is suppressed
in the planar approximation. Black and gray lines represent active and spectator partons, respectively.}
\end{figure}

Before quoting the results of Ref.\ \cite{Chicherin:2015edu}, let us briefly comment on their dependence on the R-charge $K$. This brings us back
to an observation dubbed {\sl uniformity} in that reference: in its simplest form, all planar integrals appearing in high-weight correlators first occur
in the simplest case, $G_{2,2,2,2}$. More specifically, a remarkable universality emerges at weak coupling in the planar limit: the $L$-loop corrections
$\cF^{(L)}$ have the same form for all correlators of charge $K>L$. This property is easy to understand from general features of the contributing
Feynman diagrams and their color ordering. Let us take $G_{2,K,K,2}$ as a case study (see Fig.\ \ref{KdepFig}). At tree level, all propagators are
perfectly paired and form the Born {\sl skeleton}, or in the present case the {\sl frame}, of the correlator. Since this tree-level structure is factored
out of the loop function $\cF$, no intrinsic $K$ dependence remains, and the term is normalized to one. At one loop, only the innermost and
outermost lines in the weight-$K$ operator can interact with the other operators in the correlator; Fig.\ \ref{KdepFig}(b) shows an example.
In the planar limit, gluon attachments to the interior propagators exchanged between the two heavy operators are suppressed. After division by
the Born term, all spectator propagators cancel from the quotient, leaving no $K$ dependence: the one-loop correction is the same for all correlators.
At two loops, see Fig.\ \ref{KdepFig}(c), the one-loop graph must be dressed by an extra gluon exchange. The only graph capable of inducing
$K$ dependence is the one in which the inner scalar (in our illustration) interacts with its nearest neighbor inside the trace, as shown. One might
think that all other nearest-neighbor planar exchanges inside the long bridge could be included and would sum to zero by virtue of the BPS nature
of the operator at the bottom-right corner of the graph. The correlator topology, however, prevents a complete cancellation because the exchange
between the innermost and outermost scalars is suppressed in the planar approximation. The graph is therefore nonzero and induces nontrivial
$K$ dependence whenever $K>2$. Higher loops follow the same logic.

\subsection[Twist-K--twist-2 correlator]{Twist-\textbf{\textit K}--twist-2 correlator}

We start with the conformally invariant function $\cF_{[K,2]}$. As explained at the end of the previous section, its one-loop form is the
same for all values of $K\geq 2$,
\begin{align}
\cF_{[K,2]}^{(1)}= x_{13}^2 x_{24}^2 g_{1234}
\, .
\end{align}
At two loops, the corrections read
\begin{align}
 \label{eq:G2222-two}
 \cF_{[2,2]}^{(2)}
 =
 &
 \, 
x_{13}^2 x_{24}^2
 \bigg[
 2h_{12;34}+2h_{13;24}+2h_{14;23}
 +
 \frac{1}{2}
 \left(x_{12}^2x_{34}^2 + x_{13}^2 x_{24}^2 + x_{14}^2x_{23}^2\right) [g_{1234}]^2
 \bigg]
 \, , \\                 
 \label{eq:G2KK2-two}                                 
 \cF_{[K>2,2]}^{(2)}
 =
 &
 \, 
 x_{13}^2 x_{24}^2
 \bigg[
 2h_{12;34}+2h_{13;24}+h_{14;23}
 +
 \frac{1}{2} \left(x_{12}^2x_{34}^2+x_{13}^2 x_{24}^2 \right) [g_{1234}]^2
 \bigg]
 \, .                                   
\end{align}
At three loops, they are
\begin{align}
\label{eq:G2222-three}
 \cF_{[2,2]}^{(3)}
 =& \, 
 x_{13}^2 x_{24}^2
\bigg[
2 x_{12}^2x_{34}^2(g\!\times\!h)_{12;34}
+
2 x_{13}^2 x_{24}^2(g\!\times\!h)_{13;24}
+
2x_{14}^2x_{23}^2(g\!\times\!h)_{14;23} \nonumber\\
&\quad
+6L_{12;34}+6L_{13;24}+6L_{14;23} +4E_{12;34}+4E_{13;24}+4E_{14;23} 
\nonumber\\
&\quad
+\left(1+\frac{1}{v}\right)H_{12;34} +\left(1+\frac{u}{v}\right)H_{13;24} +(1+u)H_{14;23}
\bigg]
\, , \\
\label{eq:G2332-three}
 \cF_{[3,2]}^{(3)}
 =
&\,
x_{13}^2 x_{24}^2 
\bigg[ 
-
x_{14}^2x_{23}^2(g\!\times\!h)_{14;23}
+
2x_{13}^2 x_{24}^2(g\!\times\!h)_{13;24}
+
2x_{12}^2x_{34}^2(g\!\times\!h)_{12;34} \nonumber\\
&\quad
+6L_{12;34}+6L_{13;24}+2L_{14;23}+2E_{12;34}+2E_{13;24}+4E_{14;23} 
\nonumber\\ 
&\quad+\frac{1}{v}H_{12;34}+\frac{u}{v}H_{13;24}
+\frac{1}{2}(1+u)H_{14;23}
\bigg]
\, , \\
 \cF_{[K>3,2]}^{(3)}
 =
 &\,
 x_{13}^2 x_{24}^2
\bigg[
2x_{12}^2x_{34}^2(g\!\times\!h)_{12;34}
+2 x_{13}^2 x_{24}^2 (g\!\times\!h)_{13;24}
\nonumber\\
&\quad
+6L_{12;34}+6L_{13;24}+3L_{14;23}+2E_{12;34}+2E_{13;24}+4E_{14;23} 
\nonumber\\
&\quad+\frac{1}{v}H_{12;34}+\frac{u}{v}H_{13;24}
+\frac12(1+u)H_{14;23}
\bigg]
\, .                 
\label{eq:G2KK2-three}
\end{align}

\subsection[Twist-K--twist-K correlator]{Twist-\textbf{\textit K}--twist-\textbf{\textit K} correlator}

For $K=2$, this correlator reduces to $\cF_{[2,2]}$, already displayed above. At one and two loops, we have
\begin{align}
\cF_{[K,K]}^{(1)}
=& \, 
x_{13}^2 x_{24}^2 g_{1234}
\, ,\\
\label{eq:G2K-two}
\cF_{[K\geq 3,K\geq 3]}^{(2)}
=& \, 
x_{13}^2 x_{24}^2
\bigg[
h_{12;34}+2h_{13;24}+h_{14;23} + \frac12 x_{13}^2 x_{24}^2 [g_{1234}]^2
\bigg]
\, .                                
\end{align}
At three loops, we have to distinguish the $K=3$ case,
\begin{align}
 \label{eq:G2343-three}
 \cF_{[3,3]}^{(3)}
 =& \,
x_{13}^2 x_{24}^2
\bigg[
-
x_{12}^2x_{34}^2(g\!\times\!h)_{12;34}
+
2 x_{13}^2 x_{24}^2 (g\!\times\!h)_{13;24}
-
x_{14}^2x_{23}^2(g\!\times\!h)_{14;23} 
\nonumber\\
&\quad
+
2L_{12;34}+6L_{13;24}+2L_{14;23} +2E_{12;34}+E_{13;24}+2E_{14;23} 
\nonumber\\
&\quad
+
\frac{1}{2v}H_{12;34} + \frac{1}{2} H_{14;23}
\bigg] \, .      
\end{align}
For $K>3$, on the other hand,
\begin{align}
\cF_{[K>3,K>3]}^{(3)}
=&\,
x_{13}^2 x_{24}^2
\bigg[
2 x_{13}^2 x_{24}^2 (g\!\times\!h)_{13;24} + 3 L_{12;34} + 6L_{13;24} + 3L_{14;23} \nonumber\\
&\quad
+2E_{12;34}+E_{13;24}+2E_{14;23} +\frac{1}{2v} H_{12;34}+\frac{1}{2} H_{14;23}
\bigg]\, .             \label{eq:G2K-three}
\end{align}

\subsection{Null limit}
\label{3LdataNullSection}

Having expressed both correlators in terms of the basis integrals, we can now take their null limit, in which the two cross ratios tend to zero 
simultaneously:
\begin{align}
x_{i,i+1}^2 \to 0 \quad \leftrightarrow \quad u,v \to 0
\, .
\end{align}
In this section, the symbol $\logeq$ denotes equality after retaining all terms containing positive powers of the logarithms $\Lu$ and/or $\Lv$,
\begin{align}
\Lu = \log u
\, , \qquad
\Lv = \log v
\, ,
\end{align}
while discarding the log-free term and power corrections. We present results for the logarithm of the reduced correlator including its
Born contribution,
\begin{align}
\cG \equiv 1+ 2 \cF
\, ,
\end{align}
which is normalized to one. Through three loops, we have
\begin{align}
 \log \cG
 =
&\, 
2 g^2\cF^{(1)} +g^4\left(2\cF^{(2)}-2[\cF^{(1)}]^2\right)
\nonumber\\
&\,
+g^6\left(2\cF^{(3)}-4\cF^{(1)}\cF^{(2)} +\frac83[\cF^{(1)}]^3\right)+O(g^8)
\, .
\label{eq:log-from-reduced-functions}
\end{align}
For each correlator, we define its logarithmic coefficients by
\begin{equation}
 \log \cG(u,v)\logeq
 \sum_{\substack{i,j\geq0\\i+j>0}} c_{ij}(g) \Lu^i\Lv^j
 \, .
 \label{eq:cij-definition}
\end{equation}
We stress that the log-free coefficient $c_{00}$ is not included. This form makes partial exponentiation manifest and explicitly
exhibits the superleading logarithms induced by the higher weights of the BPS operators involved. 

Using the results of Appendix \ref{app:null-integrals}, we obtain the following complete list of nonzero coefficients through three loops. For 
$\cG_{[2,2]}$, we have
\begin{align}
 c_{11}^{[2,2]}={}&-2g^2+\frac{2\pi^2}{3}g^4-\frac{62\pi^4}{45}g^6,\notag\\
 c_{20}^{[2,2]}=c_{02}^{[2,2]}={}&
 \frac{2\pi^2}{3}g^4-\frac{4\pi^4}{9}g^6,\notag\\
 c_{10}^{[2,2]}=c_{01}^{[2,2]}={}&
 -12\zeta_3g^4+
 \left(\frac{16\pi^2\zeta_3}{3}+80\zeta_5\right)g^6,\notag\\
c_{30}^{[2,2]}=c_{03}^{[2,2]}={}&-\frac{16}{3}\zeta_3g^6
 \, .
 \label{eq:cij-2222}
\end{align}
For $\cG_{[3,2]}$, the result is
\begin{align}
 c_{11}^{[3,2]}={}&-2g^2+\frac{2\pi^2}{3}g^4-\frac{52\pi^4}{45}g^6,\notag\\
 c_{20}^{[3,2]}={}&
 \frac{\pi^2}{2}g^4-\frac{16\pi^4}{45}g^6,&
 c_{02}^{[3,2]}={}&
 \frac{2\pi^2}{3}g^4-\frac{4\pi^4}{9}g^6,\notag\\
 c_{10}^{[3,2]}={}&
 -6\zeta_3g^4+\left(4\pi^2\zeta_3+20\zeta_5\right)g^6,&
 c_{01}^{[3,2]}={}&
 -12\zeta_3g^4+
 \left(\frac{8\pi^2\zeta_3}{3}+80\zeta_5\right)g^6,\notag\\
 c_{30}^{[3,2]}={}&-3\zeta_3g^6,&
 c_{03}^{[3,2]}={}&-\frac{16}{3}\zeta_3g^6
 \, .
 \label{eq:cij-2332}
\end{align}
Finally, for $\cG_{[K>3,2]}$,
\begin{align}
 c_{11}^{[K,2]}={}&-2g^2+\frac{2\pi^2}{3}g^4-\frac{52\pi^4}{45}g^6,\notag\\
 c_{20}^{[K,2]}={}&
 \frac{\pi^2}{2}g^4-\frac{2\pi^4}{5}g^6,&
 c_{02}^{[K,2]}={}&
 \frac{2\pi^2}{3}g^4-\frac{4\pi^4}{9}g^6,\notag\\
 c_{10}^{[K,2]}={}&
 -6\zeta_3g^4+\left(4\pi^2\zeta_3+40\zeta_5\right)g^6,&
 c_{01}^{[K,2]}={}&
 -12\zeta_3g^4+
 \left(\frac{8\pi^2\zeta_3}{3}+80\zeta_5\right)g^6,\notag\\
 c_{30}^{[K,2]}={}&-\frac{8}{3}\zeta_3g^6,&
 c_{03}^{[K,2]}={}&-\frac{16}{3}\zeta_3g^6
 \, .
 \label{eq:cij-2KK2}
\end{align}

For the other correlator, the $\cG_{[3,3]}$ coefficients are
\begin{align}
 c_{11}^{[3,3]}={}&-2g^2+\frac{2\pi^2}{3}g^4-\frac{89\pi^4}{90}g^6,\notag\\
 c_{20}^{[3,3]}=c_{02}^{[3,3]}={}&
 \frac{\pi^2}{2}g^4-\frac{16\pi^4}{45}g^6,\notag\\
 c_{10}^{[3,3]}=c_{01}^{[3,3]}={}&
 -6\zeta_3g^4+\left(2\pi^2\zeta_3+20\zeta_5\right)g^6,\notag\\
 c_{30}^{[3,3]}=c_{03}^{[3,3]}={}&-3\zeta_3g^6
 \, .
 \label{eq:cij-2343}
\end{align}
For $K>3$, they read
\begin{align}
 c_{11}^{[K,K]}={}&
 -2g^2+\frac{2\pi^2}{3}g^4-\frac{89\pi^4}{90}g^6,\notag\\
 c_{20}^{[K,K]}=c_{02}^{[K,K]}={}&
 \frac{\pi^2}{2}g^4-\frac{2\pi^4}{5}g^6,\notag\\
 c_{10}^{[K,K]}=c_{01}^{[K,K]}={}&
 -6\zeta_3g^4+\left(2\pi^2\zeta_3+40\zeta_5\right)g^6,\notag\\
 c_{30}^{[K,K]}=c_{03}^{[K,K]}={}&
 -\frac{8}{3}\zeta_3g^6.
 \label{eq:cij-2K}
\end{align}
All coefficients not displayed vanish through $g^6$.

\section{Conformal partial-wave expansion}

As is clear from the perturbative data introduced above, the null limit of $\cG$ is singular, whereas the Born approximation 
is finite at $u = v = 0$. The origin of this behavior is well understood from the perspective of the light-cone OPE. Let us briefly 
recall the argument, since the resulting formulas will be used in the conformal bootstrap equations for several correlators. For the 
correlator components considered here, the external dimensions obey $\Delta_1+\Delta_3=\Delta_2+\Delta_4$. To streamline the formulas below,
we parametrize the dimensions, together with the chosen frame R-symmetry structure, by bridge labels $\{K_i\}$ according to
\begin{align}
\Delta_1=K_1+K_4\,,\qquad \Delta_2=K_1+K_2\,,\qquad \Delta_3=K_2+K_3\,,\qquad \Delta_4=K_3+K_4
\, .
\end{align}
The $K_i$ count the propagators forming the bridges around the correlator frame. A choice of frame fixes the otherwise redundant
parametrization. No propagators are left for the diagonals, which we exclude from consideration.

In this section, we strip the R-charge dependence from the correlators, since we are interested only in their dependence on
the space-time cross ratios. To make contact with the classic analyses of conformal partial-wave expansions
in Refs.\ \cite{Dolan:2000ut,Dolan:2003hv}, let us spell out the relation between the functions involved. We can then 
borrow their results directly. According to Ref.\ \cite{Dolan:2011dv}, the correlation function \re{GenericCorrelator} admits the
general form 
\begin{align}
G_{\Delta_1, \Delta_2, \Delta_3, \Delta_4} 
= \lr{x_{24}^2\over x_{14}^2}^{(\Delta_1-\Delta_2)/2} \lr{x_{14}^2\over x_{13}^2}^{(\Delta_3-\Delta_4)/2}
{G(u,v) \over (x_{12}^2)^{(\Delta_1+\Delta_2)/2}(x_{34}^2)^{(\Delta_3+\Delta_4)/2} }
\, ,
\end{align}
where $G(u,v)$ is a conformally invariant function of the cross ratios. This should be compared with our convention,
\begin{align}
G_{\Delta_1, \Delta_2, \Delta_3, \Delta_4} 
= 
{F (u,v) \over (x_{12}^2)^{K_1}(x_{23}^2)^{K_2}(x_{34}^2)^{K_3}(x_{41}^2)^{K_4}} 
\, ,
\end{align}
which we use in this section. For the normalized single-structure correlators studied elsewhere in the paper, this Born-inclusive
function coincides with $\cG$; we write $F$ here to keep the discussion of external dimensions generic. The two functions differ
only by powers of $u$ and $v$:
\begin{align}
F (u,v) = u^{-(K_2+K_4)/2}  v^{K_2} G(u,v)
\, .
\end{align}
Having established this dictionary, we can proceed.

The $s$-channel OPE, $\cO_{\Delta_1} \cO_{\Delta_2} \to \cO_s \to \cO_{\Delta_3} \cO_{\Delta_4}$, is given by an infinite 
sum over conformal primaries $\cO_s = \cO_{\Delta_s, S_s}$ of allowed dimension $\Delta_s$ and spin $S_s$  
\cite{Dolan:2000ut,Dolan:2003hv,Dolan:2011dv}
\begin{align}
\label{sChannelOPE}
F (u,v)
=
u^{-(K_2+K_4)/2}  v^{K_2} 
\sum_{s} a_s
\,
g^{\Delta_{12},\Delta_{34}}_{\Delta_s, S_s}(u,v)
\, .
\end{align}
All of their descendants are encoded in the conformal blocks $g^{\Delta_{12},\Delta_{34}}_{\Delta_s, S_s}$, which depend on
the dimension differences $\Delta_{ij} = \Delta_i - \Delta_j$. The accompanying coefficients $a_s$ are products of the two 
three-point coefficients for transitions into and out of the intermediate states:
\begin{align}
\label{sChannelOPEcoeff}
a_s
=
\vev{\cO_{\Delta_1} \cO_{\Delta_2} | \cO_{\Delta_s, S_s}} \vev{ \cO_{\Delta_s, S_s} | \cO_{\Delta_3} \cO_{\Delta_4}}
\, .
\end{align}
They depend on the spin $S_s$ and dimension $\Delta_s$, but not on $u$ or $v$.

Similarly, one can write the OPE in the $t$-channel,
\begin{align}
\label{tChannelFullCPWE}
F (v,u)
=
v^{-(K_1+K_3)/2} u^{K_1} 
\sum_{t} a_t
\,
g^{\Delta_{32},\Delta_{14}}_{\Delta_t, S_t}(v,u)
\, .
\end{align}
It is obtained from Eq.\ \re{sChannelOPE} by exchanging operators sitting at the space-time points $x_1$ and $x_3$.
This {\sl crossing} permutes the $K$-labels as follows:
\begin{align}
\label{ExchangeCrossing}
\cO_{\Delta_1} (x_1) \leftrightarrow \cO_{\Delta_3} (x_3):
\qquad
{K_2 \leftrightarrow K_1 \atop K_4 \leftrightarrow K_3}
\, .
\end{align}
In writing $F(v,u)$, we suppress the correspondingly permuted external labels, which are understood to transform as in
Eq.\ \re{ExchangeCrossing}.

In the light-cone limit, the conformal blocks simplify considerably. In the $s$-channel, corresponding to 
$x_{12}^2, x_{34}^2 \to 0$, a conformal block factorizes into a power of the small cross ratio $u$,
\begin{align}
g^{\Delta_{12},\Delta_{34}}_{\Delta_s,S_s}(u,v)
= 
u^{\tau_s/2}
\,
f^{\Delta_{12},\Delta_{34}}_{\tau_s,S_s} (v) + O(u^{\tau_s/2+1})
\, ,
\end{align}
where the collinear block is \cite{Ferrara:1973vz,Dolan:2011dv}
\begin{align}
f^{\Delta_{12},\Delta_{34}}_{\tau,S}(v)
=
(1-v)^S\,
{}_2F_1
\!\left(
S+\frac{\tau-\Delta_{12}}{2},
S+\frac{\tau+\Delta_{34}}{2},
2S+\tau;
1-v
\right)
\, .
\label{eq:coll-block}
\end{align}
Here we have introduced the twist 
\begin{align}
\tau_s = \Delta_s - S_s
\, ,
\end{align}
of the intermediate primary. For the moment, we suppress the additional labels that enumerate operators of a given spin and dimension.
Such operators have nontrivial multiplicities because their bare dimensions are degenerate, although the degeneracies are lifted once
interactions are turned on. We will reinstate this dependence later; it will take center stage in the rest of the paper.

The same formula applies in the $t$-channel, after the replacements
\begin{equation}
u \leftrightarrow v
\, ,\qquad
(\Delta_{12},\Delta_{34}) \rightarrow (\Delta_{32},\Delta_{14})
\, .
\end{equation}

\subsection{Null limit and large-spin approximation}

In the double light-cone, or null, limit, both $u$ and $v$ tend to zero, and Eq.\ \re{eq:coll-block} simplifies further. Let us begin 
at fixed $S_s$. Connection formulas for the Gauss hypergeometric function around its regular singular point at 
$v=0$ reveal two distinct types of local behavior:
\begin{equation}
f^{\Delta_{12},\Delta_{34}}_{\tau,S}(v) |_{v \to 0}
=
A^{\Delta_{12},\Delta_{34}}_{\tau,S}\,[1+O(v)]
+
B^{\Delta_{12},\Delta_{34}}_{\tau,S}\,v^{(\Delta_{12}-\Delta_{34})/2} [1+O(v)]
\, ,
\end{equation}
where, for the moment, we assume that $(\Delta_{12}-\Delta_{34})/2$ is not an integer. The accompanying coefficients are
\begin{align}
 A^{\Delta_{12},\Delta_{34}}_{\tau,S}
 &=
\frac{\Gamma(2 S + \tau)\Gamma \big(\frac{\Delta_{12} - \Delta_{34}}{2}\big)
}{
\Gamma\big(S + \frac{\tau + \Delta_{12}}{2}\big)\Gamma\big(S+\frac{\tau - \Delta_{34}}{2}\big)}
\, , \\
B^{\Delta_{12},\Delta_{34}}_{\tau,S}
&=
\frac{\Gamma(2S + \tau) \Gamma \big(\frac{\Delta_{34} - \Delta_{12}}{2}\big)
}{
\Gamma\big(S + \frac{\tau - \Delta_{12}}{2}\big)
\Gamma\big(S + \frac{\tau + \Delta_{34}}{2}\big)
}
\, .
\end{align}
When $\Delta_{12} = \Delta_{34}$, the second local behavior becomes logarithmic, while if $(\Delta_{12}-\Delta_{34})/2$ is an
integer, the two local solutions mix in the usual way, producing powers possibly accompanied by logarithms. 

Thus, a single fixed-spin collinear block has endpoint behavior governed by the two hypergeometric exponents, 
$0$ and $(\Delta_{12}-\Delta_{34})/2$, with the possible logarithmic modifications just mentioned. A stronger power-law singularity 
of the sum defining the full function $F (u,v)$ (notice the factor of $v^{K_2}$ on the right-hand side of Eq.\ \re{sChannelOPE}, which must 
be compensated as $v \to 0$) is therefore not a property of any particular fixed-spin block. Rather, it arises because, as $v\to0$, 
infinitely many spins contribute with $S v^\alpha = O(1)$, so the spin sum is not uniformly convergent near the endpoint 
\cite{Alday:2013cwa,Alday:2016njk}.
  
Let us now evaluate this large-spin tail more precisely and show how the spin integral reproduces the finite Born-level double-null limit of $F$. 
This behavior emerges after multiplying the large-spin asymptotic form of the collinear 
block by the free-field coefficient and performing the resulting spin integral. 

We consider the simultaneous limits
\begin{align}
\label{ScalingVariablex}
v \to 0 \, , \qquad S \to \infty \, , \qquad x \equiv S v^\alpha = O(1)
\, ,
\end{align}
in the $s$-channel. Here we have introduced a scaling variable $x$ for later use. The simplest way to impose \re{ScalingVariablex} 
on the collinear blocks is by using the Euler representation of the hypergeometric function and expanding its integrand around the endpoint 
\cite{Komargodski:2012ek,Li:2015rfa}. The existence of this double scaling limit fixes the exponent to be $\alpha=1/2$ and produces the 
following large-spin approximation for the collinear block
\begin{equation}
f^{\Delta_{12},\Delta_{34}}_{\tau,S}(v)
\logsim
v^{-(K_2-K_4)/2} 4^{S+\tau/2} \sqrt{S/\pi} \, \mathsf{K}_{K_2-K_4} (2S\sqrt{v})
\label{eq:block-bessel}
\, , 
\end{equation}
in terms of the modified Bessel function $\mathsf{K}$.

A quick naked-eye inspection shows, however, that the collinear block grows exponentially as $4^S$. This growth can only be compensated by the OPE 
coefficient $a_s$ in Eq.\ \re{sChannelOPEcoeff}. This is a zeroth-order effect in the coupling, so it must already be present in free theory. The proper scaling
of $a_s$ at zero coupling follows immediately from the requirement that $F$ approach a finite, nonzero value as $u, v \to 0$. 
Combining the results obtained so far, we find for $F$
\begin{align}
F (u,v)
\logsim
v^{(K_2 + K_4)/2}
\sum_{S_s} \left( 4^{S_s+\tau_s/2} \sqrt{S_s/\pi} \,  a_s \right)
u^{(\tau_s - K_2 - K_4)/2} 
\, \mathsf{K}_{K_2-K_4} (2S_s\sqrt{v})
\, .
\end{align}
Here we have reduced the sum in Eq.\ \re{sChannelOPE} to a sum over the spin $S_s$, since we retain only leading-twist operators in this OPE 
channel, $\tau_s = K_2 + K_4 + O(g^2)$. Quantum corrections generate nontrivial anomalous dimensions and hence the logarithmic spin 
dependence already mentioned in Eq.\ \re{ADsLogS} of the Introduction. Notice that the $u$-dependence disappears at $g=0$, as anticipated. 
As for the $v$-dependence, if
\begin{align}
a^{(0)}_s \sim 4^{-S_s}S_s^{K_2 + K_4 - 3/2} 
\, ,
\end{align}
then, after fixing the overall normalization, the Born approximation\footnote{Here we used the Euler--Maclaurin formula and replaced the summation over spin by an integral over the
scaling variable $x$ defined in Eq.\ \re{ScalingVariablex}.}
\begin{align}
F^{(0)} (u,v)
\logsim \frac{4}{\Gamma(K_2)\Gamma(K_4)}
\int_0^\infty dx \, x^{K_2 + K_4 - 1} \mathsf{K}_{K_2-K_4} (2 x)
=1
\, ,
\end{align}
is indeed finite when both cross ratios vanish.

\subsection{Null correlator}

Once the coupling is turned on, the twist of the exchanged operators becomes coupling dependent,
\begin{align}
\tau_s = K + \gamma_s (S_s, I; g)
\, ,
\end{align}
through the anomalous dimension $\gamma_s (S_s, I; g)$ of operators of bare twist $K = K_2 + K_4$. We write the corresponding 
OPE coefficients in terms of reduced coefficient functions $C_s$ as
\begin{align}
a_s = a^{(0)}_s C_s (S_s, I; g) 
\, .
\end{align}
We have also restored the dependence on an additional label $I$, which enumerates operators with degenerate bare dimensions.
We normalize the reduced coefficient function to its Born-level value,
\begin{align}
C_s (S_s, I; 0) = 1
\, .
\end{align}
With these ingredients, the large-spin $s$-channel conformal partial-wave expansion in the null limit of the interacting theory becomes
\footnote{To avoid cluttering formulas with powers of $\log 2$ and $\gamma_{\rm E}$, we set them to zero throughout this paper, except in the next section. They do not affect 
the transcendental structure of the null correlators studied here.}
\begin{align}
\label{FsChannel}
F (u,v)
\logeq 4 
\int_0^\infty dx \, \frac{x^{K_2 + K_4 - 1} \mathsf{K}_{K_2-K_4} (2 x)}{\Gamma (K_2)\Gamma(K_4)}  
\VEV{(4u)^{\ft12 \gamma_s (x/\sqrt{v}; g)} C_s (x/\sqrt{v}; g)}
\, ,
\end{align}
where we have introduced an average over the degenerate operators,
\begin{align}
\label{DefAverage}
\VEV{(4u)^{\ft12 \gamma_s (x/\sqrt{v}; g)} C_s (x/\sqrt{v}; g)}
\equiv \frac{1}{d_s} \sum_{I = 1}^{d_s} (4u)^{\ft12 \gamma_s (x/\sqrt{v}, I; g)} C_s (x/\sqrt{v}, I; g)
\, .
\end{align}
A similar partial-wave expansion holds in the $t$-channel. It can be obtained using Eq.\ \re{ExchangeCrossing}.
Below we apply this generic representation to the Born-normalized functions $\cG$ introduced in Sect.\ \ref{3LdataSection}.

\section{Large-spin anomalous dimensions}
\label{LargeSpinADsSection}

The null correlator \re{FsChannel} contains two ingredients that depend on the coupling: $\gamma_s (g)$ and
$C_s(g)$. While the coefficient function $C_s (g)$ is not universal, as it is sensitive to the external quantum numbers of the 
half-BPS operators, the anomalous dimension $\gamma_s (g)$ is. Thus, we focus on $\gamma_s (g)$ first. 

The anomalous dimensions of bare twist-$K$ operators, $\gamma_s \equiv \gamma_{K,I}(S;g)$ from now on, have received considerable 
attention in gauge theories, particularly in QCD, because they govern scaling violations in high-energy scattering involving 
hadrons; see, for instance, Ref.\ \cite{Belitsky:2005qn}. The discovery of integrability in QCD evolution equations 
\cite{Braun:1998id,Braun:1999te,Belitsky:1999qh,Belitsky:1999bf} allowed us to evaluate these anomalous 
dimensions by reformulating the setup as a spectral problem for noncompact spin chains. Crucially, the large-spin limit streamlined  
the underlying calculations by reducing them to a WKB-type analysis. We found that the anomalous dimensions of twist-$K$ operators 
occupy a band of width $(K-2)$ in units of the $\Gamma_{\rm cusp} (g) \log S$, i.e., 
\begin{align}
\label{BandWidth}
2 \Gamma_{\rm cusp} (g) \log S \leq  \gamma_{K,I}(S;g) \leq K \Gamma_{\rm cusp} (g) \log S
\, .
\end{align}
Here, $\Gamma_{\rm cusp} (g)$ is the so-called cusp anomalous dimension. The label $I$ enumerates the eigenvalues of the higher
conserved charges that encode the fine structure of the band.

In the present case of $\mathcal{N} = 4$ sYM, high-twist large-spin anomalous dimensions are known even at finite coupling. They inherit the 
structure found in QCD and take the general form
\begin{align}
\label{TwistKads}
\gamma_{K,I}(S;g) = 2 \Gamma_{\rm cusp} (g) \log (S \e^{\gamma_{\rm E}})
+ 
\Gamma_{\rm virt} (g) 
+
\Gamma_{K - 2,I} (g)
\, ,
\end{align}
where the cusp and virtual anomalous dimensions are known to all orders in perturbation theory 
\cite{Beisert:2006ez,Freyhult:2009my,Fioravanti:2009xt}. The first few terms needed below are
\begin{align}
\label{CuspVirtADs}
\Gamma_{\rm cusp} (g)
&=
\sum_{L = 1}g^{2L} \Gamma_{\rm cusp}^{(L)}
\nonumber\\
&
=
4g^2
-
\frac{4\pi^2}{3}g^4
+
\frac{44\pi^4}{45}g^6
+
g^8
\left(
-\frac{292\pi^6}{315}
-
32\zeta_3^2
\right)
+
\dots
\, ,
\\
\Gamma_{\rm virt} (g)
&=
\sum_{L = 2}g^{2L} \Gamma_{\rm virt}^{(L)}
\nonumber\\
&=
-24g^4\zeta_3
+
g^6
\left(
\frac{16\pi^2}{3}\zeta_3
+
160\zeta_5
\right)
+
g^8
\left(
-\frac{56\pi^4}{15}\zeta_3
-
\frac{80\pi^2}{3}\zeta_5
-
1400\zeta_7
\right)
+ \dots
\, . \nonumber
\end{align}
The last term in Eq.\ \re{TwistKads} encodes the dependence of the anomalous dimensions on the additional label $I$. For
twist-two operators, $\Gamma_0$ vanishes:
\begin{align}
\Gamma_0 = 0
\, .
\end{align}
For $K>2$, by contrast, $\Gamma_{K-2,I}$ is generally nonzero.

To grasp its physical content, let us offer a qualitative discussion of high-twist large-spin operators within the spin-chain paradigm.
We start with the lowest twist-two operator of spin $S = s_1 + s_2$,
\begin{align}
\label{Tw2State}
\ket{s_1, s_2} = {\rm tr} \big[ (i D_+)^{s_1} Z (i D_+)^{s_2} Z \big]
\, .
\end{align}
The left-hand side denotes a spin-chain state with $s_1$ and $s_2$ magnon excitations on the length-two noncompact chain. 
The right-hand side is a schematic representation\footnote{To belong to an irrep of the $\mathrm{SL}(2)$ collinear subgroup of the $\mathrm{SO}(2,4)$ conformal group,
additional terms must be included so that the covariant derivatives become arguments of Gegenbauer polynomials. This detail is
irrelevant to the intuitive picture we wish to convey.} of the single-trace Wilson operator formed by two complex 
elementary scalars $Z = \phi_1 + i \phi_2$ and the null projection of covariant derivatives $D_+ \equiv n \cdot D$. As the Lorentz spins $s_i$ 
are taken to infinity, these $Z$ fields can be viewed as creating or absorbing the gauge-field flux \cite{Mueller:1979ih,Radyushkin:1997ki}:
\begin{align}
\vev{0| (i D_+)^\infty Z |p} \sim \vev{0| W_{[\infty, 0]} Z |p}  = \vev{0| P \exp \left( i g_{\rm YM} \int_{[0, \infty]} d \sigma A_+ (n \sigma) \right) Z |p} 
\, .
\end{align}
Here $W_{[\infty, 0]}$ is the Wilson line along the light ray defined by the null vector $n$. In the limit $S \to \infty$, the particle-hole transformation 
maps the highly excited state \re{Tw2State} to a state with no excitations of the new type, the so-called {\sl holes}. From the spin-chain 
perspective, holes are gaps in the distribution of Bethe roots. From the gauge-theory vantage point, only a color flux of the gauge 
field $A$ remains between the scalar source and sink. This defines the new vacuum, $\ket{\rm GKP} = \lim_{s_\ell \to \infty} \ket{s_1, s_2}$. 
Its energy is the anomalous dimension $\gamma_{2}$ introduced in Eq.\ \re{TwistKads}. From the string perspective of AdS/CFT duality, $R = 2 \log S$ 
is the length of the rapidly rotating, folded closed string with orbital angular momentum $S$, known as the GKP string \cite{GubKlePol02}. 
Accordingly, $\Gamma_{\rm cusp} (g)$ acquires the meaning of its energy density, and the GKP string is the vacuum of the flux tube.

Moving on to twist three, the spin-chain state in the large-spin limit is mapped to a gauge flux with an insertion of a scalar $Z_{\rm ex}$,
\begin{align}
\label{FluxTubeOperators}
\ket{s_1, s_2, s_3} \stackrel{s_\ell \to \infty}{\to} \tr \big[ Z_{\rm sink} W_{[{\rm sink}, {\rm ex}]} Z_{\rm ex} W_{[{\rm ex}, {\rm source}]} Z_{\rm source} \big]
\, .
\end{align}
Thus, under the particle-hole transformation, $Z_{\rm ex}$ becomes a {\sl hole} excitation on top of the vacuum flux created by
$Z_{\rm source}$ and absorbed by $Z_{\rm sink}$. The energy of this state is $\gamma_3$, with $\Gamma_1$ in Eq.\ \re{TwistKads} being the energy of this hole,
\begin{align}
\Gamma_1 = \varepsilon (u)
\, .
\end{align}
The hole rapidity $u$ is related to the eigenvalues of a higher conserved charge in the large-spin limit. At leading order in 
perturbation theory, the momentum of the hole is twice its rapidity, $p = 2 u + O(g^2)$. The energy $\varepsilon$ of the hole excitation is 
known from the literature. It was determined at one loop in the mid-1990s in studies that uncovered the integrability of the QCD dilatation 
operator \cite{Braun:1999te,Belitsky:1999bf}. Within the current context of $\mathcal{N} = 4$ sYM, its leading-order form was discussed in
Ref.\ \cite{Belitsky:2006en}, generalized to three loops in Ref.\ \cite{Belitsky:2008mg}, and then nonperturbatively in Ref.\ 
\cite{Basso:2010in}. Its perturbative expansion reads
\begin{align}
\varepsilon (u) = \Gamma_{\rm cusp} (g) \varepsilon_{1}(u) + g^4 \varepsilon_{2}(u) + g^6 \varepsilon_{3}(u) +  \dots
\, , 
\end{align}
where \cite{Basso:2010in}
\begin{align}
\varepsilon_{1}(u)
&=
\frac{1}{2}
\left[
\psi\!\left(\ft{1}{2}+iu\right)
+
\psi\!\left(\ft{1}{2}-iu\right)
-
2\psi(1)
\right],
\\
\varepsilon_{2}(u)
&=
-\left[
\psi^{(2)}\!\left(\ft{1}{2}+iu\right)
+
\psi^{(2)}\!\left(\ft{1}{2}-iu\right)
+
12\zeta_{3}
\right],
\\
\varepsilon_{3}(u)
&=
\frac{1}{3}
\left[
\ft{1}{2}
[
\psi^{(4)}\!\left(\ft{1}{2}+iu\right)
+
\psi^{(4)}\!\left(\ft{1}{2}-iu\right)
]
+
\pi^{2}
\left(
\psi^{(2)}\!\left(\ft{1}{2}+iu\right)
+
\psi^{(2)}\!\left(\ft{1}{2}-iu\right)
\right)
\right.
\\
&\hspace{2.5cm}
\left.
+
12\zeta_{3}
\left(
\psi^{(1)}\!\left(\ft{1}{2}+iu\right)
+
\psi^{(1)}\!\left(\ft{1}{2}-iu\right)
\right)
+
8\left(\pi^{2}\zeta_{3}+30\zeta_{5}\right)
\right]\, . \nonumber 
\end{align}

At leading order in the large-spin expansion, the integrable dynamics of hole excitations on top of the GKP vacuum gives the $(K-2)$-hole energy as a sum of the individual energies,
\begin{align}
\Gamma_{K - 2} (g)
=
\sum_{\ell=1}^{K-2} \varepsilon (u_\ell)
\, .
\end{align}
This expression passes a number of checks. First, setting all hole rapidities to zero recovers the lowest trajectory in the spectrum,
\begin{align}
\gamma_K 
&
= 2 \Gamma_{\rm cusp} \log (S \e^{\gamma_{\rm E}})
+ 
\Gamma_{\rm virt} (g) 
\nonumber\\
&
+
(K-2)
\left[
- 2 \Gamma_{\rm cusp} (g) \log 2 + 16 g^4 \zeta_3 - \ft83 g^6 [\pi^2 \zeta_3 + 63 \zeta_5]
+ \dots
\right]
\, .
\end{align}
Second, in the large-$u$ limit, the hole energy grows as
\begin{align}
\varepsilon (u) \stackrel{u \to \infty}{=} \Gamma_{\rm cusp} (g) \log (u \e^{\gamma_{\rm E}}) + \ft12 \Gamma_{\rm virt} (g) + O(u^{-2})
\, .
\end{align}
In this kinematic region, the hole enters the quasiclassical domain and moves the anomalous dimension one step into the interior of the 
band \re{BandWidth}:
\begin{align}
\gamma_K \stackrel{u_{K-2}\sim S\to\infty}{=} 3 \Gamma_{\rm cusp} (g) \log (S \e^{\gamma_{\rm E}})
+ 
\ft32 \Gamma_{\rm virt} (g) 
+
\sum_{\ell=1}^{K-3} \varepsilon (u_\ell)
\, .
\end{align}
If all rapidities scale with $S$, we reach the top of the spectrum.

To conclude this brief recollection, the distribution of anomalous dimensions in the band becomes dense at large spin. We can then
pass from the discrete quantum numbers $I$ that enumerate the operators to a continuous description in terms of hole rapidities, 
$I \leftrightarrow \{ u_\ell\}$: $\Gamma_{K-2, I} \leftrightarrow \Gamma_{K-2} (u_1, \dots, u_{K-2})$. We expect the main contribution 
to the averages \re{DefAverage}, introduced in the previous section, to come from operators near the lower boundary. Their contribution 
is governed by an appropriate integration measure over hole rapidities. This measure is not known and must be determined. We do so 
perturbatively in this work.

\section{Crossing}

The second ingredient entering the averages is the set of coefficient functions $C_{i}$. Since each arises from a product of 
three-point functions involving the external states and the propagating state, these functions are not universal and must be determined case by case.
In this section, we elucidate their structure using only minimal assumptions about their behavior at large exchanged spin, together with the 
conformal bootstrap. The central equation is the crossing relation \re{GenericCrossing}.
In the null limit, it takes the form
\begin{align}
\label{CrossingEq}
&
\int_0^\infty dx \, \frac{x^{K_2 + K_4 - 1} \mathsf{K}_{K_2-K_4} (2 x)}{\Gamma (K_2)\Gamma(K_4)}  
\VEV{(4u)^{\ft12 \gamma_s (x/\sqrt{v}; g)} C_s (x/\sqrt{v}; g)}
\\
&
=
\int_0^\infty dx \, \frac{x^{K_1 + K_3 - 1} \mathsf{K}_{K_1-K_3} (2 x)}{\Gamma (K_1)\Gamma(K_3)}  
\VEV{(4v)^{\ft12 \gamma_t (x/\sqrt{u}; g)} C_t (x/\sqrt{u}; g)}
\, . \nonumber
\end{align}

\subsection{Large-spin coefficient function}

At each perturbative order, the expansion of the Sudakov exponent inside the average yields a finite polynomial in the logarithm
of the argument of the anomalous dimension. It is therefore natural to assume the same structure for the coefficient 
function, since the two behaviors must be commensurate. Thus,
\begin{align}
\label{CoeffFunctPT}
C_i (x; g) = 1 + \sum_{p = 1}^L g^{2 p} \sum_{r = 0}^p C_i^{(p, r)} \log^r x + O(g^{2L+2})
\, .
\end{align}
Here and below, it is understood that the expansion coefficients $C_i^{(p, r)}$ inherit the label $I$ that counts
degenerate operators for a given value of spin.

Finally, we introduce the Bessel moments of logarithms,
\begin{align}
m_K (p) = \int_0^\infty dx\, \frac{4 x^{K-1}}{\Gamma (K-1)} \log^{p - 1} (2 x) \mathsf{K}_{K-2} (2 x)
\, ,
\end{align}
since these will be generated by perturbative expansions. Their explicit form is deferred to Appendix \ref{BesselMomApp}.

We now present solutions to the crossing equation, beginning with the simplest case, the $\cG_{[2,2]}$ correlator,
which was analyzed in full in Ref.\ \cite{Alday:2013cwa}. We then move to the more involved cases $\cG_{[K,2]}$ 
and $\cG_{[K,K]}$. For the twist-two case, we display the solution through three loops. For the higher-twist cases,
we show the first two loops here and relegate higher orders to the corresponding appendices, since the resulting formulas are rather lengthy and not particularly illuminating.

\subsection{Twist-2--twist-2 correlator}

We begin with $\cG_{[2,2]}$. The partial-wave expansion of this null correlator contains only twist-two operators in both the
$s$- and $t$-channels. These are not degenerate, and no averages show up in the crossing equation. 

\subsubsection{Solution to crossing}

At one loop, we find
\begin{align}
C_s^{(1,1)}
&=-2\Gamma_{\rm cusp}^{(1)}m_2(2),
\\
C_t^{(1,1)}
&=-2\Gamma_{\rm cusp}^{(1)}m_2(2)
\, .
\end{align}
At two loops, the solution is
\begin{align}
C_s^{(2,1)}
&=
-\Gamma_{\rm virt}^{(2)}
-2C_s^{(1,0)}\Gamma_{\rm cusp}^{(1)}m_2(2)
-2\Gamma_{\rm cusp}^{(2)}m_2(2),
\\
C_s^{(2,2)}
&=
2\bigl(\Gamma_{\rm cusp}^{(1)}\bigr)^2m_2(3),
\\
C_t^{(2,1)}
&=
-\Gamma_{\rm virt}^{(2)}
-2C_s^{(1,0)}\Gamma_{\rm cusp}^{(1)}m_2(2)
-2\Gamma_{\rm cusp}^{(2)}m_2(2),
\\
C_t^{(2,2)}
&=
2\bigl(\Gamma_{\rm cusp}^{(1)}\bigr)^2m_2(3).
\end{align}
At three loops, it becomes
\begin{align}
C_s^{(3,0)}
&=C_t^{(3,0)},
\\
C_s^{(3,1)}
&=
-
C_s^{(1,0)}\Gamma_{\rm virt}^{(2)}
-
2C_t^{(2,0)}\Gamma_{\rm cusp}^{(1)}m_2(2)
-
2C_s^{(1,0)}\Gamma_{\rm cusp}^{(2)}m_2(2)
\nonumber\\
&
-
\Gamma_{\rm virt}^{(3)}
-
2\Gamma_{\rm cusp}^{(3)}m_2(2)
-
2\Gamma_{\rm virt}^{(2)} \Gamma_{\rm cusp}^{(1)}m_2^2(2)
+
2\Gamma_{\rm virt}^{(2)} \Gamma_{\rm cusp}^{(1)}m_2(3),
\\
C_s^{(3,2)}
&=
2C_s^{(1,0)}  \bigl(\Gamma_{\rm cusp}^{(1)}\bigr)^2m_2(3)
+
2\Gamma_{\rm virt}^{(2)} \Gamma_{\rm cusp}^{(1)}m_2(2)
+
4\Gamma_{\rm cusp}^{(1)} \Gamma_{\rm cusp}^{(2)}m_2(3),
\nonumber\\
C_s^{(3,3)}
&=
-
\ft{4}{3} \bigl(\Gamma_{\rm cusp}^{(1)}\bigr)^3m_2(4),
\\
C_t^{(3,1)}
&=
-
C_s^{(1,0)}\Gamma_{\rm virt}^{(2)}
-
2C_t^{(2,0)}\Gamma_{\rm cusp}^{(1)}m_2(2)
-
2C_s^{(1,0)}\Gamma_{\rm cusp}^{(2)}m_2(2)
\nonumber\\
&\quad
-\Gamma_{\rm virt}^{(3)}
-
2 \Gamma_{\rm cusp}^{(3)}m_2(2)
-
2\Gamma_{\rm virt}^{(2)} \Gamma_{\rm cusp}^{(1)}m_2^2(2)
+
2\Gamma_{\rm virt}^{(2)} \Gamma_{\rm cusp}^{(1)}m_2(3),
\\
C_t^{(3,2)}
&=
2C_s^{(1,0)}
 \bigl(\Gamma_{\rm cusp}^{(1)}\bigr)^2m_2(3)
 +
2\Gamma_{\rm virt}^{(2)}
 \Gamma_{\rm cusp}^{(1)}m_2(2)
+
4\Gamma_{\rm cusp}^{(1)}
 \Gamma_{\rm cusp}^{(2)}m_2(3),
\\
C_t^{(3,3)}
&=
-\ft{4}{3}
 \bigl(\Gamma_{\rm cusp}^{(1)}\bigr)^3m_2(4)
 \, .
\end{align}
Crossing fixes only the equality of the log-free terms; it does not determine their common value:
\begin{align}
C_s^{(L,0)} = C_t^{(L,0)}
\, .
\end{align}
With these results in hand, we can determine the coefficients accompanying the logarithms of the cross ratios in \re{eq:cij-definition}. 
In the following formulas, all right-hand sides are understood through $O(g^6)$:
\begin{align}
c_{30}^{[2,2]}=c_{03}^{[2,2]}
&=
\ft{1}{6}\Gamma_{\rm cusp}^{\,3}
\left[
 2m_2^3(2)
 -3m_2(2)m_2(3)
 +m_2(4)
\right],
\\[4pt]
c_{20}^{[2,2]}=c_{02}^{[2,2]}
&=
-\ft{1}{2}\Gamma_{\rm cusp}^{\,2}
\left[
 m_2^2(2)-m_2(3)
\right]
\nonumber\\
&\quad
-\ft{1}{2}\Gamma_{\rm cusp}^{\,3}
\left[
 4m_2^4(2)
 -6m_2^2(2)m_2(3)
 +2m_2(2)m_2(4)
\right],
\\
c_{10}^{[2,2]}=c_{01}^{[2,2]}
&=
\ft{1}{2}\Gamma_{\rm virt}
+
\Gamma_{\rm cusp}m_2(2)
+
\ft{1}{2}\Gamma_{\rm virt}\Gamma_{\rm cusp}
\left[
 2m_2^2(2)-2m_2(3)
\right]
\nonumber\\
&\quad
+\ft{1}{2}\Gamma_{\rm cusp}^{\,2}
\left[
 4m_2^3(2)-4m_2(2)m_2(3)
\right]
\nonumber\\
&\quad
+\ft{1}{2}\Gamma_{\rm cusp}^{\,3}
\left[
 8m_2^5(2)
 -8m_2^3(2)m_2(3)
 -4m_2(2)m_2^2(3)
 +4m_2(3)m_2(4)
\right],
\\[4pt]
c_{11}^{[2,2]}
&=
-\ft{1}{2}\Gamma_{\rm cusp}
-\ft{1}{2}\Gamma_{\rm cusp}^{\,3}
\left[
 4m_2^4(2)
 -8m_2^2(2)m_2(3)
 +4m_2^2(3)
\right].
\end{align}
Substituting the values of the Bessel moments from Appendix \ref{BesselMomApp} and the anomalous dimensions \re{CuspVirtADs}, 
we reproduce the data \re{eq:cij-2222}.

\subsubsection{Resummed formula}

Based on the explicit solution above, an educated guess was put forward \cite{Alday:2010zy,Alday:2013cwa,Korchemsky:2019nzm} 
for the all-order resummed form of the $\cG_{[2,2]}$ correlator. It takes the concise form \cite{Korchemsky:2019nzm}
\begin{equation}
\cG_{[2,2]}
=
H (g)
\int_0^\infty \frac{dy_1}{y_1}
\int_0^\infty \frac{dy_2}{y_2}\,
S\!\left(\frac{u}{y_1},\frac{v}{y_2}\right)
f_2(y_1)f_2(y_2)
\, .
\end{equation}
Here $S(u,v)$ is the Sudakov form factor,
\begin{equation}
S(u,v)
=
\exp\!\left(
-\frac{1}{2}\Gamma_{\rm cusp}\log u\log v
+\frac{1}{2}\Gamma_{\rm virt}\log (u v)
\right) ,
\end{equation}
while $f_2(y)$ is the large-spin limit of the collinear conformal blocks \re{eq:block-bessel}, with all dependence on the
conformal cross ratios stripped off:
\begin{equation}
f_2(y)=2y\mathsf{K}_0(2\sqrt{y})
\, .
\end{equation}
The overall function of the 't Hooft coupling, $H(g)$, accounts for log-free contributions to
the correlator in question. Taking into account the identity
\begin{equation}
y^{-u\partial_u}h(u)=h(u/y),
\end{equation}
and the Mellin moments of the recoil functions,
\begin{align}
\qquad
\int_0^\infty\frac{dy}{y}\,y^{-j}f_2(y)=\Gamma^2(1-j)
\, ,
\end{align}
we can equivalently write the relation above as \cite{Alday:2013cwa}
\begin{align}
\cG_{[2,2]}
=
H (g)
\Gamma^2(1-u\partial_u)\Gamma^2(1-v\partial_v)S(u,v)
\, .
\end{align}
Finally, employing the integral representation of the Euler gamma function,
\begin{equation}
\Gamma(1-u\partial_u)
=
\int_0^\infty ds\,e^{-s}s^{-u\partial_u}
=
\int_{-\infty}^{\infty}d\sigma\,
\Phi_1(\sigma)e^{\sigma u\partial_u},
\end{equation}
where $\sigma=-\log s$ and $\Phi_1(\sigma)=e^{-\sigma-e^{-\sigma}}$, we obtain another representation for the same null correlator \cite{Alday:2010zy}
\begin{align}
\cG_{[2,2]}
=
H (g)
\int_{-\infty}^\infty d\sigma_1\,d\sigma_2
\int_{-\infty}^\infty d\tau_1\,d\tau_2\,
\Phi_1(\sigma_1)\Phi_1(\sigma_2)
\Phi_1(\tau_1)\Phi_1(\tau_2)
S\!\left(e^{\sigma_1+\sigma_2}u,
e^{\tau_1+\tau_2}v\right)
\, .
\end{align}
Here the $\Phi$ functions encode the recoil, or backreaction, of the gauge-field flux sourced by the highly energetic particles forming the
correlator frame on those same particles.

\subsection[Twist-K--twist-2 correlator]{Twist-\textbf{\textit K}--twist-2 correlator}

We next turn to the correlator $\cG_{[K,2]}$, which allows higher-twist operators to propagate in the $s$-channel. The $t$-channel,
as before, is dominated by the lowest twist-two operators. This correlator was considered in Ref.\ \cite{Alday:2013cwa}; however, as we 
will now demonstrate, the conclusions reached there are incomplete. The new feature relative to the previous subsection is
that the $s$-channel operators are degenerate at zero coupling and must be enumerated by an additional label $I$. 
Consequently, when solving the crossing equation \re{CrossingEq}, the $s$-channel coefficients enter the averages \re{DefAverage}. 
The solution to crossing is straightforward, but it comes with a few subtleties.

\subsubsection{Solution to crossing}

At one loop, nothing unusual happens. The only difference from the twist-two case is that both the $s$-channel coefficient function
and the corresponding higher-twist anomalous dimension $\Gamma_{K-2}$ enter under averages. This immediately tells us that 
crossing cannot directly constrain the microscopic dynamics governing them: it has access only to their averaged characteristics.
We find
\begin{align}
\vev{C_s^{(1, 1)}} 
= 
-2 \Gamma_{\rm cusp}^{(1)} m_2 (2), 
\quad 
C_t^{(1, 1)} = -\vev{\Gamma_{K-2}^{(1)}} - 2 \Gamma_{\rm cusp}^{(1)} m_K (2)
\, .
\end{align}
As before, crossing constrains only the difference $\vev{C_s^{(1, 0)}}-C_t^{(1, 0)}$:
\begin{align}
C_t^{(1,0)} - \vev{C_s^{(1,0)}}
&=
\vev{\Gamma_{K-2}^{(1)}} m_2(2)
\, .
\end{align}
Moving on to two loops, we find 
\begin{align}
C_t^{(2, 2)}
&
= 
\ft{1}{2} \vev{\big(\Gamma_{K-2}^{(1)}\big)^2} 
+ 
2 \vev{\Gamma_{K-2}^{(1)}} \Gamma_{\rm cusp}^{(1)} m_K (2) + 2 \left(\Gamma_{\rm cusp}^{(1)}\right)^2 m_K (3)
\, , \\
C_t^{(2, 1)}
&
= 
- \vev{C_s^{(1, 0)} \Gamma_{K-2}^{(1)}} 
- 
\vev{C_s^{(1, 1)} \Gamma_{K-2}^{(1)}} m_K (2) 
- 
\vev{\big(\Gamma_{K-2}^{(1)}\big)^2} m_2 (2) 
\nonumber\\
&
- 4 \vev{\Gamma_{K-2}^{(1)}} \Gamma_{\rm cusp}^{(1)} m_2 (2) m_K (2)
- 
2 \Gamma_{\rm cusp}^{(2)} m_K (2) 
- 
\vev{\Gamma_{K-2}^{(2)}} - \Gamma_{\rm virt}^{(2)} 
\, , \\
\vev{C_s^{(2, 2)}} 
&= 
2 \left(\Gamma_{\rm cusp}^{(1)}\right)^2 m_2 (3)
\, , \\
\vev{C_s^{(2, 1)}} 
&
= - \Gamma_{\rm virt}^{(2)} - 2 \Gamma_{\rm cusp}^{(2)} m_2 (2) 
- 
2 \vev{\Gamma_{K-2}^{(1)}} \Gamma_{\rm cusp}^{(1)} \left(m_2 (2)\right)^2 
+ 
2 \vev{\Gamma_{K-2}^{(1)}} \Gamma_{\rm cusp}^{(1)} m_2 (3)
\, .
\end{align}
In addition, one equation fixes the mixed average,
\begin{align}
\vev{C_s^{(1, 1)} \Gamma_{K-2}^{(1)}} 
&
= - 2 \vev{\Gamma_{K-2}^{(1)}} \Gamma_{\rm cusp}^{(1)} m_2 (2)
\, .
\end{align}
Finally, just as at one loop, crossing fixes only the difference between the constants $\vev{C_s^{(2, 0)}}$ and $C_t^{(2, 0)}$:
\begin{align}
C_t^{(2,0)}
-
\vev{C_s^{(2,0)}}
&=
\vev{C_s^{(1,0)}\Gamma_{K-2}^{(1)}}m_2(2)
+
\vev{\Gamma_{K-2}^{(2)}}m_2(2)
+
\Gamma_{\rm virt}^{(2)}m_2(2)
+
\vev{\big(\Gamma_{K-2}^{(1)}\big)^2}m_2^2(2)
\nonumber\\
&\quad
-
\ft{1}{2}\vev{\big(\Gamma_{K-2}^{(1)}\big)^2}m_2(3)
-
\Gamma_{\rm virt}^{(2)}m_K(2)
\, .
\end{align}
The three- and four-loop solutions are relegated to Appendix \ref{2Kcorr34loopsApp} because of their length. 

A quick inspection reveals a useful general pattern: the solution contains only averages of the expansion coefficients 
$C_{s,t}^{(L,\ell \neq 0)}$ and mixed averages with anomalous dimensions from the set
\begin{align}
\{
\vev{C_s^{(1, 1)} \big(\Gamma_{K-2}^{(1)}\big)^{n_1}} , \vev{C_s^{(2, \ell_2)}  \big(\Gamma_{K-2}^{(1)} \big)^{n_2}}, \vev{C_s^{(3, \ell_3)} \Gamma_{K-2}^{(1)}} 
\}
\, ,
\end{align}
with
\begin{align}
\ell_2 = 1,2, \qquad \ell_3 = 1,2,3
\, .
\end{align}
At four-loop order, the possible exponents are $n_1=1,2,3$ and $n_2=1,2$. What the crossing 
relations do not provide are (i) separate solutions for the lowest expansion coefficients $C_{s,t}^{(L,0)}$ at each loop order and (ii) constraints 
on their averages with powers of $\Gamma_{K-2}$. Thus, they allow a common constant shift of $C_{s}^{(L,0)}$ and 
$C_{t}^{(L,0)}$. Moreover, they permit shifts of $C_{s}^{(L,0)}$ that vanish upon averaging.

\subsubsection{Gauge freedom}

Let us take a closer look at the freedom left by crossing. It fixes the averages of the logarithmic expansion coefficients uniquely, including
\begin{align}
\vev{C_s^{(L, \ell)} \big( \Gamma_{K-2}^{(1)} \big)^n}|_{0 < \ell \leq L}
=
\mbox{fixed}
\, ,
\end{align}
but leaves the average of the lowest coefficient $C_s^{(L, 0)}$ at each loop order unconstrained. It enters only through its difference
from the lowest expansion coefficient in the other channel. Even if one insists that the latter be fixed by $s$-channel data, all one can say about
$C_s^{(L, 0)}$ is that
\begin{align}
\vev{C_s^{(L, 0)}} = \vev{C_s^{\prime (L, 0)}}
\, ,
\end{align}
under a {\sl minimal gauge} transformation,
\begin{align}
\label{MinimalUTgauge}
C_s^{(L, 0)} 
\to C_s^{\prime (L, 0)} = C_s^{(L, 0)} 
+ 
\sum_{k = 1}^{k_{\rm max}} \textcolor{red}{r_{k, L}} \big[( \Gamma^{(1)}_{K-2})^k - \vev{( \Gamma^{(1)}_{K-2})^k}\big]
\, ,
\end{align}
for arbitrary values of the gauge parameters $r_{k, L}$. This is fully consistent with crossing. The number $k_{\rm max}$ of gauge 
parameters is not fixed a priori. Below we propose a criterion based on the proper transcendental structure of the correlators. We call the resulting choice the {\sl uniform
transcendental gauge}, or UT gauge, and also discuss its limitations. The good news is that only a few such parameters are needed at each perturbative order.

\subsubsection{Four-loop correlator}

With the gauge freedom understood, we can now give a four-loop solution for $\cG_{[K,2]}$. To this end, we choose a reference frame
in which the two-loop coefficient $C_s^{(2,0)}$ factorizes and parametrize gauge transformations away from it as
\begin{align}
\vev{C_s^{(2,0)}( \Gamma_{K-2}^{(1)})^m}
\to
\vev{C_s^{\prime (2,0)}(\Gamma_{K-2}^{(1)})^m}
&
=
\vev{C_s^{(2,0)}} \vev{ (\Gamma_{K-2}^{(1)})^m} 
\\
&
+ 
\sum_{k = 1}^2 r_{k, 2} \left[
\vev{( \Gamma^{(1)}_{K-2})^{k+m}}
- \vev{( \Gamma^{(1)}_{K-2})^k} \vev{( \Gamma^{(1)}_{K-2})^m}
\right]
\, . \nonumber
\end{align}
As we will advocate in Sect.\ \ref{4LG2KminimalUTsection}, it suffices to keep only two gauge parameters, $r_{1,2}$ and $r_{2,2}$, which 
form a minimal set. In this way, the freedom in the lowest coefficients $C_{s}^{(L,0)}$, left unfixed by crossing, is parametrized by the numerical
parameters $r_{k,2}$. We will see below that agreement with data requires $C_s^{(1,\ell)}$ to be independent of $I$ for $\ell=0,1$, while at two loops only 
$C_s^{(2,\ell)}$ with $\ell=1,2$ have this property. No further constraints on higher-order coefficients are needed to the order at which we work,
although additional gauge freedom would enter at five loops and beyond.

To cast the solution for $\cG_{[K,2]}$ in a reasonably concise form, it is useful to introduce cumulants. The first four
needed for our purposes are
\begin{align}
\vev{\vev{f}}_1
&=
\vev{f},
\\
\vev{\vev{f}}_2
&=
\vev{f^2}
-\vev{f}^2,
\\
\vev{\vev{f}}_3
&=
\vev{f^3}
-3\vev{f^2}\vev{f}
+2\vev{f}^3,
\\
\vev{\vev{f}}_4
&=
\vev{f^4}
-4\vev{f^3}\vev{f}
-3\vev{f^2}^2
+12\vev{f^2}\vev{f}^2
-6\vev{f}^4
\, ,
\end{align}
for a test function $f$.

Returning to the logarithm of the correlator in \re{eq:cij-definition}, we find the following nonvanishing 
coefficients through four loops:
\begin{align}
c_{21}^{[K,2]}
&=
\ft{1}{8}\Gamma_{\rm cusp}
\left(
m_2^2(2)-m_2(3)
\right)
\left[
\vev{\vev{\Gamma_{K-2}}}_3
+
8\Gamma_{\rm cusp}^3
\left(
2m_K^3(2)
-
3m_K(2)m_K(3)
+
m_K(4)
\right)
\right]
\, ,
\\
c_{12}^{[K,2]}
&=
-\ft{1}{4}\Gamma_{\rm cusp}^2
\left(
2m_2^3(2)
-
3m_2(2)m_2(3)
+
m_2(4)
\right)
\left[
\vev{\vev{\Gamma_{K-2}}}_2
+
4\Gamma_{\rm cusp}^2
\left(
-
m_K^2(2)
+
m_K(3)
\right)
\right]
\, ,
\\
c_{40}^{[K,2]}
&=
\ft{1}{384}\vev{\vev{\Gamma_{K-2}}}_4
\\
&
-
\ft{1}{24}\Gamma_{\rm cusp}^4
\left(
6m_K^4(2)
-
12m_K^2(2)m_K(3)
+
3m_K^2(3)
+
4m_K(2)m_K(4)
-
m_K(5)
\right)
\, ,
\nonumber\\
c_{04}^{[K,2]}
&=
\ft{1}{24}\Gamma_{\rm cusp}^4
\left(
-
6m_2^4(2)
+
12m_2^2(2)m_2(3)
-
3m_2^2(3)
-
4m_2(2)m_2(4)
+
m_2(5)
\right)
\, ,
\\
c_{30}^{[K,2]}
&=
\ft{1}{48}\vev{\vev{\Gamma_{K-2}}}_3
+
\ft{1}{6}\Gamma_{\rm cusp}^3
\left(
2m_K^3(2)
-
3m_K(2)m_K(3)
+
m_K(4)
\right)
\nonumber\\
&
+
\ft{1}{6}\Gamma_{\rm cusp}^4
\left[
12m_2(2)m_K^4(2)
-
24m_2(2)m_K^2(2)m_K(3)
+
8m_2(2)m_K(2)m_K(4)
\right.
\nonumber\\
&\qquad\left.
+
2m_2(2)
\left(
3m_K^2(3)
-
m_K(5)
\right)
\right]
\, ,
\\
c_{03}^{[K,2]}
&=
\ft{1}{6}\Gamma_{\rm cusp}^3
\left(
2m_2^3(2)
-
3m_2(2)m_2(3)
+
m_2(4)
\right)
\nonumber\\
&
+
\ft{1}{6}\Gamma_{\rm cusp}^3
\vev{\vev{\Gamma_{K-2}}}_1
\left(
6m_2^4(2)
-
12m_2^2(2)m_2(3)
+
3m_2^2(3)
+
4m_2(2)m_2(4)
-
m_2(5)
\right)
\nonumber\\
&\quad
+
\ft{1}{6}\Gamma_{\rm cusp}^4
\left(
12m_2^4(2)m_K(2)
-
24m_2^2(2)m_2(3)m_K(2)
+
6m_2^2(3)m_K(2)
\right.
\nonumber\\
&\qquad\left.
+
8m_2(2)m_2(4)m_K(2)
-
2m_2(5)m_K(2)
\right)
\, , \\
c_{20}^{[K,2]}
&=
\ft{1}{8}\vev{\vev{\Gamma_{K-2}}}_2
+
\left(
m_2^2(2)-m_2(3)
\right)
\left(
m_K^2(2)-m_K(3)
\right)
\vev{\vev{\Gamma_{K-2}}}_2\Gamma_{\rm cusp}^2
\nonumber\\
&
+
\ft{1}{4}
\left(
-m_2^2(2)m_K(2)
+
m_2(3)m_K(2)
\right)
\Gamma_{\rm cusp}\vev{\vev{\Gamma_{K-2}}}_3
\nonumber\\
&
-
\left(
m_2^2(2)-m_2(3)
\right)
\left(
2m_K^3(2)
-
3m_K(2)m_K(3)
+
m_K(4)
\right)
\vev{\vev{\Gamma_{K-2}}}_1\Gamma_{\rm cusp}^3
\nonumber\\
&
+
\Gamma_{\rm cusp}^2
\left(
-\ft{1}{2}m_K^2(2)
+
\ft{1}{2}m_K(3)
\right)
+
\Gamma_{\rm virt}\Gamma_{\rm cusp}^2
\left(
-m_K^3(2)
+
\ft{3}{2}m_K(2)m_K(3)
-
\ft{1}{2}m_K(4)
\right)
\nonumber\\
&
+
\Gamma_{\rm cusp}^3
\left(
-2m_2(2)m_K^3(2)
+
3m_2(2)m_K(2)m_K(3)
-
m_2(2)m_K(4)
\right)
\nonumber\\
&
+
\Gamma_{\rm cusp}^4
\left(
-6m_2^2(2)m_K^4(2)
+
10m_2^2(2)m_K^2(2)m_K(3)
+
2m_2(3)m_K^2(2)m_K(3)
\right.
\nonumber\\
&\qquad\left.
-
2m_2^2(2)m_K^2(3)
-
m_2(3)m_K^2(3)
-
2m_2^2(2)m_K(2)m_K(4)
\right.
\nonumber\\
&\qquad\left.
-
2m_2(3)m_K(2)m_K(4)
+
m_2(3)m_K(5)
\right)
\nonumber\\
&
+
\ft{1}{8}g^2\vev{\vev{\Gamma_{K-2}}}_3 \textcolor{red}{r_{1,2}}
+
\ft{1}{8}
\left(
\vev{\vev{\Gamma_{K-2}}}_4
+
2\vev{\vev{\Gamma_{K-2}}}_2^2
+
2\vev{\vev{\Gamma_{K-2}}}_1\vev{\vev{\Gamma_{K-2}}}_3
\right)\textcolor{red}{r_{2,2}}
\, ,
\\
c_{02}^{[K,2]}
&=
\Gamma_{\rm cusp}^2
\left(
-\ft{1}{2}m_2^2(2)
+
\ft{1}{2}m_2(3)
\right)
+
\Gamma_{\rm virt}\Gamma_{\rm cusp}^2
\left(
-m_2^3(2)
+
\ft{3}{2}m_2(2)m_2(3)
-
\ft{1}{2}m_2(4)
\right)
\nonumber\\
&
+
\Gamma_{\rm cusp}^2\vev{\vev{\Gamma_{K-2}}}_1^2
\left(
-\ft{3}{2}m_2^4(2)
+
3m_2^2(2)m_2(3)
-
\ft{3}{4}m_2^2(3)
-
m_2(2)m_2(4)
+
\ft{1}{4}m_2(5)
\right)
\nonumber\\
&
+
\Gamma_{\rm cusp}^2\vev{\vev{\Gamma_{K-2}}}_2
\left(
-m_2^4(2)
+
2m_2^2(2)m_2(3)
-
\ft{1}{4}m_2^2(3)
-
m_2(2)m_2(4)
+
\ft{1}{4}m_2(5)
\right)
\nonumber\\
&
+
\Gamma_{\rm cusp}^3
\left(
-2m_2^3(2)m_K(2)
+
3m_2(2)m_2(3)m_K(2)
-
m_2(4)m_K(2)
\right)
\nonumber\\
&
+
\vev{\vev{\Gamma_{K-2}}}_1
\left[
\Gamma_{\rm cusp}^2
\left(
-m_2^3(2)
+
\ft{3}{2}m_2(2)m_2(3)
-
\ft{1}{2}m_2(4)
\right)
\right.
\nonumber\\
&\qquad\left.
+
\Gamma_{\rm cusp}^3
\left(
-6m_2^4(2)m_K(2)
+
12m_2^2(2)m_2(3)m_K(2)
-
3m_2^2(3)m_K(2)
\right.
\right.
\nonumber\\
&\qquad\left.\left.
-
4m_2(2)m_2(4)m_K(2)
+
m_2(5)m_K(2)
\right)
\right]
\nonumber\\
&
+
\Gamma_{\rm cusp}^4
\left(
-6m_2^4(2)m_K^2(2)
+
10m_2^2(2)m_2(3)m_K^2(2)
-
2m_2^2(3)m_K^2(2)
\right.
\nonumber\\
&\qquad\left.
-
2m_2(2)m_2(4)m_K^2(2)
+
2m_2^2(2)m_2(3)m_K(3)
-
m_2^2(3)m_K(3)
\right.
\nonumber\\
&\qquad\left.
-
2m_2(2)m_2(4)m_K(3)
+
m_2(5)m_K(3)
\right)
\, , \\
c_{01}^{[K,2]}
&=
\ft{1}{2} \Gamma_{\rm virt}
+
\Gamma_{\rm cusp}m_2(2)
+
\Gamma_{\rm cusp}
\big[
\Gamma_{\rm virt}
+
\vev{\vev{\Gamma_{K-2}}}_1
+
2 \Gamma_{\rm cusp}
m_K(2)
\big]
\left(
m_2^2(2)-m_2(3)
\right)
\nonumber\\
&
+
\Gamma_{\rm cusp}
\big[
\vev{\vev{\Gamma_{K-2}}}_2
+
\vev{\vev{\Gamma_{K-2}}}_1^2
+
2
\Gamma_{\rm virt}\vev{\vev{\Gamma_{K-2}}}_1
\big]
\left(
m_2^3(2)
-
\ft{3}{2}m_2(2)m_2(3)
+
\ft{1}{2}m_2(4)
\right)
\nonumber\\
&
+
\Gamma_{\rm cusp}\vev{\vev{\Gamma_{K-2}}}_1\vev{\vev{\Gamma_{K-2}}}_2
\big(
3m_2^4(2)
-
6m_2^2(2)m_2(3)
\nonumber\\
&\qquad
+
\ft{3}{2}m_2^2(3)
+
2m_2(2)m_2(4)
-
\ft{1}{2}m_2(5)
\big)
\nonumber\\
&
+
\Gamma_{\rm cusp}
\big[
\vev{\vev{\Gamma_{K-2}}}_3
+
\vev{\vev{\Gamma_{K-2}}}_1^3
\big]
\big(
m_2^4(2)
-
2m_2^2(2)m_2(3)
\nonumber\\
&\qquad
+
\ft{1}{2}m_2^2(3)
+
\ft{2}{3}m_2(2)m_2(4)
-
\ft{1}{6}m_2(5)
\big)
\nonumber\\
&
+
\Gamma_{\rm cusp}^2\vev{\vev{\Gamma_{K-2}}}_1
\left(
4m_2^3(2)m_K(2)
-
6m_2(2)m_2(3)m_K(2)
+
2m_2(4)m_K(2)
\right)
\nonumber\\
&
+
\Gamma_{\rm cusp}^2\vev{\vev{\Gamma_{K-2}}}_2
\left(
4m_2^4(2)m_K(2)
-
8m_2^2(2)m_2(3)m_K(2)
+
m_2^2(3)m_K(2)
\right.
\nonumber\\
&\qquad
\left.
+
4m_2(2)m_2(4)m_K(2)
-
m_2(5)m_K(2)
\right)
\nonumber\\
&
+
\Gamma_{\rm cusp}^2\vev{\vev{\Gamma_{K-2}}}_1^2
\left(
6m_2^4(2)m_K(2)
-
12m_2^2(2)m_2(3)m_K(2)
+
3m_2^2(3)m_K(2)
\right.
\nonumber\\
&\qquad
\left.
+
4m_2(2)m_2(4)m_K(2)
-
m_2(5)m_K(2)
\right)
\nonumber\\
&
+
\Gamma_{\rm virt}\Gamma_{\rm cusp}^2
\left(
4m_2^3(2)m_K(2)
-
6m_2(2)m_2(3)m_K(2)
+
2m_2(4)m_K(2)
+
2m_2^2(2)m_K^2(2)
\right.
\nonumber\\
&\qquad\left.
-
2m_2(3)m_K^2(2)
-
2m_2^2(2)m_K(3)
+
2m_2(3)m_K(3)
\right)
\nonumber\\
&
+
\Gamma_{\rm cusp}^3
\left(
4m_2^3(2)m_K^2(2)
-
4m_2(2)m_2(3)m_K^2(2)
-
2m_2(2)m_2(3)m_K(3)
+
2m_2(4)m_K(3)
\right)
\nonumber\\
&
+
\Gamma_{\rm cusp}^3\vev{\vev{\Gamma_{K-2}}}_1
\left(
12m_2^4(2)m_K^2(2)
-
20m_2^2(2)m_2(3)m_K^2(2)
+
4m_2^2(3)m_K^2(2)
\right.
\nonumber\\
&\qquad\left.
+
4m_2(2)m_2(4)m_K^2(2)
-
4m_2^2(2)m_2(3)m_K(3)
+
2m_2^2(3)m_K(3)
+
4m_2(2)m_2(4)m_K(3)
\right.
\nonumber\\
&\qquad\left.
-
2m_2(5)m_K(3)
\right)
\nonumber\\
&
+
\Gamma_{\rm cusp}^4
\left(
8m_2^4(2)m_K^3(2)
-
8m_2^2(2)m_2(3)m_K^3(2)
-
8m_2^2(2)m_2(3)m_K(2)m_K(3)
\right.
\nonumber\\
&\qquad\left.
+
4m_2^2(3)m_K(2)m_K(3)
+
4m_2(2)m_2(4)m_K(2)m_K(3)
+
\ft{4}{3}m_2(2)m_2(4)m_K(4)
\right.
\nonumber\\
&\qquad\left.
-
\ft{4}{3}m_2(5)m_K(4)
\right)
\nonumber\\
&
+
g^2\Gamma_{\rm cusp}\vev{\vev{\Gamma_{K-2}}}_2
\left(
m_2^2(2)-m_2(3)
\right) \textcolor{red}{r_{1,2}}
\nonumber\\
&\quad
+
\big[
2\vev{\vev{\Gamma_{K-2}}}_1\vev{\vev{\Gamma_{K-2}}}_2
+
\vev{\vev{\Gamma_{K-2}}}_3
\big]
\Gamma_{\rm cusp}
\left(
m_2^2(2)-m_2(3)
\right)
\textcolor{red}{r_{2,2}}
\, , \\
c_{11}^{[K,2]}
&=
-
\ft{1}{2}\Gamma_{\rm cusp}
+
\Gamma_{\rm cusp}\vev{\vev{\Gamma_{K-2}}}_3
\left(
\ft{1}{2}m_2^3(2)
-
\ft{3}{4}m_2(2)m_2(3)
+
\ft{1}{4}m_2(4)
\right)
\nonumber\\
&
+
\vev{\vev{\Gamma_{K-2}}}_2
\left[
\Gamma_{\rm cusp}
\left(
\ft{1}{2}m_2^2(2)
-
\ft{1}{2}m_2(3)
\right)
\right.
\nonumber\\
&\qquad\left.
+
\Gamma_{\rm cusp}\vev{\vev{\Gamma_{K-2}}}_1
\left(
m_2^3(2)
-
\ft{3}{2}m_2(2)m_2(3)
+
\ft{1}{2}m_2(4)
\right)
\right.
\nonumber\\
&\qquad\left.
+
\Gamma_{\rm cusp}^2
\left(
2m_2^3(2)m_K(2)
-
3m_2(2)m_2(3)m_K(2)
+
m_2(4)m_K(2)
\right)
\right]
\nonumber\\
&
+
\Gamma_{\rm cusp}^3
\left(
-2m_2^2(2)m_K^2(2)
+
2m_2(3)m_K^2(2)
+
2m_2^2(2)m_K(3)
-
2m_2(3)m_K(3)
\right)
\nonumber\\
&
+
\Gamma_{\rm cusp}^3\vev{\vev{\Gamma_{K-2}}}_1
\left(
-4m_2^3(2)m_K^2(2)
+
6m_2(2)m_2(3)m_K^2(2)
-
2m_2(4)m_K^2(2)
\right.
\nonumber\\
&\qquad\left.
+
4m_2^3(2)m_K(3)
-
6m_2(2)m_2(3)m_K(3)
+
2m_2(4)m_K(3)
\right)
\nonumber\\
&
+
\Gamma_{\rm cusp}^4
\left(
-8m_2^3(2)m_K^3(2)
+
8m_2(2)m_2(3)m_K^3(2)
+
8m_2^3(2)m_K(2)m_K(3)
\right.
\nonumber\\
&\qquad\left.
-
6m_2(2)m_2(3)m_K(2)m_K(3)
-
2m_2(4)m_K(2)m_K(3)
-
2m_2(2)m_2(3)m_K(4)
\right.
\nonumber\\
&\qquad\left.
+
2m_2(4)m_K(4)
\right)
\, .
\end{align}
Finally, the coefficient $c^{[K,2]}_{10}$ is known only through three loops because it involves $\vev{\vev{\Gamma_{K-2}}}_1$,
which the available data determine only to this order; see Sect.\ \ref{3LdataSection}. It reads
\begin{align}
\label{c10forCumulant1}
c_{10}^{[K,2]}
&=
\ft{1}{2}\vev{\vev{\Gamma_{K-2}}}_1
+
\Gamma_{\rm cusp}m_K(2)
+
\ft{1}{2}\Gamma_{\rm virt}
+
\left(
m_K^2(2)-m_K(3)
\right)
\Gamma_{\rm virt}\Gamma_{\rm cusp}
\nonumber\\
&
+
\left(
2m_2(2)m_K^2(2)
-
2m_2(2)m_K(3)
\right)
\Gamma_{\rm cusp}^2
\nonumber\\
&
+
\left(
4m_2^2(2)m_K^3(2)
-
4m_2^2(2)m_K(2)m_K(3)
-
2m_2(3)m_K(2)m_K(3)
+
2m_2(3)m_K(4)
\right)
\Gamma_{\rm cusp}^3
\nonumber\\
&
+
\left(
2m_2^2(2)m_K^2(2)
-
2m_2(3)m_K^2(2)
-
2m_2^2(2)m_K(3)
+
2m_2(3)m_K(3)
\right)
\vev{\vev{\Gamma_{K-2}}}_1\Gamma_{\rm cusp}^2
\nonumber\\
&
+
\left(
-m_2^2(2)
+
m_2(3)
\right) m_K(2)
\Gamma_{\rm cusp}\vev{\vev{\Gamma_{K-2}}}_2
\nonumber\\
&
+
\ft{1}{2} g^2\vev{\vev{\Gamma_{K-2}}}_2 \textcolor{red}{r_{1,2}}
+
\ft{1}{2}
\left(
\vev{\vev{\Gamma_{K-2}}}_3
+
2\vev{\vev{\Gamma_{K-2}}}_1\vev{\vev{\Gamma_{K-2}}}_2
\right)
\textcolor{red}{r_{2,2}}
\, .
\end{align}
At this stage, we do not give numerical values for these coefficients, since the cumulants remain largely unconstrained. Although we could 
compare the expressions with available data and extract predictions for some of the contributing cumulants, we instead take a more systematic 
route and derive a microscopic description of the averages from the physics of flux-tube excitations.

\subsubsection{Resummed formula}

The solutions above can be summarized by the concise formula
\begin{equation}
\label{G2KresummedEq}
\cG_{[K,2]}
=
H (g)
\int_0^\infty \frac{dy_1}{y_1}
\int_0^\infty \frac{dy_2}{y_2}\,
S\!\left(\frac{u}{y_1},\frac{v}{y_2}\right)
f_2(y_1)f_{K}(y_2)
\vev{w \e^{\ft12 \Gamma_{K-2} \log(u/y_1)}}
\, ,
\end{equation}
where, in addition to the Sudakov form factor and the twist-two recoil function $f_2$, we need the higher-twist recoil function
\begin{align}
f_K (y) = \frac{2 y^{K/2} \mathsf{K}_{K-2} (2 \sqrt{y})}{\Gamma (K-1)}
\, .
\end{align}
The most consequential nonkinematic ingredient in this formula is the last average, which denotes a sum 
over the degeneracy label $I$ of twist-$K$ operators; for example,
\begin{align}
\vev{w \Gamma_{K - 2}} = \frac{1}{d_s} \sum_{I = 1}^{d_s} w_I \Gamma_{K - 2,I}
\, .
\end{align}
As already noted in Sect.\ \ref{LargeSpinADsSection}, the degeneracy is lifted by the anomalous dimension 
$\Gamma_{K-2} = \Gamma_{K-2,I}$. In addition, there is a weight $w_I(g)$ that depends on both the degeneracy label and the 't Hooft coupling. 
This weight is not fixed by crossing, but explicit data show that it is not simply unity. This is the most significant departure
of our analysis from Ref.\ \cite{Alday:2013cwa}.

Let us match this expression onto the solution from the previous subsection. The coefficients in the perturbative
expansion of $w$,
\begin{align}
w = 1 + g^2 w_1 + g^4 w_2 + \dots
\, ,
\end{align}
have the following expectation values:
\begin{align}
\vev{w_1}
&=
\vev{\Gamma^{(1)}_{K-2}}\,m_2(2)
\, , \\
\vev{w_1\Gamma^{(1)}_{K-2}}
&=
\vev{\big(\Gamma^{(1)}_{K-2}\big)^2}
m_2(2)\, ,
\\
\vev{w_1\big(\Gamma^{(1)}_{K-2}\big)^2}
&=
\vev{\big(\Gamma^{(1)}_{K-2}\big)^3}
 m_2(2)\, ,
\\
\vev{w_2}
&=
\vev{w_1\Gamma^{(1)}_{K-2}}m_2(2)
+\vev{\Gamma^{(2)}_{K-2}}m_2(2)
+\Gamma_{\rm virt}^{(2)}m_2(2)
\\
&\quad
-\ft{1}{2}
 \vev{\big(\Gamma^{(1)}_{K-2}\big)^2}m_2(3)
+\Gamma_{\rm virt}^{(2)}m_K(2)\, ,
\\
\vev{w_2\Gamma^{(1)}_{K-2}}
&=
-\vev{w_1\Gamma^{(2)}_{K-2}}
+2\vev{\Gamma^{(1)}_{K-2}\Gamma^{(2)}_{K-2}}m_2(2)
+\vev{\Gamma^{(1)}_{K-2}}
 \Gamma_{\rm virt}^{(2)}m_2(2)
\\
&
+
\vev{\big(\Gamma^{(1)}_{K-2}\big)^3}m_2^2(2)
-\frac{1}{2}
 \vev{\big(\Gamma^{(1)}_{K-2}\big)^3}m_2(3)
+\vev{\Gamma^{(1)}_{K-2}}
 \Gamma_{\rm virt}^{(2)}m_K(2)
 \nonumber\\
&
+
\left[
 -\vev{\Gamma^{(1)}_{K-2}}^2
 +\vev{\big(\Gamma^{(1)}_{K-2}\big)^2}
\right]
\textcolor{red}{r_{1,2}}
+ 
\left[
 -\vev{\Gamma^{(1)}_{K-2}}
  \vev{\big(\Gamma^{(1)}_{K-2}\big)^2}
 +\vev{\big(\Gamma^{(1)}_{K-2}\big)^3}
\right]
\textcolor{red}{r_{2,2}}
 \, ,\nonumber\\
 \vev{w_3}
&=
C_t^{(3,0)}
+\Gamma_{\rm virt}^{(3)}m_K(2)
+\vev{\Gamma^{(1)}_{K-2}}
 \Gamma_{\rm virt}^{(2)}m_2(2)m_K(2)\, .
\end{align}
These relations can be solved at the first two nontrivial orders:
\begin{align}
w_1
&=
\Gamma^{(1)}_{K-2}m_2(2)\, ,
\\
w_2
&=
\Gamma^{(2)}_{K-2}m_2(2)
+\big(\Gamma^{(1)}_{K-2}\big)^2
 \left[m_2^2(2)-\ft{1}{2}m_2(3)\right]
+\Gamma_{\rm virt}^{(2)}
 \left[m_2(2)+m_K(2)\right]
\nonumber\\
&
+
 \big[
  \Gamma^{(1)}_{K-2}
  -\vev{\Gamma^{(1)}_{K-2}}
 \big]
 \textcolor{red}{r_{1,2}}
+
 \big[
  \big(\Gamma^{(1)}_{K-2}\big)^2
  -\vev{\big(\Gamma^{(1)}_{K-2}\big)^2}
 \big]
 \textcolor{red}{r_{2,2}}
 \, .
\end{align}
Thus, the corrections to $w$ are nonzero and inherit the gauge dependence left by crossing. At present, we do not know how to fix these extra 
weights to all orders in perturbation theory, largely because of the a priori ambiguity in the gauge choice. The nonperturbative principle, if any,
that governs their particular form remains obscure to us.

\subsection[Twist-K--twist-K correlator]{Twist-\textbf{\textit K}--twist-\textbf{\textit K} correlator}

Finally, we turn to the correlator $\cG_{[K,K]}$, which supports twist-$K$ excitations in both channels. There are now two sets of extra labels,
$I$ and $J$, enumerating the degeneracies of the high-twist operators. We will see that this further enhances the gauge freedom
of the OPE coefficients compared with $\cG_{[K,2]}$. Although the symmetry of the correlator implies that the $s$- and $t$-channel high-twist
anomalous dimensions $\Gamma_{K-2}$ and $\bar{\Gamma}_{K-2}$ are identical, we adorn the latter with a bar to identify its $t$-channel
origin. This is purely a bookkeeping gimmick that tells the reader in which channel an average is taken.

\subsubsection{Solution to crossing}

At one loop, the averages of the expansion coefficients are
\begin{align}
\vev{C_s^{(1,1)}}
&=
-\vev{\bar{\Gamma}_{K-2}^{(1)}}
-
2\Gamma_{\rm cusp}^{(1)} m_K(2)
\, ,
\\
\vev{C_t^{(1,1)}}
&=
-\vev{\Gamma_{K-2}^{(1)}}
-
2\Gamma_{\rm cusp}^{(1)} m_K(2)
\, .
\end{align}
As before, crossing constrains only the difference $\vev{C_{s}^{(1,0)}} - \vev{C_{t}^{(1,0)}}$, namely,
\begin{align}
\label{Cst10eq}
\vev{C_t^{(1,0)}}
-
\vev{C_s^{(1,0)}}
&=
\vev{\Gamma^{(1)}_{K-2}}m_K(2)
-
\vev{\bar{\Gamma}^{(1)}_{K-2}}m_K(2)
\, .
\end{align}
Notice that, had we imposed symmetry between the $s$- and $t$-channels at this stage, the two coefficients $C_{s,t}^{(1,0)}$ would have been
completely unconstrained.

At two loops, crossing gives
\begin{align}
\vev{C_t^{(2,2)}}
&=
\ft{1}{2}
\vev{\big(\Gamma_{K-2}^{(1)}\big)^2}
+
2\vev{\Gamma_{K-2}^{(1)}}\Gamma_{\rm cusp}^{(1)}m_K(2)
+
2\big(\Gamma_{\rm cusp}^{(1)}\big)^2 m_K(3)
\, ,
\\
\vev{C_t^{(2,1)}}
&=
-\vev{C_s^{(1,0)}\Gamma_{K-2}^{(1)}}
-
\vev{\Gamma_{K-2}^{(2)}}
-
\Gamma_{\rm virt}^{(2)}
-
\vev{C_s^{(1,1)}\Gamma_{K-2}^{(1)}}m_K(2)
-
\vev{\big(\Gamma_{K-2}^{(1)}\big)^2}m_K(2)
\nonumber\\
&\quad
-
2\Gamma_{\rm cusp}^{(2)}m_K(2)
-
4\vev{\Gamma_{K-2}^{(1)}}\Gamma_{\rm cusp}^{(1)}m_K^2(2)
\nonumber\\
&\quad
-
2\vev{\bar{\Gamma}_{K-2}^{(1)}}\Gamma_{\rm cusp}^{(1)}m_K^2(2)
+
2\vev{\bar{\Gamma}_{K-2}^{(1)}}\Gamma_{\rm cusp}^{(1)}m_K(3)
\, ,
\\
\vev{C_s^{(2,2)}}
&=
\ft{1}{2}
\vev{\big(\bar{\Gamma}_{K-2}^{(1)}\big)^2}
+
2\vev{\bar{\Gamma}_{K-2}^{(1)}}\Gamma_{\rm cusp}^{(1)}m_K(2)
+
2\big(\Gamma_{\rm cusp}^{(1)}\big)^2m_K(3)
\, ,
\\
\vev{C_s^{(2,1)}}
&=
-\vev{C_t^{(1,0)}\bar{\Gamma}_{K-2}^{(1)}}
-
\vev{\bar{\Gamma}_{K-2}^{(2)}}
-
\Gamma_{\rm virt}^{(2)}
-
\vev{C_t^{(1,1)}\bar{\Gamma}_{K-2}^{(1)}}m_K(2)
-
\vev{\big(\bar{\Gamma}_{K-2}^{(1)}\big)^2}m_K(2)
\nonumber\\
&\quad
-
2\Gamma_{\rm cusp}^{(2)}m_K(2)
-
2\vev{\Gamma_{K-2}^{(1)}}\Gamma_{\rm cusp}^{(1)}m_K^2(2)
\nonumber\\
&\quad
-
4\vev{\bar{\Gamma}_{K-2}^{(1)}}\Gamma_{\rm cusp}^{(1)}m_K^2(2)
+
2\vev{\Gamma_{K-2}^{(1)}}\Gamma_{\rm cusp}^{(1)}m_K(3)
\, ,
\end{align}
while their difference obeys
\begin{align}
\vev{C_t^{(2,0)}}
&
-
\vev{C_s^{(2,0)}}
=
\vev{C_s^{(1,0)}\Gamma^{(1)}_{K-2}}m_K(2)
+
\vev{\Gamma^{(2)}_{K-2}}m_K(2)
\nonumber\\
&\quad
-
\vev{C_t^{(1,0)}\bar{\Gamma}^{(1)}_{K-2}}m_K(2)
-
\vev{\bar{\Gamma}^{(2)}_{K-2}}m_K(2)
\nonumber\\
&
+
\vev{C_s^{(1,1)}\Gamma^{(1)}_{K-2}}m_K^2(2)
+
\vev{\big(\Gamma^{(1)}_{K-2}\big)^2}m_K^2(2)
-
\vev{C_t^{(1,1)}\bar{\Gamma}^{(1)}_{K-2}}m_K^2(2)
\nonumber\\
&
-
\vev{\big(\bar{\Gamma}^{(1)}_{K-2}\big)^2}m_K^2(2)
+
2\vev{\Gamma^{(1)}_{K-2}}\Gamma_{\rm cusp}^{(1)}m_K^3(2)
-
2\vev{\bar{\Gamma}^{(1)}_{K-2}}\Gamma_{\rm cusp}^{(1)}m_K^3(2)
\nonumber\\
&
-
\ft{1}{2}\vev{\big(\Gamma^{(1)}_{K-2}\big)^2}m_K(3)
+
\ft{1}{2}\vev{\big(\bar{\Gamma}^{(1)}_{K-2}\big)^2}m_K(3)
\, .
\end{align}
Finally, crossing requires
\begin{align}
\vev{C_t^{(1,1)}\bar{\Gamma}^{(1)}_{K-2}}
-
\vev{C_s^{(1,1)}\Gamma^{(1)}_{K-2}}
&=
2\Gamma_{\rm cusp}^{(1)}m_K(2) [\vev{\Gamma^{(1)}_{K-2}} - \vev{\bar{\Gamma}^{(1)}_{K-2}}]
\, .
\end{align}
In a nutshell, this implies symmetry between the two channels:
\begin{align}
\label{SymEqs}
\vev{\Gamma_{K-2}^{(1)}}
&=
\vev{\bar{\Gamma}_{K-2}^{(1)}}\, ,
\\
\vev{C_s^{(1,1)}\Gamma_{K-2}^{(1)}}
&=
\vev{C_t^{(1,1)}\bar{\Gamma}_{K-2}^{(1)}}\, .
\end{align}
This trend continues through three loops, with explicit results collected in Appendix \ref{KKcorr3loopsApp}.

\subsubsection{Minimal and non-minimal gauge fixing}

The gauge parameters introduced below are specific to $\cG_{[K,K]}$. Although we reuse the notation of the preceding subsection,
they are independent of the corresponding parameters for $\cG_{[K,2]}$.

The solution above reveals an enhanced gauge freedom, relative to the $\cG_{[K,2]}$ case, in the choice of OPE coefficients consistent with
their prescribed averages. As we will see, agreement with perturbative data requires the one-loop coefficients to be mere numbers. At two loops,
however, the coefficient functions depend intrinsically on the degeneracy labels and are therefore not defined unambiguously. This ambiguity
deserves special attention at the order considered here. The explicit solutions for $\vev{C_s^{(2,2)}}$,
$\vev{C_s^{(2,2)}\Gamma_{K-2}^{(1)}}$, $\vev{C_t^{(2,2)}}$, and
$\vev{C_t^{(2,2)}\bar{\Gamma}_{K-2}^{(1)}}$ show that $C_{s,t}^{(2,2)}$ have no gauge freedom. By contrast,
$C_{s,t}^{(2,\ell)}$ with $\ell=0,1$ do\footnote{Recall that for the $\cG_{[K,2]}$ correlator, $C_s^{(2,1)}$ was unique, i.e., a number.}.
They can be gauge-transformed to a frame in which the resulting coefficients are numbers, without affecting their averages. We will see that the
following {\sl minimal} shifts indeed yield null correlators with {\sl uniform transcendentality}:
\begin{align}
\label{GaugeTransformationC1}
C_{s,t}^{(2,1)}
&
\to C_{s,t}^{\prime (2,1)} = C_{s,t}^{(2,1)}
+ 
\textcolor{red}{r_{0,2}} [\Gamma_{K-2}^{(1)} - \vev{\Gamma_{K-2}^{(1)}}]
\, , \\
\label{GaugeTransformationC0}
C_{s,t}^{(2,0)}
&
\to C_{s,t}^{\prime (2,0)} = C_{s,t}^{(2,0)} 
+
\sum_{k = 1}^{k_{\rm max}} \textcolor{red}{r_{k, 2}} \big[( \Gamma^{(1)}_{K-2})^k - \vev{( \Gamma^{(1)}_{K-2})^k}\big]
\, ,
\end{align}
with $k_{\rm max}=2$. The coefficients $c_{01}^{[K,K]}$ and $c_{10}^{[K,K]}$, however, inherit a dependence on $C_{s,t}^{(2,0)}$, which can be
modified further in a non-minimal fashion. Such modifications can generate additional, separately uniform-transcendental contributions, so we
supplement the right-hand side of Eq.\ \re{GaugeTransformationC0} with the following {\sl non-minimal} terms\footnote{The $q_{2,2}$ term is retained
for now; its incompatible transcendental weight will force us to discard it below.}\footnote{Correlators involve differences of expectation
values containing unconstrained coefficients, so a constant shift of those coefficients is invisible. The expectation-value subtractions in
the gauge transformation may therefore be omitted.}
\begin{align}
\label{NonMinimalGT}
C_{s,t}^{(2,0)} \to C_{s,t}^{\prime (2,0)} = C_{s,t}^{(2,0)}
+
\dots
&
+
\textcolor{red}{q_{1, 2}} \big[ \tilde\Gamma^{(1)}_{K-2} - \vev{\tilde\Gamma^{(1)}_{K-2}} \big]
\\
&
+
\textcolor{red}{q_{2, 2}} \big[ \Gamma^{(2)}_{K-2} - \vev{ \Gamma^{(2)}_{K-2}} \big]
\, . \nonumber
\end{align}
For the $t$-channel transformations in Eqs.\ \re{GaugeTransformationC1}, \re{GaugeTransformationC0}, and \re{NonMinimalGT},
the quantities $\Gamma^{(n)}_{K-2}$ and $\tilde\Gamma^{(1)}_{K-2}$ on the right-hand sides are replaced by their barred
counterparts, and the averages run over the degeneracy label $J$.
Here, we have introduced a complementary quantity $\tilde\Gamma$ that depends on the degeneracy label $I$. Its meaning will become
clear when we provide the microscopic flux-tube theory for the averages involved. The added terms drop out of
\begin{align}
\vev{C_{s,t}^{(2,n)}} = \vev{C_{s,t}^{\prime (2,n)}}
\, ,
\end{align}
but their averages with powers of $\Gamma_{K-2}^{(1)}$ can and do change: the kinematic
constraints from crossing are completely oblivious to them.

\subsubsection{Three-loop correlator}
\label{3LGKKsection}

Using the crossing solutions, we can construct the perturbative expansion of $\cG_{[K,K]}$. We display only half of the
logarithmic coefficients, since the other half is identical by the $s\leftrightarrow t$ symmetry of this correlator. We impose this
symmetry from the get-go, so neither $C_t$ nor $\bar{\Gamma}$ appears in the formulas. At first, however, we refrain from making
gauge transformations and display everything in terms of the averages produced directly by crossing. This presentation
highlights the failure of crossing symmetry alone to fix this more intricate correlator.

Let us first examine the `superleading' logarithms. The coefficient $c_{21}^{[K,K]}$ can become nonzero at three loops,
\begin{align}
c_{21}^{[K,K]}
=
\ft{1}{16} g^6
\Big(
&
2\vev{\Gamma_{K-2}^{(1)}}^3
+
\vev{\Gamma_{K-2}^{(1)}}
\left[
2\vev{C_s^{(1,1)}\Gamma_{K-2}^{(1)}}
-
\vev{\big(\Gamma_{K-2}^{(1)}\big)^2}
\right]
-
\vev{C_s^{(1,1)}\big(\Gamma_{K-2}^{(1)}\big)^2}
\nonumber\\
&\quad
+
4\vev{\Gamma_{K-2}^{(1)}}^2\Gamma_{\rm cusp}^{(1)}m_K(2)
-
2\vev{\big(\Gamma_{K-2}^{(1)}\big)^2}\Gamma_{\rm cusp}^{(1)}m_K(2)
\Big)
\, .
\end{align}
Crossing alone does not constrain $\vev{C_s^{(1,1)}\big(\Gamma_{K-2}^{(1)}\big)^m}$, only $\vev{C_s^{(1,1)}}$.
Agreement with the perturbative data of Sect.\ \ref{3LdataNullSection}, however, requires $C_s^{(1,1)}$ to be a number, so that
\begin{align}
\vev{C_s^{(1,1)}\big(\Gamma_{K-2}^{(1)}\big)^m}
=
C_s^{(1,1)} \vev{\big(\Gamma_{K-2}^{(1)}\big)^m}
\, ,
\end{align}
where $C_s^{(1,1)}$ is set equal to the value of $\vev{C_s^{(1,1)}}$ found from crossing. Under this condition,
$c_{21}^{[K,K]}$ vanishes identically at three loops, in agreement with the data.

The other `superleading' coefficient, $c_{30}^{[K,K]}$, contains no unconstrained OPE coefficients and is simply
given by
\begin{align}
c_{30}^{[K,K]}
=
\ft{1}{48}\vev{\vev{\Gamma_{K-2}}}_{3}
+
\ft{1}{6}\Gamma_{\rm cusp}^{3}
\left(
2m_K^3(2)
-
3m_K(2)m_K(3)
+
m_K(4)
\right)
\, .
\end{align}

Moving on to $c_{20}^{[K,K]}$, we encounter unfixed averages of powers of $\Gamma_{K-2}$ weighted by
$C_s^{(1,\ell)}$ ($\ell=0,1$), namely,
\begin{align}
c_{20}^{[K,K]}
&=
-
\ft{1}{4}g^6\vev{\Gamma_{K-2}^{(1)}}\vev{C_s^{(1,0)}\Gamma_{K-2}^{(1)}}
+
\ft{1}{8}g^6\vev{C_s^{(1,0)}\big(\Gamma_{K-2}^{(1)}\big)^2}
+
\ft{1}{8}g^6\vev{C_s^{(1,1)}\big(\Gamma_{K-2}^{(1)}\big)^2}m_K(2)
\nonumber\\
&\quad
-
\ft{1}{4}g^6
\vev{C_s^{(1,1)}\Gamma_{K-2}^{(1)}}
\left[
\vev{\Gamma_{K-2}^{(1)}}m_K(2)
+
2
\Gamma_{\rm cusp}^{(1)}
\left(
m_K^2(2)
-
m_K(3)
\right)
\right]
\nonumber\\
&\quad
+
\ft{1}{8}g^4
\vev{\big(\Gamma_{K-2}^{(1)}\big)^2}
\left(
1
+
2 g^2 
\Gamma_{\rm cusp}^{(1)}m_K^2(2)
\right)
\nonumber\\
&\quad
-
\frac{1}{2}g^4
\Gamma_{\rm cusp}^{(1)} \big( \Gamma_{\rm cusp}^{(1)}+ 2 g^2 \Gamma_{\rm cusp}^{(2)}\big)
\left(
m_K^2(2)
-
m_K(3)
\right)
\nonumber\\
&\quad
-
\ft{1}{8} g^4
\vev{\Gamma_{K-2}^{(1)}}^2
\left[
1
+
4 g^2
\Gamma_{\rm cusp}^{(1)}
\big(
2 
m_K^2(2)
-
m_K(3)
\big)
\right]
+
\ft{1}{4}
g^6\vev{\Gamma_{K-2}^{(1)}\Gamma_{K-2}^{(2)}}
\nonumber\\
&\quad
-
g^6
\big(\Gamma_{\rm cusp}^{(1)}\big)^3
\big[
2m_K^4(2)
-
3m_K^2(2)m_K(3)
+
m_K(2)m_K(4)
\big]
\nonumber\\
&\quad
-
\ft{1}{4} g^6
\vev{\Gamma_{K-2}^{(1)}}
\left[
\vev{\Gamma_{K-2}^{(2)}}
+
2
\big(\Gamma_{\rm cusp}^{(1)}\big)^2
\left(
4 m_K^3(2)
-
5 m_K(2)m_K(3)
+
m_K(4)
\right)
\right]
\, .
\end{align}
Consistency with the three-loop data also requires $C_s^{(1,0)}$ to be a number; $C_s^{(1,1)}$ was fixed in the same way
from $c_{21}^{[K,K]}$ above. This condition alone reduces the result to the simpler expression
\begin{align}
c_{20}^{[K,K]}
&
=
\ft{1}{8}\vev{\vev{\Gamma_{K-2}}}_{2}
-
\ft{1}{2}\Gamma_{\rm cusp}^{2}\vev{\vev{\Gamma_{K-2}}}_{1}
\left(
2m_K^3(2)
-
3m_K(2)m_K(3)
+
m_K(4)
\right)
\\
&-
\ft{1}{2}\Gamma_{\rm cusp}^{2}
\left(
m_K^2(2)
-
m_K(3)
\right)
-
\Gamma_{\rm cusp}^{3}m_K(2)
\left(
2m_K^3(2)
-
3m_K(2)m_K(3)
+
m_K(4)
\right)
\, . \nonumber
\end{align}

The crossing solution for the double-logarithmic coefficient $c_{11}^{[K,K]}$ carries a greater ambiguity:
\begin{align}
c_{11}^{[K,K]}
&
=
-
\ft{1}{4}g^6\vev{C_s^{(2,1)}\Gamma_{K-2}^{(1)}}
-
\ft{1}{4}g^6\vev{C_s^{(1,1)}\Gamma_{K-2}^{(2)}}
+
\ft{1}{4}g^6\vev{C_s^{(1,1)}\big(\Gamma_{K-2}^{(1)}\big)^2} m_K(2)
\nonumber\\
&
-
\ft{1}{4} g^4
\vev{C_s^{(1,1)} \Gamma_{K-2}^{(1)}}
\big[
1 + 2 g^2 \vev{\Gamma_{K-2}^{(1)}}m_K(2)
\big]
-
\ft{1}{2}g^6
\vev{C_s^{(1,0)}\Gamma_{K-2}^{(1)}}
\big[
\vev{\Gamma_{K-2}^{(1)}}
+
\Gamma_{\rm cusp}^{(1)}m_K(2)
\big]
\nonumber\\
&
-
\ft{1}{2}g^2 \left[ \Gamma_{\rm cusp}^{(1)} + g^2 \Gamma_{\rm cusp}^{(2)} + g^4\Gamma_{\rm cusp}^{(3)} \right]
-
\ft{1}{4}g^4\vev{\Gamma_{K-2}^{(1)}}^2
-
\ft{1}{2}g^4\vev{\Gamma_{K-2}^{(1)}} \big[ \Gamma_{\rm cusp}^{(1)} + g^2 \Gamma_{\rm cusp}^{(2)} \big] m_K(2)
\nonumber\\
&
-
\ft{1}{2}g^6\vev{\Gamma_{K-2}^{(1)}}\vev{\Gamma_{K-2}^{(2)}}
-
\ft{1}{4}g^6\vev{\Gamma_{K-2}^{(1)}}\Gamma_{\rm virt}^{(2)}
-
\ft{1}{2}g^6\vev{\Gamma_{K-2}^{(2)}}\Gamma_{\rm cusp}^{(1)}m_K(2)
\nonumber\\
&
+
\ft{1}{2} g^6 \Gamma_{\rm cusp}^{(1)} \big[ \vev{\big(\Gamma_{K-2}^{(1)}\big)^2} - 2 \vev{\Gamma_{K-2}^{(1)}}^2 \big]
( 2 m_K^2(2) - m_K(3))
\nonumber\\
&
-
2 g^6\left(\Gamma_{\rm cusp}^{(1)}\right)^3
\big[
m_K^4(2)
- 
2 m_K^2(2)m_K(3)
+
m_K^2(3)
\big]
\, .
\end{align}
It involves the following averages:
\begin{align}
\big\{
\vev{C_s^{(2,1)}\Gamma_{K-2}^{(1)}}\, ,
\vev{C_s^{(1,0)}\Gamma_{K-2}^{(1)}}\, ,
\vev{C_s^{(1,1)}\Gamma_{K-2}^{(1)}}\, ,
\vev{C_s^{(1,1)}\Gamma_{K-2}^{(2)}}\, ,
\vev{C_s^{(1,1)}\big(\Gamma_{K-2}^{(1)}\big)^2}
\big\}
\, .
\end{align}
Thus, in addition to higher powers of $\Gamma_{K-2}$ accompanying one-loop coefficients, it contains a new coefficient,
$C_s^{(2,1)}$. Unlike its counterpart for the $\cG_{[K,2]}$ correlator, it is not a number. To account for its ambiguity,
we perform the {\sl minimal gauge} transformation in Eq.\ \re{GaugeTransformationC1}, moving to a frame in which it
factorizes out of the average up to a gauge function:
\begin{align}
\vev{C_s^{(2,1)}\Gamma_{K-2}^{(1)}}
\to
\vev{C_s^{\prime (2,1)}\Gamma_{K-2}^{(1)}}
=
\vev{C_s^{(2,1)}} \vev{ \Gamma_{K-2}^{(1)}} + \vev{\vev{\Gamma_{K-2}^{(1)}}}_2 \textcolor{red}{r_{0,2}} 
\, .
\end{align}
We then find
\begin{align}
c_{11}^{[K,K]}
&
=
-
\ft{1}{2} \Gamma_{\rm cusp}
-
\Gamma_{\rm cusp}^{3}
\left(
2m_K^4(2)
-
4m_K^2(2)m_K(3)
+
2m_K^2(3)
\right)
\nonumber\\
&
+
\ft{1}{2}
\Gamma_{\rm cusp}\vev{\vev{\Gamma_{K-2}}}_{2}
\left(
m_K^2(2)
-
m_K(3)
\right)
-
\ft{1}{4} g^2 \vev{\vev{\Gamma_{K-2}}}_{2} \textcolor{red}{r_{0,2}}
\, .
\end{align}

Finally, we come to $c_{10}^{[K,K]}$. Its expression is quite lengthy and contains the following unfixed averages
\begin{align}
\big\{
&
\vev{C_s^{(2,0)}\Gamma_{K-2}^{(1)}}\, ,
\vev{C_s^{(2,1)}\Gamma_{K-2}^{(1)}}\, ,
\\
&
\vev{C_s^{(1,0)}\Gamma_{K-2}^{(1)}}\, ,
\vev{C_s^{(1,0)}\Gamma_{K-2}^{(2)}}\, ,
\vev{C_s^{(1,1)}\Gamma_{K-2}^{(1)}}\, ,
\vev{C_s^{(1,1)}\Gamma_{K-2}^{(2)}}\, ,
\vev{C_s^{(1,1)}\big(\Gamma_{K-2}^{(1)}\big)^2}
\big\}
\, . \nonumber
\end{align}
Apart from the first line of this list, we may safely assume that $C_s$ factorizes out of each average. Moreover, the
nonfactorizable part of $\vev{C_s^{(2,1)}\Gamma_{K-2}^{(1)}}$ has already been parametrized by $r_{0,2}$, leaving only
the gauge freedom in $C_s^{(2,0)}$. Keeping $\vev{C_s^{(2,0)}\Gamma_{K-2}^{(1)}}$ intact for now, we obtain
\begin{align}
c_{10}^{[K,K]}
&
=
\ft{1}{2}\Gamma_{\rm virt}
+
\Gamma_{\rm cusp} \Gamma_{\rm virt}
\left(
m_K^2(2)
-
m_K(3)
\right)
+
\ft{1}{2}
\big[
\vev{\vev{\Gamma_{K-2}}}_1
+
2 \Gamma_{\rm cusp} m_K(2)
\big]
\nonumber\\
&\quad
+
\Gamma_{\rm cusp}
\big[
\vev{\vev{\Gamma_{K-2}}}_1
+
2
\Gamma_{\rm cusp}m_K(2)
\big]
\left(
m_K^2(2)
-
m_K(3)
\right)
\nonumber\\
&\quad
+
2
\Gamma_{\rm cusp}^3
\big[
2m_K^5(2)
-
2m_K^3(2)m_K(3)
-
m_K(2)m_K^2(3)
+
m_K(3)m_K(4)
\big]
\nonumber\\
&\quad
+
\ft{1}{2}
\Gamma_{\rm cusp}
\big[
\vev{\vev{\Gamma_{K-2}}}_2
+
\vev{\vev{\Gamma_{K-2}}}_1^2
\big]
\big[
2m_K^3(2)
-
3m_K(2)m_K(3)
+
m_K(4)
\big]
\nonumber\\
&\quad
+
2
\Gamma_{\rm cusp}^2
\vev{\vev{\Gamma_{K-2}}}_1
\big[
3m_K^4(2)
-
5m_K^2(2)m_K(3)
+
m_K^2(3)
+
m_K(2)m_K(4)
\big]
\nonumber\\
&\quad
+
\ft{1}{2}g^6
\big[
\vev{C_s^{(2,0)}\Gamma^{(1)}_{K-2}}
-
\vev{C_s^{(2,0)}}\vev{\Gamma^{(1)}_{K-2}}
\big]
+
\ft{1}{2}g^2
\vev{\vev{\Gamma_{K-2}}}_2
m_K(2)
\textcolor{red}{r_{0,2}}
\, .
\end{align}
A few comments are in order. First, so far we have gauge-transformed only $C_s^{(2,1)}$. Second, as noted above, the
physical correlation function depends only on the combination
$\vev{C_s^{(2,0)}\Gamma^{(1)}_{K-2}}-\vev{C_s^{(2,0)}}\vev{\Gamma^{(1)}_{K-2}}$, so any constant shift of
$C_s^{(2,0)}$ is irrelevant. Third, the unfixed coefficient $C_s^{(2,0)}$ brings additional, more extensive gauge freedom,
making the penultimate term in the preceding equation pure gauge. Applying the non-minimal gauge transformations in
Eqs.\ \re{GaugeTransformationC0} and \re{NonMinimalGT}, we find
\begin{align}
\label{C20gaugetransform}
\vev{C_s^{(2,0)}\Gamma^{(1)}_{K-2}}
&
-
\vev{C_s^{(2,0)}}\vev{\Gamma^{(1)}_{K-2}}
=
\vev{\vev{\Gamma^{(1)}_{K-2}}}_2
\textcolor{red}{r_{1,2}}
\nonumber\\
&
+
[
\vev{\vev{\Gamma^{(1)}_{K-2}}}_3
+
2 \vev{\vev{\Gamma^{(1)}_{K-2}}}_1 \vev{\vev{\Gamma^{(1)}_{K-2}}}_2
]
\textcolor{red}{r_{2,2}}
+
\vev{\vev{\vev{\Gamma_{K-2}^{(1)} \tilde\Gamma_{K-2}^{(1)}}}}
\textcolor{red}{q_{1,2}}
\nonumber\\
&
+
\vev{\vev{\vev{\Gamma_{K-2}^{(1)} \Gamma_{K-2}^{(2)}}}}
\textcolor{red}{q_{2,2}}
\, ,
\end{align}
where the first two terms are expressed in terms of cumulants and the last two in terms of cross-correlations,
\begin{align}
\vev{\vev{\vev{\Gamma_{K-2}^{(1)} \tilde\Gamma_{K-2}^{(1)}}}}
&\equiv
\vev{\Gamma_{K-2}^{(1)} \tilde\Gamma_{K-2}^{(1)}}
-
\vev{\Gamma_{K-2}^{(1)}} \vev{\tilde\Gamma_{K-2}^{(1)}}
\, , \\
\vev{\vev{\vev{\Gamma_{K-2}^{(1)} \Gamma_{K-2}^{(2)}}}}
&\equiv
\vev{\Gamma_{K-2}^{(1)} \Gamma_{K-2}^{(2)}}
-
\vev{\Gamma_{K-2}^{(1)}} \vev{\Gamma_{K-2}^{(2)}}
\, .
\end{align}
These will be determined below.

\subsubsection{Resummed formula}

The crossing solutions above can be cast in the compact form
\begin{align}
\label{GKKresummedEq}
\cG_{[K,K]}
=
H (g)
\int_0^\infty \frac{dy_1}{y_1}
\int_0^\infty \frac{dy_2}{y_2}\,
S\!\left(\frac{u}{y_1},\frac{v}{y_2}\right)
f_K (y_1) f_{K} (y_2)
\vev{\e^{\ft12 \Gamma_{K-2} \log(u/y_1)} w \e^{\ft12 \bar{\Gamma}_{K-2} \log(v/y_2)}}
\, ,
\end{align}
where 
\begin{align}
w = w_{IJ} (g)
\, ,
\end{align}
is a new unconstrained ingredient akin to $w$ in Eq.\ \re{G2KresummedEq}. Here $w$ depends on the degeneracy indices in
both channels. These indices are naturally contracted in matrix fashion with
$\Gamma_{K-2}=\Gamma_{K-2,I}$ and $\bar{\Gamma}_{K-2}=\bar{\Gamma}_{K-2,J}$; for example,
\begin{align}
\vev{\Gamma_{K-2} w \bar{\Gamma}_{K-2}} = \frac{1}{d_s d_t} \sum_{I,J} \Gamma_{K-2,I} w_{IJ} \bar{\Gamma}_{K-2,J}
\, .
\end{align}
Interestingly, we will establish below that $w_{IJ}$ does not factorize in the indices $I$ and $J$.

Let us compare the perturbative expansion generated by the resummed formula \re{GKKresummedEq} with the explicit crossing
solution of the preceding subsection. To ensure consistency of the resummed formula with the data, we impose factorization on averages involving
one-loop OPE coefficients and powers of $\Gamma_{K-2}$:
\begin{align*}
\vev{C_s^{(1,\ell)} \Gamma_{K-2}^m} = \vev{C_s^{(1,\ell)}} \vev{\Gamma_{K-2}^m}
\, .
\end{align*}
The only nonfactorized contributions remaining at three loops are
\begin{align*}
\big\{
\vev{C_s^{(2,0)}\Gamma_{K-2}^{(1)}}\, ,
\vev{C_s^{(2,1)}\Gamma_{K-2}^{(1)}}
\big\}
\, .
\end{align*}
As before, we gauge-transform $C_s^{(2,\ell)}$ to a frame in which they factorize, at the price of residual gauge parameters.
Before doing so, let us present the matching conditions for the averages involving the coefficients $w_\ell$. At one, two, and three loops, we find
\begin{align}
\label{Eq1}
\vev{w_1}
&=
2\vev{\Gamma_{K-2}^{(1)}}m_K(2)
\, ,
\\
\vev{\Gamma_{K-2}^{(1)} w_1}
&=
\big[
\vev{\Gamma_{K-2}^{(1)}}^2
+
\vev{\big(\Gamma_{K-2}^{(1)}\big)^2}
\big] m_K(2)
\, ,
\\
\vev{w_1\bar{\Gamma}_{K-2}^{(1)}}
&=
\big[
\vev{\Gamma_{K-2}^{(1)}}^2
+
\vev{\big(\Gamma_{K-2}^{(1)}\big)^2}
\big]
m_K(2)
\, , \\
\vev{\big(\Gamma^{(1)}_{K-2}\big)^2 w_1}
&=
\big[
\vev{\Gamma^{(1)}_{K-2}}
\vev{\big(\Gamma^{(1)}_{K-2}\big)^2}
+
\vev{\big(\Gamma^{(1)}_{K-2}\big)^3}
\big]
m_K(2)
\label{Eq5}
\, , \\
\vev{w_1\big(\bar{\Gamma}^{(1)}_{K-2}\big)^2}
&=
\big[
\vev{\Gamma^{(1)}_{K-2}}
\vev{\big(\Gamma^{(1)}_{K-2}\big)^2}
+
\vev{\big(\Gamma^{(1)}_{K-2}\big)^3}
\big]
m_K(2)
\label{WbarGammaSquaredEq}
\, , \\
\vev{\Gamma^{(1)}_{K-2}w_1 \bar{\Gamma}^{(1)}_{K-2}}
&=
-
\vev{C_s^{(2,1)}\Gamma^{(1)}_{K-2}}
-
\vev{\Gamma^{(1)}_{K-2}}
\vev{\Gamma^{(2)}_{K-2}}
-
\vev{\Gamma^{(1)}_{K-2}}\Gamma_{\rm virt}^{(2)}
\nonumber\\
&
+
\big[
\vev{\Gamma^{(1)}_{K-2}}
\vev{\big(\Gamma^{(1)}_{K-2}\big)^2}
-
2\vev{\Gamma^{(1)}_{K-2}}\Gamma_{\rm cusp}^{(2)}
\big]
m_K(2)
\nonumber\\
&
-
2
\big[
\vev{\Gamma^{(1)}_{K-2}}^2
\Gamma_{\rm cusp}^{(1)}
+
\vev{\big(\Gamma^{(1)}_{K-2}\big)^2}
\Gamma_{\rm cusp}^{(1)}
\big]
m_K^2(2)
\nonumber\\
&
+
2 \vev{\big(\Gamma^{(1)}_{K-2}\big)^2}
\Gamma_{\rm cusp}^{(1)}m_K(3)
\, ,
\label{GammaWbarGammaEq}
\end{align}
and 
\begin{align}
\vev{w_2}
&=
\big[
\vev{\Gamma_{K-2}^{(1)} w_1}
+
\vev{w_1\bar{\Gamma}_{K-2}^{(1)}}
+
2\vev{\Gamma_{K-2}^{(2)}}
+
\Gamma_{\rm virt}^{(2)}
\big]
m_K(2)
\nonumber\\
&
-
\vev{\Gamma_{K-2}^{(1)}}^2m_K^2(2)
-
\vev{\big(\Gamma_{K-2}^{(1)}\big)^2}m_K(3)
\, ,
\\
\vev{\Gamma^{(1)}_{K-2} w_2}
&=
\vev{C_s^{(2,0)}\Gamma^{(1)}_{K-2}}
-
\vev{\Gamma^{(2)}_{K-2} w_1}
\nonumber\\
&
+
\big[
\vev{\Gamma^{(1)}_{K-2}}\vev{\Gamma^{(2)}_{K-2}}
+
2\vev{\Gamma^{(1)}_{K-2}\Gamma^{(2)}_{K-2}}
+
\vev{\Gamma^{(1)}_{K-2}}\Gamma_{\rm virt}^{(2)}
\big]
m_K(2)
\nonumber\\
&
+
\big[
\vev{\Gamma^{(1)}_{K-2}}
\vev{\big(\Gamma^{(1)}_{K-2}\big)^2}
+
\vev{\big(\Gamma^{(1)}_{K-2}\big)^3}
\big]
m_K^2(2)
-
\ft{1}{2}\vev{\big(\Gamma^{(1)}_{K-2}\big)^3}m_K(3)
\, , \\
\vev{w_2\bar{\Gamma}^{(1)}_{K-2}}
&=
\vev{C_s^{(2,0)}\Gamma^{(1)}_{K-2}}
-
\vev{w_1\bar{\Gamma}^{(2)}_{K-2}}
\nonumber\\
&
+
\big[
\vev{\Gamma^{(1)}_{K-2}}\vev{\Gamma^{(2)}_{K-2}}
+
2\vev{\Gamma^{(1)}_{K-2}\Gamma^{(2)}_{K-2}}
+
\vev{\Gamma^{(1)}_{K-2}}\Gamma_{\rm virt}^{(2)}
\big]
m_K(2)
\nonumber\\
&
+
\big[
\vev{\Gamma^{(1)}_{K-2}}
\vev{\big(\Gamma^{(1)}_{K-2}\big)^2}
+
\vev{\big(\Gamma^{(1)}_{K-2}\big)^3}
\big]
m_K^2(2)
-
\ft{1}{2}\vev{\big(\Gamma^{(1)}_{K-2}\big)^3}m_K(3)
\, ,
\end{align}
and 
\begin{align}
\vev{w_3}
&=
\vev{C_s^{(3,0)}}
\nonumber\\
&
+\big[
 \vev{C_s^{(2,0)}\Gamma_{K-2}^{(1)}}
 +\vev{\Gamma_{K-2}^{(3)}}
 +\Gamma_{\rm virt}^{(3)}
 \big]m_K(2)
\nonumber\\
&
+\big[
 \vev{\Gamma_{K-2}^{(1)}}
 \vev{\Gamma_{K-2}^{(2)}}
 +2\vev{\Gamma_{K-2}^{(1)}\Gamma_{K-2}^{(2)}}
 +3\vev{\Gamma_{K-2}^{(1)}}\Gamma_{\rm virt}^{(2)}
 \big]m_K^2(2)
\nonumber\\
&
+\big[
 \vev{\Gamma_{K-2}^{(1)}}
 \vev{\bigl(\Gamma_{K-2}^{(1)}\bigr)^2}
 +\vev{\bigl(\Gamma_{K-2}^{(1)}\bigr)^3}
 \big]m_K^3(2)
\nonumber\\
&
-\big[
 \vev{\Gamma_{K-2}^{(1)}\Gamma_{K-2}^{(2)}}
 +\vev{\Gamma_{K-2}^{(1)}}\Gamma_{\rm virt}^{(2)}
 \big]m_K(3)
\nonumber\\
&
-\big[
 \ft{1}{2}
 \vev{\Gamma_{K-2}^{(1)}}
 \vev{\bigl(\Gamma_{K-2}^{(1)}\bigr)^2}
 +\vev{\bigl(\Gamma_{K-2}^{(1)}\bigr)^3}
 \big]m_K(2)m_K(3)
\nonumber\\
&
+\ft{1}{6}
 \vev{\bigl(\Gamma_{K-2}^{(1)}\bigr)^3}m_K(4)
 \, ,
\end{align}
respectively.

Let us now find the coefficients $w_\ell$ themselves, subject to these constraints. Consider $w_1$ first. Its only ambiguity inherited from
the OPE coefficients comes from the gauge parameter associated with $C_s^{(2,1)}$ and resides in Eq.\ \re{GammaWbarGammaEq}:
\begin{align}
\vev{\Gamma^{(1)}_{K-2} w_1 \bar{\Gamma}^{(1)}_{K-2}}
&
=
-
\vev{\Gamma^{(1)}_{K-2}}^3m_K(2)
+
2\vev{\Gamma^{(1)}_{K-2}}
\vev{\big(\Gamma^{(1)}_{K-2}\big)^2}m_K(2)
\\
&
-
2\vev{\big(\Gamma^{(1)}_{K-2}\big)^2}
\Gamma_{\rm cusp}^{(1)}
\left(
m_K^2(2)
-
m_K(3)
\right)
-
2\vev{\Gamma^{(1)}_{K-2}}^2
\Gamma_{\rm cusp}^{(1)}m_K(3)
\nonumber\\
&
+
\big[
\vev{\Gamma^{(1)}_{K-2}}^2
-
\vev{\big(\Gamma^{(1)}_{K-2}\big)^2}
\big]
\textcolor{red}{r_{0,2}}
+
\vev{C_s^{(1,0)}}
\big[
\vev{\Gamma^{(1)}_{K-2}}^2
+
2\vev{\Gamma^{(1)}_{K-2}}\Gamma_{\rm cusp}^{(1)}m_K(2)
\big]
\, . \nonumber
\end{align}
Here, neither $r_{0,2}$ nor $\vev{C_s^{(1,0)}}$ is fixed by crossing. For the latter, let us make the educated guess
\begin{align}
\vev{C_s^{(1,0)}}
&=
m_K(2)\vev{\bar{\Gamma}^{(1)}_{K-2}}
\, .
\end{align}
This choice is natural in light of the symmetry between the $s$- and $t$-channel expansions; see Eq.\ \re{Cst10eq}.
Then we find
\begin{align}
\vev{\Gamma^{(1)}_{K-2} w_1 \bar{\Gamma}^{(1)}_{K-2}}
&=
2\vev{\Gamma^{(1)}_{K-2}}
\vev{\big(\Gamma^{(1)}_{K-2}\big)^2}
m_K(2)
\nonumber\\
&\quad
-
\vev{\vev{\Gamma^{(1)}_{K-2}}}_2 \textcolor{red}{r_{0,2}} 
-
2\vev{\vev{\Gamma^{(1)}_{K-2}}}_2
\Gamma_{\rm cusp}^{(1)}
\left(
m_K^2(2)
-
m_K(3)
\right)
\, .
\end{align}
The first term on the right-hand side arises naturally from
\begin{align}
\label{w1Solution}
w_1|_{\rm fixed}
&=
m_K(2)
\big(
\Gamma_{K-2}^{(1)}
+
\bar{\Gamma}_{K-2}^{(1)}
\big)
\, ,
\end{align}
which obeys all of Eqs.\ \re{Eq1}--\re{WbarGammaSquaredEq}. The remaining terms are gauge degrees of freedom.
Reflecting this freedom, $w_1$ can be gauge-transformed by adding a term invisible to all the equations above:
\begin{align}
w_1
=
w_1|_{\rm fixed}
+ \lambda (\Gamma_{K-2}^{(1)} - \vev{\Gamma_{K-2}^{(1)}}) (\bar{\Gamma}_{K-2}^{(1)} - \vev{\bar{\Gamma}_{K-2}^{(1)}})
\, .
\end{align}
Consistency requires
\begin{align}
\label{GKKconsistencyW1}
\lambda \vev{\vev{\Gamma^{(1)}_{K-2}}}_2
+
2\Gamma_{\rm cusp}^{(1)}
\left(
m_K^2(2)
-
m_K(3)
\right)
+
\textcolor{red}{r_{0,2}}
=
0
\, .
\end{align}
We will verify in Sect.\ \ref{GKK3Lsection} that this condition is indeed satisfied for an appropriate choice of $r_{0,2}$ and
$\lambda=-\ft18$.

Similarly, we find
\begin{align}
\label{w2Solution}
w_2|_{\rm fixed}
&=
\Gamma_{\rm virt}^{(2)} m_K(2)
\\
&
+
m_K(2)
\big(
\Gamma_{K-2}^{(2)}
+
\bar{\Gamma}_{K-2}^{(2)}
\big)
+
\left(
m_K^2(2)
-
\ft12 m_K(3)
\right)
\big(\Gamma_{K-2}^{(1)}\big)^2
\nonumber\\
&
+
\left(
m_K^2(2)
-
\ft12 m_K(3)
\right)
\big(\bar{\Gamma}_{K-2}^{(1)}\big)^2
+
m_K^2 (2)
\bar{\Gamma}_{K-2}^{(1)}
\Gamma_{K-2}^{(1)}
\, , \nonumber
\end{align}
up to gauge transformations that reflect the shifts of the OPE coefficient functions in Eq.\ \re{C20gaugetransform}.
The number of gauge parameters proliferates with perturbative order. Although $w$ carries dynamical
information on par with the anomalous dimensions themselves, we cannot currently determine it in closed form. Crossing symmetry
is oblivious to its existence, leaving wide open the question of whether $w$ obeys equations that could fix its form.

With this cliffhanger, we leave crossing behind and develop a microscopic theory of the averages.

\section{Flux tube}

\begin{figure}[t]
\begin{center}
\mbox{
\begin{picture}(0,135)(105,0)
\put(0,10){\insertfig{7}{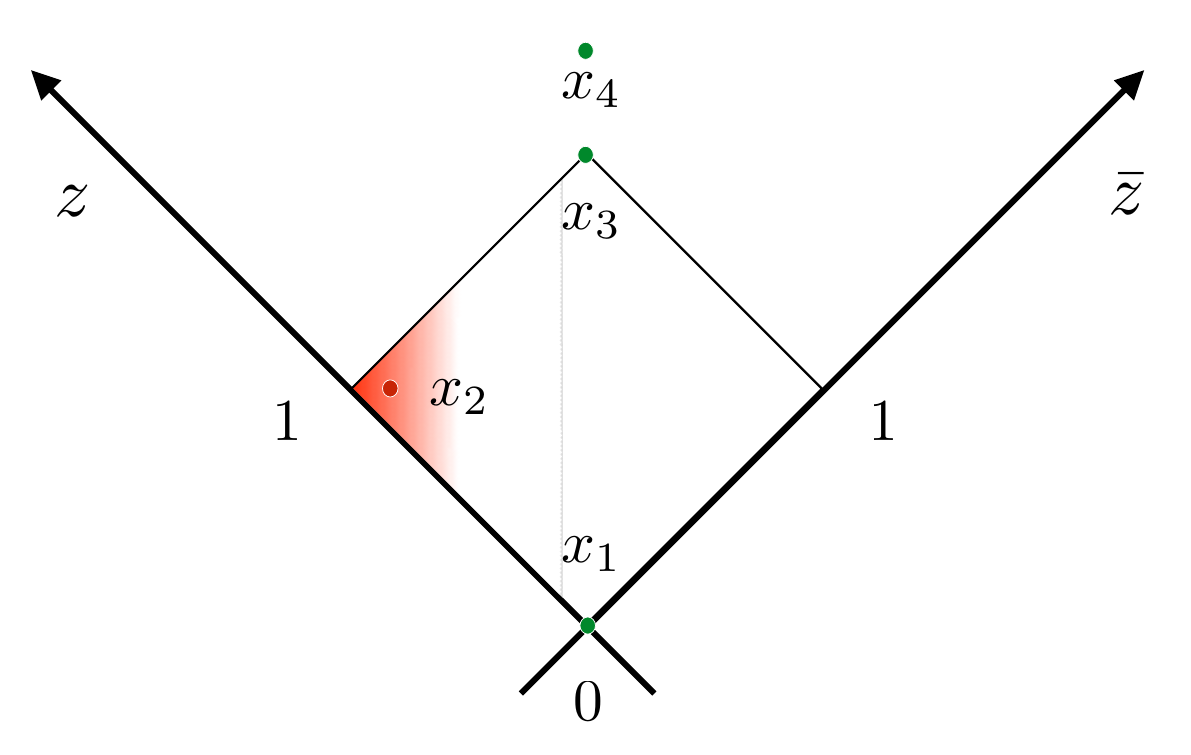}}
\end{picture}
}
\end{center}
\caption{\label{ConformalFrameFig} Two-dimensional conformal plane. The red wedge marks one of the two corners relevant
to the null limit of correlators.}
\end{figure}

We have already taken the first step toward understanding high-twist operators in terms of hole dynamics on top of the
GKP background: their energies are known to all orders. We must now determine how the holes couple to the frame of a
given correlator. Let us proceed systematically and briefly recall what we have learned. Consider $\cG$ in a special
conformal frame. Using conformal invariance to fix three of the four points and taking all points to lie in a light-cone
$(x_+,x_-)$ plane, we choose
\begin{align}
x_1 = (0, 0) \, , \qquad x_3 = (1,1) \, , \qquad x_4 = (\infty, \infty)
\, ,
\end{align}
so that the dependence on the cross ratios is encoded in $x_2\equiv Z=(z,\bar z)$ through the `rationalizing' variables
$z$ and $\bar z$:
\begin{align}
u = z \bar{z}
\, , \qquad
v = (1 - z) (1 - \bar{z})
\, .
\end{align}
The corner of $Z$ space studied in this paper is shown in Fig.\ \ref{ConformalFrameFig}: $z\to1$, $\bar z\to0$.
In this domain, the $s$-channel OPE expansion is dominated by the large-spin tail of the collinear conformal blocks 
\begin{align*}
\cG
\logsim
\int dS
\sum_{I} \vev{\cO_{k_1} (0) \cO_{k_2} (Z)  | S, I } \vev{ S, I | \cO_{k_3} (1) \cO_{k_4} (\infty)}
\, , 
\end{align*}
entering through the matrix element $\vev{\cO_{k_1} (0) \cO_{k_2} (Z)  | S, I }$. The holes emerge in this setup once we
resolve the identity in terms of their states, as in Eq.\ \re{CompletenessHoles}, and sum over $I$:
\begin{align*}
\cG
\logsim
\int dS
\SumInt
\vev{\cO_{k_1} (0) \cO_{k_2} (Z)  | S} \otimes | \mbox{holes} \rangle \langle \mbox{holes} | \otimes \vev{ S | \cO_{k_3} (1) \cO_{k_4} (\infty)}
\, .
\end{align*}
This representation makes clear that the new ingredient in the flux-tube picture of null correlators is the overlap of hole
states with the high-twist operators entering the OPE of the half-BPS operators. We turn to it next.

\subsection{One-hole measure: Born and one loop}

As reviewed in Sect.\ \ref{LargeSpinADsSection}, the number of holes $n_{\rm h}$ on top of the GKP vacuum is tied to the
twist $K$ of the intermediate operator: $n_{\rm h}=K-2$. To determine their overlap, we begin with a single hole. Once
the cross-ratio dependence has been stripped off, the couplings of the high-twist operator to the initial and final
$s$-channel frames are the same, up to a possible phase. We are therefore sensitive only to the absolute square of the
one-hole transition, which we dub the {\sl measure}. It can be extracted from the next-to-simplest correlator,
$\cG_{[3,2]}$.

We determine the measure for a single-hole excitation order by order in the 't Hooft coupling:
\begin{align}
\mu (u) = \mu_{1}(u) + g^2 \mu_{2}(u) + g^4 \mu_{3}(u) + \dots
\, .
\end{align}
We choose it to have unit normalization,
\begin{align}
\int \frac{du}{2 \pi} \mu (u) = 1
\, ,
\end{align}
which implies for the individual perturbative terms that
\begin{align}
\label{NormalizationMeasure}
\int \frac{du}{2 \pi} \mu_1 (u) = 1
\, , \qquad
\int \frac{du}{2 \pi} \mu_{k>1} (u) = 0
\, .
\end{align}

The coefficient in the logarithm of the correlation function that provides our initial constraint on the Born term $\mu_1$ is
$c_{10}^{[3,2]}$, since it receives a leading contribution from $\vev{\vev{\Gamma_1}}_1$; see Eq.\ \re{c10forCumulant1}.
With a little guesswork, we arrive at
\begin{align}
\mu_{1}(u)
&=
\pi^{2}\sech^{2}(\pi u)
\, .
\end{align}
This measure does the job: it reproduces Eq.\ \re{eq:cij-2332}. More precisely, the one-loop value of the
first cumulant completely cancels the leading contribution from $\Gamma_{\rm cusp}$ in the combination
$\vev{\vev{\Gamma_{K-2}}}_1 + 2 \Gamma_{\rm cusp}m_3(2) \sim O(g^4)$. Every other coefficient of the correlator
$\cG_{[3,2]}$ likewise agrees perfectly at this order in the 't Hooft coupling.

Three features of $\mu_1$ are worth emphasizing. First, it is sharply peaked near zero hole rapidity and has exponentially suppressed
tails. This confirms our expectation that the null limit is dominated by high-twist operators near the lower edge of the spectrum in
Eq.\ \re{BandWidth}. Second, $\mu_1$ is $i$-periodic, a property that persists at higher orders: $\mu(u+i)=\mu(u)$. Third, an attentive
reader familiar with the pentagon OPE for scattering amplitudes will probably have noticed that $\mu_1$ is the square of the one-loop
measure of a flux-tube scalar propagating on a null polygonal Wilson loop; see, e.g., \cite{Basso:2013aha,Belitsky:2014sla}. In retrospect,
this is perhaps not entirely surprising: from the stringy viewpoint, the closed worldsheet of a correlator is, in a sense, a double copy
of the open worldsheet of an amplitude. The precise dynamical mechanism behind this coincidence nevertheless remains obscure. There
are two settings in which correlators cleanly split into amplitudes: (i) the light-cone limit of correlators of the stress-tensor multiplet
containing $\cO_2$ \cite{Alday:2010zy}, and (ii) the infinite-R-charge limit of polarized correlators of $\cO_k$ operators,
$k\to\infty$ \cite{Caron-Huot:2021usw,Belitsky:2025bgb}. At present, only the first admits a robust framework based on flux-tube dynamics.
Our study may be viewed as a first step toward such a framework for the second. A proper understanding will require incorporating the
various R-symmetry representations propagating in the OPE channels, most likely through an analogue of the so-called matrix part of
nonsinglet pentagons; see, e.g., \cite{Basso:2015uxa,Belitsky:2016vyq}. We leave this interesting problem for the future.

We next move to one loop. To begin with, $\mu_2$ must have zero norm. A suitable ansatz is a linear combination of the first
and second powers of $\mu_1$. Equation \re{NormalizationMeasure} fixes the relative coefficient, while agreement of the first
cumulant with the data fixes the overall normalization. The one-loop correction to the measure is
\begin{align}
\mu_{2}(u)
&=
\mu_{1}^{2}(u)
\left[-2+\cosh(2\pi u)\right]
\, .
\end{align}

\begin{figure}[t]
\begin{center}
\mbox{
\begin{picture}(0,145)(85,0)
\IfFileExists{Spoiler.pdf}{%
  \put(0,10){\insertfig{4.9}{Spoiler}}%
}{%
  \put(0,10){\fbox{\parbox[c][3cm][c]{4.7cm}{\centering Missing figure\\[2pt]\texttt{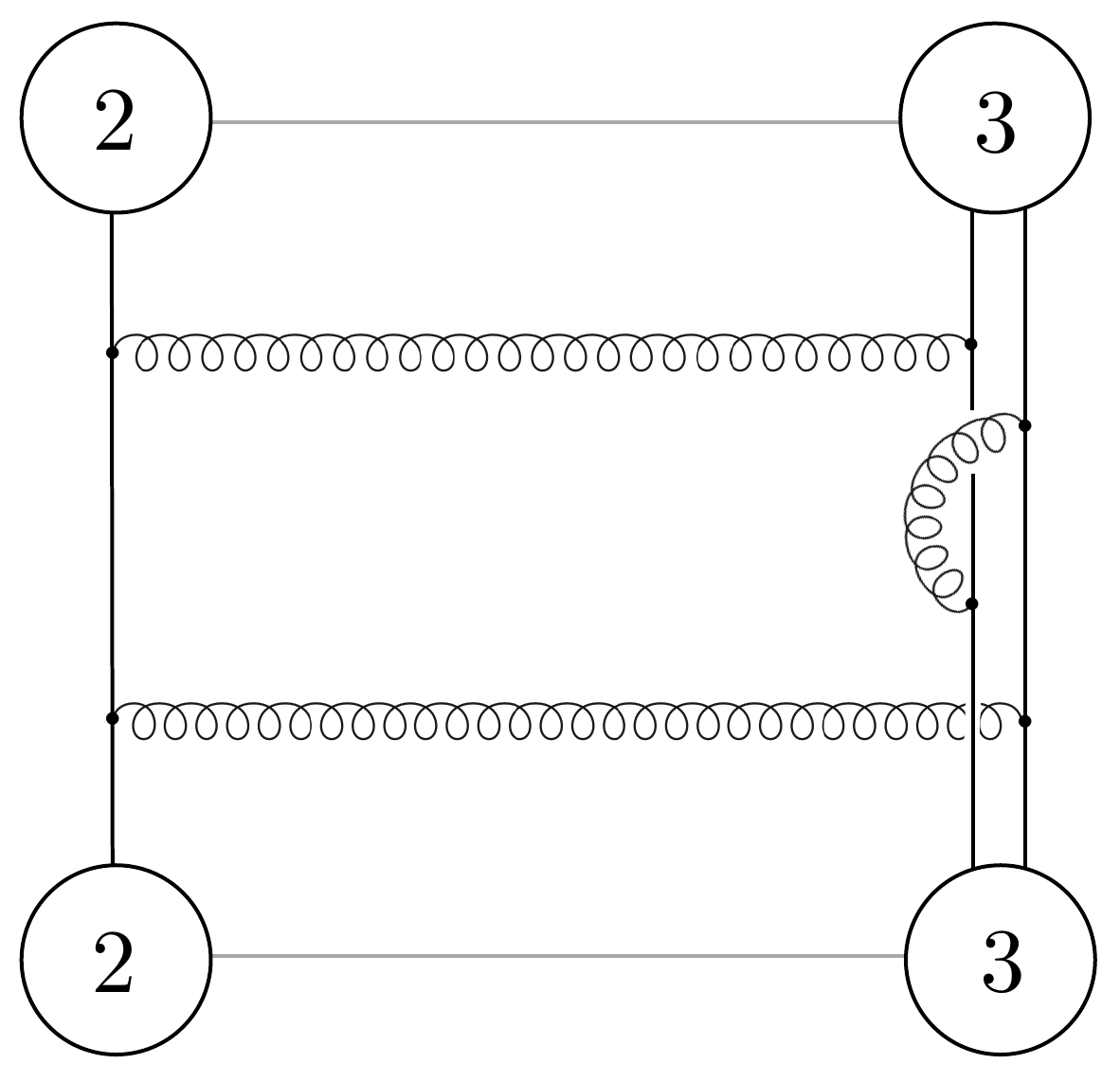}}}}%
}
\end{picture}
}
\end{center}
\caption{\label{SpoilerFig} Three-loop graph for the $\cG_{[3,2]}$ correlator that invalidates the flux-tube picture: gluons hop from one side
of the correlator to the other. For $K>3$, this does not happen at the same order in the 't Hooft coupling.}
\end{figure}

One might now try to reconstruct the next correction, $\mu_3$, by the same method. This attempt fails. Extracting a candidate
$\mu_3$ from $\cG_{[3,2]}$ is relatively easy, but using it for $\cG_{[K,2]}$ with $K>3$ does not reproduce the data. The culprit
is that, starting at three loops, gluon exchanges can hop across the long bridge of length two, from one side of the correlator to
the other, as shown in Fig.\ \ref{SpoilerFig}. This effect invalidates the flux-tube picture, which relies on effective {\sl open}-string
dynamics. We expect the problem to become less severe with increasing $K$ and eventually disappear as $K\to\infty$. We should
therefore study $\cG_{[K,2]}$ with $K>3$. Already at $K=4$, however, two flux-tube excitations are emitted simultaneously, and
their production need not factorize into a product of one-hole measures. The holes can `feel' each other's presence; their interaction
must be described by an additional form factor. Like the measure, this form factor must be constructed perturbatively. We turn to it next.

\subsection{Two-hole form factor: Born and one loop}

As anticipated at the end of the previous section, the correlator $\cG_{[4,2]}$ introduces a new building block in the flux-tube
picture: the two-hole creation/absorption form factor $F(u_1,u_2)$. This local notation should not be confused with the generic
reduced correlator $F(u,v)$ used in the conformal partial-wave discussion. The form factor has the perturbative expansion
\begin{align}
F (u_{1},u_{2}) = F_{1}(u_{1},u_{2}) + g^2 F_{2}(u_{1},u_{2})+ g^4 F_{3}(u_{1},u_{2}) + \dots
\, .
\end{align}
What properties should it have? First, it is symmetric in its arguments\footnote{Strictly speaking, this quantity is the product of
creation and absorption form factors for flux-tube excitations emitted by the correlator frame. Bootstrapping it to all loops would require
a prescription for taking its `square root', thereby recovering a possible Watson equation. For the product itself, that equation appears
to be superfluous.}:
\begin{align}
F (u_{1},u_{2}) = F (u_{2},u_{1})
\, ,
\end{align}
and it is invariant under simultaneous reversal of the rapidities:
\begin{align}
F (- u_{1},- u_{2}) = F (u_{1},u_{2})
\, .
\end{align}
Its integral with the two-hole measure is normalized to unity
\begin{align}
\int d^2 \mathcal{M} \, F (u_{1},u_{2}) = 1
\, .
\end{align}
Since the flux-tube excitations are not relativistic, however, $F$ need not depend only on the rapidity difference.

As with the one-hole measure, we draw inspiration from the Wilson-loop flux tube. For the Born term, we take the absolute
square of the leading-order inverse pentagon transition\footnote{A double Wick rotation \cite{Basso:2011rc} relates it to the
inverse creation form factor.} \cite{Basso:2013aha,Belitsky:2014rba}:
\begin{align}
F_1 (u_1, u_2) 
= \frac{1}{|P_{\rm h|h} (u_1|u_2)|^2} 
&
= \left| \frac{\Gamma (\ft12 + i u_1) \Gamma (\ft12 - i u_2)}{\Gamma (i u_1 - i u_2)} \right|^2
\nonumber\\
&
=
\pi (u_1 - u_2) \left[ \tanh(\pi u_1) - \tanh(\pi u_2) \right]
\, .
\end{align}
This hits the nail on the head: the expression has all the required properties and yields cumulants in agreement with the
lowest-order data.

At one loop, following the same strategy as for the measure, we use powers of the Born form factor to span the functions entering $F_2$. From these
we construct two linear combinations that (i) have zero norm and (ii) produce the correct transcendental structure when
integrated against the leading-order hole energies and combined with the contribution of the one-loop measure found above.
At this order, the result must be proportional to $\zeta_3$. We find
\begin{align*}
F_{2}(u_{1},u_{2})
&=
\pi^2 
F_1(u_1,u_2)
\left[
\alpha_1
\!
\left(
2
-
3
\sech^2(\pi u_1)
-
3
\sech^2(\pi u_2)
\right)
+
3\alpha_2
\tanh(\pi u_1)\tanh(\pi u_2)
\right]
\, .
\end{align*}
This leaves one free constant, since $\alpha_2=1-\alpha_1$. Comparison with the data fixes $\alpha_1=\ft13$, and
we obtain
\begin{align}
F_{2}(u_{1},u_{2})
&=
\pi^{2} F_{1}(u_{1},u_{2})
\left[
\ft{2}{3}
-
\sech^{2}(\pi u_{1})
-
\sech^{2}(\pi u_{2})
+
2\tanh(\pi u_{1})\tanh(\pi u_{2})
\right]
\, .
\end{align}

Interestingly, the form factors and measures are not entirely independent:
\begin{align}
\lim_{u_2 \to u_1} F_1 (u_1, u_2)/(u_1-u_2)^2 = \mu_1 (u_1) 
\, , \qquad
\lim_{u_2 \to u_1} F_2 (u_1, u_2)/(u_1-u_2)^2 = \ft43 \mu_2 (u_1)
\, . 
\end{align}
The leading term in the coincident-limit expansion of the form factor is proportional to the measure---though, unlike for the
pentagon flux tube, the two are not identical at $O(g^2)$. It would be interesting to understand the origin of this relation.

\subsection{Two loops: measure and multiparticle form factors}
\label{TwoLoopMeasureFFsection}

Because of the hopping interactions at three loops, we now use the data for $K=4,5,6$, excluding $K=3$, to fix the
two-loop measure and form factors {\sl simultaneously}. We then test the result at $K=7,8$.

The functional space entering the ans\"atze is built from powers of lower-loop expressions. For the two-loop quantities, we take
\begin{align}
\mu_3 (u) =
\mu_1^3 (u)
[
\alpha_1
+
\alpha_2\cosh(2\pi u)
+
\alpha_3\cosh(4\pi u)
]
\, ,
\end{align}
and 
\begin{align}
F_3(u_1,u_2)
&
=
\pi^4 F_1(u_1,u_2)
\Big[
\beta_1
+
\beta_2
\left[
\sech^2(\pi u_1)
+
\sech^2(\pi u_2)
\right]
+
\beta_3
\left[
\sech^4(\pi u_1)
+
\sech^4(\pi u_2)
\right]
\nonumber\\
&
+
\beta_4\tanh(\pi u_1)\tanh(\pi u_2)
+
\beta_5
\left[
\sech^2(\pi u_1)
+
\sech^2(\pi u_2)
\right]
\tanh(\pi u_1)\tanh(\pi u_2)
\nonumber\\
&\qquad\qquad\qquad
+
\beta_6 \tanh^2(\pi u_1)\tanh^2(\pi u_2)
\Big]
\, .
\end{align}

Without using perturbative data, we can require only that $\mu_3$ have zero norm. This eliminates one degree of freedom:
\begin{align}
\alpha_3
&=
-\frac{\alpha_1}{6}
-
\frac{\alpha_2}{4}
\, .
\end{align}
As expected, the resulting structures multiplying $\alpha_1$ and $\alpha_2$ have transcendental weight five when integrated
against the leading-order hole energy $\varepsilon_1$.

We next consider two, three, and four excitations and impose vanishing norm on the $O(g^4)$ correction to the averaging measure:
\begin{align}
\left.\int d^\ell \mathcal{M}\,F (u_1, \dots, u_\ell)\right|_{g^4} = 0
\, ,
\end{align}
where the multiparticle measure is defined by
\begin{align}
d^{\ell} \mathcal{M} \equiv \prod_{k = 1}^{\ell} \frac{du_k}{2\pi} \mu (u_k)
\, .
\end{align}
At present, we know only the two-hole form factor. Anticipating integrable flux-tube dynamics, we assume that the multiparticle
form factor factorizes into two-particle ones:
\begin{align}
F (u_1, \dots, u_\ell) 
&
= \prod_{1\leq i<j\leq\ell} F (u_i, u_j)
\, .
\end{align}
The conditions for $\ell=2,3,4$ then eliminate three unknowns from the ansatz for $F_3$:
\begin{align}
\beta_4
&=
-\frac{448}{221}
+
\frac{498}{221}\beta_1
-
\frac{3865}{1989}\alpha_1
+
\frac{553}{221}\beta_2
-
\frac{4091}{1326}\alpha_2
+
\frac{1310}{663}\beta_3
\, ,
\\
\beta_5
&=
\frac{32}{221}
-
\frac{225}{221}\beta_1
-
\frac{250}{663}\alpha_1
-
\frac{150}{221}\beta_2
+
\frac{125}{221}\alpha_2
-
\frac{168}{221}\beta_3
\, ,
\\
\beta_6
&=
\frac{424}{221}
-
\frac{495}{221}\beta_1
+
\frac{5}{1326}\alpha_1
-
\frac{330}{221}\beta_2
-
\frac{5}{884}\alpha_2
-
\frac{237}{221}\beta_3
\, .
\end{align}

We can now calculate the first cumulants. Having introduced all the main players, it is time to give the microscopic definition
of the averages:
\begin{align}
\label{FluxTubeAveraging}
\vev{f} \equiv \int d^{K-2} \mathcal{M} \, F (u_1, \dots, u_{K-2}) f ( u_1, \dots, u_{K-2})
\, ,
\end{align}
for any function $f$ of the hole rapidities. The averages of the hole energies then depend on the still-unfixed constants. The
lowest few needed to determine them completely are
\begin{align}
\label{vevG1}
\vev{\vev{\Gamma_{1}}}_{1}
&=
-\Gamma_{\rm cusp} H_1^{(1)}
+
3g^2\Gamma_{\rm cusp}\zeta_3
+
g^6
\left(
\frac{4}{3}\pi^2\zeta_3
-
80\zeta_5
\right)
\nonumber\\
&\quad
+
g^4\Gamma_{\rm cusp}
\left[
\left(
-\frac{13}{6}\alpha_1
-
\frac{11}{4}\alpha_2
\right) \pi^2 \zeta_3
+
\left(
-\frac{5}{6}\alpha_1
+
\frac{5}{4}\alpha_2
\right)\zeta_5
\right]
\, ,
\\
\vev{\vev{\Gamma_{2}}}_{1}
&=
- 
\Gamma_{\rm cusp} H_2^{(1)}
-
12g^4\zeta_3
+
\frac{16}{3}g^6\pi^2\zeta_3
+
6g^2\Gamma_{\rm cusp}\zeta_3
\nonumber\\
&\quad
+
g^4\Gamma_{\rm cusp}
\left[
\left(
-\frac{12}{221}
-
\frac{993}{221}\beta_1
-
\frac{466}{221}\alpha_1
-
\frac{1325}{221}\beta_2
-
\frac{627}{221}\alpha_2
-
\frac{1042}{221}\beta_3
\right)\pi^2\zeta_3
\right.
\nonumber\\
&\qquad\left.
+
\left(
-\frac{32}{221}
+
\frac{225}{221}\beta_1
-
\frac{285}{221}\alpha_1
+
\frac{150}{221}\beta_2
+
\frac{855}{442}\alpha_2
-
\frac{274}{221}\beta_3
\right)\zeta_5
\right]
\, ,
\\
\vev{\vev{\Gamma_{3}}}_{1}
&=
-
\Gamma_{\rm cusp} H_3^{(1)}
-
20g^4\zeta_3
+
8g^2\Gamma_{\rm cusp}\zeta_3
+
g^6
\left(
8\pi^2\zeta_3
+
20\zeta_5
\right)
\nonumber\\
&\quad
+
g^4\Gamma_{\rm cusp}
\left[
\left(
-\frac{12}{17}
-
\frac{126}{17}\beta_1
-
\frac{106}{51}\alpha_1
-
\frac{169}{17}\beta_2
-
\frac{49}{17}\alpha_2
-
\frac{124}{17}\beta_3
\right)\pi^2\zeta_3
\right.
\nonumber\\
&\qquad\left.
+
\left(
\frac{1465}{221}
-
\frac{45}{221}\beta_1
-
\frac{385}{221}\alpha_1
-
\frac{30}{221}\beta_2
+
\frac{1155}{442}\alpha_2
-
\frac{1227}{221}\beta_3
\right)\zeta_5
\right]
\, ,
\\
\vev{\vev{\Gamma_{4}}}_{1}
&=
-
\Gamma_{\rm cusp} H_4^{(1)}
-
28g^4\zeta_3
+
10g^2\Gamma_{\rm cusp}\zeta_3
+
g^6
\left(
\frac{32}{3}\pi^2\zeta_3
+
40\zeta_5
\right)
\nonumber\\
&\quad
+
g^4\Gamma_{\rm cusp}
\left[
\left(
-\frac{180}{221}
-
\frac{2298}{221}\beta_1
-
\frac{4124}{1989}\alpha_1
-
\frac{3079}{221}\beta_2
-
\frac{1916}{663}\alpha_2
-
\frac{6668}{663}\beta_3
\right)\pi^2\zeta_3
\right.
\nonumber\\
&\qquad\left.
+
\left(
\frac{2930}{221}
-
\frac{90}{221}\beta_1
-
\frac{1205}{663}\alpha_1
-
\frac{60}{221}\beta_2
+
\frac{1205}{442}\alpha_2
-
\frac{2454}{221}\beta_3
\right)\zeta_5
\right]
\, . \label{vevG4}
\end{align}
Here $H_\ell^{(1)}$ are harmonic numbers,
\begin{align}
H^{(1)}_{\ell} = \left\{ 1, \ft32, \ft{11}6 ,  \ft{25}{12}, \dots \right\}
\, ,
\end{align}
which are themselves related to values of the Euler digamma function.

The remaining constants are fixed by matching the coefficient $c_{10}^{[K,2]}$ for $K=4,5,6$ to the three-loop data of
Sect.\ \ref{3LdataNullSection}. Using the Bessel moments of logarithms collected in Appendix \ref{BesselMomApp}, we find
\begin{align}
c_{10}^{[K,2]}
&
=
\ft{1}{2}\Gamma_{\rm virt}
-
\ft{1}{24} \pi^2\Gamma_{\rm virt}\Gamma_{\rm cusp}
-
\ft{1}{24} \pi^2 \Gamma_{\rm cusp}^3\zeta_3
+
\ft{1}{2}\vev{\vev{\Gamma_{K-2}}}_1
+
\ft{1}{144} \pi^4 \Gamma_{\rm cusp}^2\vev{\vev{\Gamma_{K-2}}}_1
\nonumber\\
&
+
\ft{1}{2}
\big[
\Gamma_{\rm cusp}
+
\ft{1}{72} \pi^4 \Gamma_{\rm cusp}^3
+
\ft{1}{12} \pi^2 \Gamma_{\rm cusp}\vev{\vev{\Gamma_{K-2}}}_2
\big]
\psi(K-1)
\nonumber\\
&
-
\ft{1}{4}
\big[
\Gamma_{\rm virt}\Gamma_{\rm cusp}
-
\ft{1}{6} \pi^2 \Gamma_{\rm cusp}^2\vev{\vev{\Gamma_{K-2}}}_1
\big]
\psi^{(1)}(K-1)
\nonumber\\
&
+
\ft{1}{24} \pi^2 \Gamma_{\rm cusp}^3\psi(K-1)\psi^{(1)}(K-1)
+
\ft{1}{48} \pi^2 \Gamma_{\rm cusp}^3\psi^{(2)}(K-1)
\nonumber\\
&
+
\ft{1}{2}g^2\vev{\vev{\Gamma_{K-2}}}_2 \textcolor{red}{r_{1,2}}
+
\ft{1}{2} \big[ \vev{\vev{\Gamma_{K-2}}}_3 + 2 \vev{\vev{\Gamma_{K-2}}}_1\vev{\vev{\Gamma_{K-2}}}_2 \big] \textcolor{red}{r_{2,2}}
\, .
\label{c102Kpregaugefixing}
\end{align}
As follows from crossing, this coefficient contains two unfixed gauge parameters and depends on $K$. Perturbation theory tells us
that, for $K>3$, the three-loop coefficient is the same for every $K$. Enforcing $K$-independence requires all polygamma terms to
cancel. This kills two birds with one stone: (i) the result becomes constant in $K$, and (ii) it acquires {\sl uniform transcendental}
(UT) weight. We call this choice of the $r$-parameters the {\sl UT gauge}:
\begin{align}
\label{G2KUTgauge}
r_{1,2}
=
3\zeta_3
\, , \qquad
r_{2,2}
=
\frac{\pi^2}{24}
\, ,
\end{align}
under which it becomes
\begin{align}
c_{10}^{[K,2]}
\!
=
&
\ft{1}{2}\Gamma_{\rm virt}
-
\ft{1}{16}\pi^2 \Gamma_{\rm virt}\Gamma_{\rm cusp}
-
\ft{1}{24}\pi^2\Gamma_{\rm cusp}^3\zeta_3
+
\ft{1}{2}
\left(
1
+
\ft{1}{48}\pi^4\Gamma_{\rm cusp}^2
\right)
\left[
\vev{\vev{\Gamma_{K-2}}}_1
+
\Gamma_{\rm cusp}\psi(K - 1)
\right]
.
\end{align}
The remaining digamma is canceled by the harmonic numbers multiplying $\Gamma_{\rm cusp}$ in the first cumulants,
Eqs.\ \re{vevG1}--\re{vevG4}. We then insert their values at $K=4,5,6$ and match Eq.\ \re{c102Kpregaugefixing} to the
three-loop data, quoted here for convenience:
\begin{align*}
c_{10}^{[K>3,2]}
=
-6g^4\zeta_3
+
g^6
\left(
4\pi^2\zeta_3
+
40\zeta_5
\right)
\, .
\end{align*}
We find the unique solution:
\begin{align}
&
\alpha_1=\frac{329}{60}
\, ,
\quad
\alpha_2=-\frac{391}{90}
\, , \\
&
\beta_1=-\frac{28}{15}\, ,
\quad
\beta_2=0\, ,
\quad
\quad
\beta_3=2\, ,
\quad
\beta_4=\frac{7}{15}\, ,
\quad
\beta_5=-4\, ,
\quad
\beta_6=4
\, .
\end{align}
Notice that at $K=3$ we obtain the wrong rational coefficient multiplying $\pi^2\zeta_3$,
\begin{align}
c_{10}^{[3,2]}
=
-6g^4\zeta_3
+
g^6
\left(
\ft{67}{15} \pi^2\zeta_3
+
20\zeta_5
\right)
\, .
\end{align}
This is why $K=3$ was excluded from our three-loop analysis, as anticipated above.

With the measure and form factors fixed, we can check that the norms indeed vanish at $K=7,8$. For both values, the first
cumulant also gives the same prediction for $c_{10}^{[K,2]}$, as required. This provides a stringent cross-check of the flux-tube
picture in general and of the measures and form factors computed here in particular. Finally, using the higher cumulants listed
explicitly in Appendix \ref{CumulantsApp}, we evaluated all remaining logarithmic coefficients and found perfect agreement with
the three-loop data.

\subsection{Integrable deformation?}

While searching for the rational prefactor multiplying the weight-five transcendental $\pi^2\zeta_3$ in $c_{10}^{[K,2]}$, we
stumbled upon a family of measures and form factors satisfying the zero-norm condition for any number of holes. This is not
entirely trivial: although the deformation enters every equation linearly, it preserves factorization of the multiparticle form factors,
a harbinger of integrability. We therefore find this observation worth recording.

Namely, if we write
\begin{align}
F_3(u_1,u_2)
&=
\pi^4 F_1(u_1,u_2)
\big[
-\ft{28}{15}
+
2\left(\sech^4(\pi u_1)+\sech^4(\pi u_2)\right)
+
\beta_4 \,\tanh(\pi u_1)\tanh(\pi u_2)
\nonumber\\
&
-
4\left(\sech^2(\pi u_1)+\sech^2(\pi u_2)\right)
\tanh(\pi u_1)\tanh(\pi u_2)
+
4\tanh^2(\pi u_1)\tanh^2(\pi u_2)
\big]
\, ,
\end{align}
with an arbitrary deformation parameter $\beta_4$, then the correlated deformation of $\mu_{3}$ is
\begin{align}
\mu_{3}(u)
=
\mu_{1}^3(u)
\left[
\alpha_1(\beta_4)
+
\alpha_2(\beta_4)\cosh(2\pi u)
+
\alpha_3(\beta_4)\cosh(4\pi u)
\right]
\, .
\end{align}
The coefficients are
\begin{align}
\alpha_1 (\beta_4)
=
\ft{28}{5}
-
\ft{1}{4} \beta_4
\, ,
\qquad
\alpha_2(\beta_4)
=
-
\ft{64}{15}
-
\ft{1}{6} \beta_4
\, ,
\qquad
\alpha_3(\beta_4)
=
\ft{2}{15}
+
\ft{1}{12} \beta_4
\, .
\end{align}
We did not explore this dependence beyond the two values $\beta_4=\ft{7}{15}$ and $\beta_4=1+\ft{7}{15}$. Both preserve the
normalization of the multiparticle measure and the constancy of $c_{10}^{[K>3,2]}$.

This nonrigidity is nevertheless a little troubling, since it suggests flexibility in the underlying integrable structure and deserves
further study.

\section[Twist-K--twist-2 null correlator]{Twist-\textbf{\textit K}--twist-2 null correlator}

We now bring together the information accumulated above to study the null limit of the $\cG_{[K,2]}$ correlator. We first use
the minimal UT gauge to predict its four-loop form and then examine the potential ambiguity from non-minimal gauge fixing.

\subsection{Four-loop prediction in minimal UT gauge}
\label{4LG2KminimalUTsection}

The known hole energies, measure, and two-particle form factor allow us to predict all four-loop logarithmic coefficients of
$\cG_{[K,2]}$, except for $c_{10}^{[K,2]}$.

For the double logarithm, we get
\begin{align}
c_{11}^{[3,2]}
&=
-
2g^2
+
\ft{2}{3} \pi^2 g^4
-
\ft{52}{45} \pi^4 g^6
+
g^8
\left(
\ft{1096}{945} \pi^6
+
88 \zeta_3^2
\right)
\, , \nonumber\\
c_{11}^{[K>3,2]}
&=
- 
2 g^2
+
\ft{2}{3} \pi^2 g^4
-
\ft{52}{45} \pi^4 g^6
+
g^8
\left(
\ft{128}{105} \pi^6
+
80\zeta_3^2
\right)
\, .
\end{align}
The `superleading' logarithms acquire the following nonvanishing coefficients:
\begin{align}
c_{21}^{[3,2]}
=
6 \pi^2\zeta_3 g^8
\, , \qquad
c_{21}^{[K>3,2]}
=
\ft{16}{3} \pi^2 \zeta_3 g^8
\, ,
\\
c_{12}^{[K\geq 3,2]}
=
8\pi^2\zeta_3 g^8
\, .
\end{align}
At quartic order in the logarithms,
\begin{align}
c_{40}^{[3,2]}
=
\ft{13}{270} \pi^4 g^8
\, , \qquad
c_{40}^{[4,2]}
=
\ft{49}{1080} \pi^4 g^8
\, , \qquad
c_{40}^{[K>4,2]}
=
\ft{2}{45} \pi^4 g^8
\, ,
\end{align}
whereas
\begin{align}
c_{04}^{[K\geq 3,2]}
&
=
\ft{4}{45} \pi^4 g^8
\, .
\end{align}
At cubic order,
\begin{align}
c_{30}^{[3,2]}
&
=
-3g^6\zeta_3
+
g^8
\left(
\ft{10}{3} \pi^2 \zeta_3
-
5\zeta_5
\right)
\, , \\
c_{30}^{[4,2]}
&
=
-\ft{8}{3}g^6\zeta_3
+
g^8
\left(
\ft{8}{3} \pi^2 \zeta_3
-
\ft{5}{3}\zeta_5
\right)
\, , \\
c_{30}^{[K>4,2]}
&
=
-\ft{8}{3}g^6\zeta_3
+
\ft{8}{3}g^8 \pi^2\zeta_3
\, ,
\end{align}
whereas
\begin{align}
c_{03}^{[K\geq 3,2]}
&
=
-
\ft{16}{3} g^6\zeta_3
+
\ft{16}{3}g^8 \pi^2 \zeta_3
\, .
\end{align}
All the coefficients above are predicted unambiguously by crossing and the flux-tube construction. Those that follow are not: they depend on
two minimal-gauge parameters. For $K=3$, the coefficient $c_{20}^{[3,2]}$ before gauge fixing is the non-UT expression
\begin{align}
c_{20}^{[3,2]}
=
\ft{1}{2} \pi^2 g^4
-
\ft{16}{45} \pi^4 g^6
+
g^8
\Big(
-16\pi^2
+
\ft{4}{9} \pi^4
+
\ft{9493}{14175} \pi^6
-
102 \zeta_3^2
+
\ft{14}{3} \zeta_3 \pi^2
+
48\zeta_3
\nonumber\\
\qquad
-
\big(
16
-
14 \zeta_3
\big)
\textcolor{red}{r_{1,2}}
+
\big(
384
-
\ft{32}{3} \pi^2
-
\ft{68}{45} \pi^4
-
112\zeta_3
\big) \textcolor{red}{r_{2,2}}
\Big)
\, .
\end{align}
With the gauge-parameter values in Eq.\ \re{G2KUTgauge}, the UT-gauge result becomes
\begin{align}
c_{20}^{[K,2]}|_{K=3,4}
=
\Big\{
&\ft{1}{2} \pi^2g^4
-
\ft{16}{45} \pi^4g^6
+
g^8
\left(
\ft{17201}{28350} \pi^6
-
60 \zeta_3^2
\right)
\, ,
\nonumber\\
&
\ft{1}{2} \pi^2 g^4
-
\ft{2}{5} \pi^4g^6
+
g^8
\left(
\ft{1399}{2100} \pi^6
-
48\zeta_3^2
\right)
\Big\}
\, , \nonumber\\
c_{20}^{[K>4,2]}
=
&
\ft{1}{2} \pi^2 g^4
-
\ft{2}{5} \pi^4 g^6
+
g^8
\left(
\ft{617}{945} \pi^6
-
48\zeta_3^2
\right)
\, .
\end{align}
Thus, $c_{20}^{[K,2]}$ becomes independent of $K$ for $K>4$, and its $\zeta_3^2$ term coincides with that of
$c_{02}^{[K\geq 3,2]}$ below.
By contrast, the coefficient $c_{02}^{[K\geq 3,2]}$ is unique and is given by
\begin{align}
c_{02}^{[K\geq 3,2]}
=
\ft{2}{3} \pi^2 g^4
-
\ft{4}{9} \pi^4 g^6
+
g^8
\left(
\ft{34}{45} \pi^6
-
48 \zeta_3^2
\right)
\, .
\end{align}
At $K=3$, the coefficient $c_{01}^{[3,2]}$ also becomes gauge-dependent at four loops:
\begin{align}
c_{01}^{[3,2]}
=
-12 g^4\zeta_3
&
+
g^6
\big(
\ft{8}{3} \pi^2 \zeta_3
+
80\zeta_5
\big)
\nonumber\\
&
+
g^8
\Big(
-\ft{32}{9} \pi^4
+
\ft{4}{27} \pi^6
+
16 \pi^2\zeta_3
-
\ft{44}{45} \pi^4 \zeta_3
-
\ft{80}{3}\pi^2 \zeta_5
-
700\zeta_7
\nonumber\\
&\quad
-
\big(
\ft{16}{3} \pi^2
-
\ft{4}{9} \pi^4 
\big)
\textcolor{red}{r_{1,2}}
+
\big(
\ft{256}{3} \pi^2
-
\ft{32}{9} \pi^4
-
\ft{112}{3}\pi^2 \zeta_3
\big)
\textcolor{red}{r_{2,2}}
\Big)
\, .
\end{align}
In the UT gauge, however, we find
\begin{align}
c_{01}^{[3,2]}
=
&
-12g^4\zeta_3
+
g^6
\left(
\ft{8}{3} \pi^2 \zeta_3
+
80\zeta_5
\right)
+
g^8
\left(
-\ft{6}{5} \pi^4 \zeta_3
-
\ft{80}{3} \pi^2 \zeta_5
-
700\zeta_7
\right)
\, , \\
c_{01}^{[K>3,2]}
=
&
-12g^4\zeta_3
+
g^6
\left(
\ft{8}{3} \pi^2 \zeta_3
+
80\zeta_5
\right)
+
g^8
\left(
-\ft{28}{45} \pi^4 \zeta_3
-
40\pi^2\zeta_5
-
700\zeta_7
\right)
\, .
\end{align}
For completeness, let us also display the coefficient $c_{10}^{[3,2]}$ before gauge fixing:
\begin{align}
c_{10}^{[3,2]}
=
- 6 g^4\zeta_3
&
+
g^6
\Big(
\ft{16}{3} \pi^2
-
\ft{2}{9} \pi^4
-
24\zeta_3
+
\ft{62\pi^2}{15}\zeta_3
+
20 \zeta_5
\nonumber\\
&\quad
+
\big(
8 
-
\ft{2}{3} \pi^2 
\big)
\textcolor{red}{r_{1,2}}
-
\big(
128
-
\ft{16\pi^2}{3}
-
56\zeta_3
\big)
\textcolor{red}{r_{2,2}}
\Big)
\, .
\end{align}
In the UT gauge, we obtain
\begin{align}
c_{10}^{[3,2]}
=
-6g^4\zeta_3
+
g^6
\left(
\ft{67\pi^2}{15}\zeta_3
+
20\zeta_5
\right)
\, , \\
c_{10}^{[K>3,2]}
=
-6g^4 \zeta_3 + g^6 \left( 4\pi^2 \zeta_3 + 40\zeta_5 \right)
\, .
\end{align}
The same reasoning that explained the failed three-loop prediction for $c_{10}^{[3,2]}$ implies that some four-loop
coefficients can be trusted only for $K>4$. At present, this remains a hand-waving argument; a robust criterion is still lacking.
We nevertheless leave these results as a challenge for future explicit four-loop calculations in $\mathcal{N} = 4$ sYM.

\subsection{Non-minimal UT gauge and connected correlations}

How robust are these four-loop predictions? One might hope that they are reliable whenever (i) the results are UT and (ii) they
plateau beyond some number of excitations. The intrinsic ambiguity of crossing solutions, however, makes these two criteria
insufficient. For the simple correlator at hand, the problem first appears at four loops; for the more complicated correlators of
the next section, it is already relevant at three loops.

The gauge transformation in Eq.\ \re{MinimalUTgauge} is {\sl minimal} in the sense that it renders the expansion coefficients
UT. At higher perturbative orders, however, infinitely many additional terms leave the expectation values unchanged while
contributing to cross-correlations with the flux-tube energies. Consider just two terms that can be added to the right-hand side
of Eq.\ \re{MinimalUTgauge}:
\begin{align}
C_s^{(2, 0)} \to C_s^{\prime (2, 0)} 
&
= C_s^{(2, 0)}
+ 
\dots 
\nonumber\\
&
+
\big[ \tilde\Gamma^{(1)}_{K-2} - \vev{\tilde\Gamma^{(1)}_{K-2}} \big] \textcolor{red}{q_{1, 2}} 
+
\big[ \Gamma^{(2)}_{K-2} - \vev{ \Gamma^{(2)}_{K-2}} \big] \textcolor{red}{q_{2, 2}} 
\, ,
\end{align}
where the ellipses stand for the minimal choice and where we have introduced
\begin{align}
\tilde\Gamma^{(1)}_{K-2}
=
\sum_{\ell = 1}^{K-2} \tilde\varepsilon (u_\ell)
\, ,
\end{align}
with
\begin{align}
\tilde\varepsilon (u)
\equiv
\psi^{(1)}\!\left(\ft{1}{2}+iu\right)
+
\psi^{(1)}\!\left(\ft{1}{2}-iu\right)
\, .
\end{align}

Under this non-minimal gauge transformation, the coefficients $c_{20}^{[K,2]}$, $c_{01}^{[K,2]}$, and $c_{10}^{[K,2]}$ acquire
additional contributions:
\begin{align}
\label{c20}
c_{20}^{[K,2]}
&
=
\dots
+
\ft{1}{8} g^8
\Big[
\vev{\vev{\vev{\tilde\Gamma^{(1)}_{K-2} \big( \Gamma^{(1)}_{K-2} \big)^2}}}
\textcolor{red}{q_{1,2}}
+
\vev{\vev{\vev{\Gamma^{(2)}_{K-2} \big( \Gamma^{(1)}_{K-2} \big)^2}}}
\textcolor{red}{q_{2,2}}
\Big]
\, , \\
c_{01}^{[K,2]}
&
=
\dots
+
g^6 \Gamma_{\rm cusp} (m_2(2)^2 - m_2(3))
\Big[
\vev{\vev{\vev{\tilde\Gamma^{(1)}_{K-2} \Gamma^{(1)}_{K-2} }}}
\textcolor{red}{q_{1,2}}
+
\vev{\vev{\vev{\Gamma^{(2)}_{K-2} \Gamma^{(1)}_{K-2}}}}
\textcolor{red}{q_{2,2}}
\Big]
\, , \\
c_{10}^{[K,2]}
&
=
\dots
+
\ft{1}{2} g^6
\Big[
\vev{\vev{\vev{\tilde\Gamma^{(1)}_{K-2} \Gamma^{(1)}_{K-2} }}}
\textcolor{red}{q_{1,2}}
+
\vev{\vev{\vev{\Gamma^{(2)}_{K-2} \Gamma^{(1)}_{K-2}}}}
\textcolor{red}{q_{2,2}}
\Big]
\, . \label{c10}
\end{align}
Here we have introduced the connected correlations $\vev{\vev{\vev{\dots}}}$, defined by
\begin{align}
\vev{\vev{\vev{\tilde\Gamma^{(1)}_{K-2} \Gamma^{(1)}_{K-2} }}}
=
\vev{\tilde\Gamma^{(1)}_{K-2}  \ \Gamma^{(1)}_{K-2}}
&
-
\vev{\tilde\Gamma^{(1)}_{K-2}} \vev{\Gamma^{(1)}_{K-2}}
\, , \\
\vev{\vev{\vev{\Gamma^{(2)}_{K-2} \Gamma^{(1)}_{K-2} }}}
=
\vev{\Gamma^{(2)}_{K-2}  \ \Gamma^{(1)}_{K-2}}
&
-
\vev{\Gamma^{(2)}_{K-2}} \vev{\Gamma^{(1)}_{K-2}}
\, , \\
\vev{\vev{\vev{\tilde\Gamma^{(1)}_{K-2} \big( \Gamma^{(1)}_{K-2} \big)^2}}}
=
\vev{\tilde\Gamma^{(1)}_{K-2}  \big( \Gamma^{(1)}_{K-2} \big)^2}
&
-
\vev{\tilde\Gamma^{(1)}_{K-2}} \vev{\big( \Gamma^{(1)}_{K-2} \big)^2}
\\
&
-
2
\vev{\tilde\Gamma^{(1)}_{K-2} \Gamma^{(1)}_{K-2}} \vev{\Gamma^{(1)}_{K-2}}
+
2
\vev{\tilde\Gamma^{(1)}_{K-2}} \vev{\Gamma^{(1)}_{K-2}}^2
\, , \nonumber\\
\vev{\vev{\vev{\Gamma^{(2)}_{K-2} \big( \Gamma^{(1)}_{K-2} \big)^2}}}
=
\vev{\Gamma^{(2)}_{K-2}  \big( \Gamma^{(1)}_{K-2} \big)^2}
&
-
\vev{\Gamma^{(2)}_{K-2}} \vev{\big( \Gamma^{(1)}_{K-2} \big)^2}
\\
&
-
2
\vev{\Gamma^{(2)}_{K-2} \Gamma^{(1)}_{K-2}} \vev{\Gamma^{(1)}_{K-2}}
+
2
\vev{\Gamma^{(2)}_{K-2}} \vev{\Gamma^{(1)}_{K-2}}^2
\, . \nonumber
\end{align}
Their analytic evaluation yields
\begin{align}
\vev{\vev{\vev{\tilde\Gamma^{(1)}_{K-2} \Gamma^{(1)}_{K-2} }}}
&
=
\{ - \zeta_3, - 2 \zeta_3, - 2 \zeta_3, - 2 \zeta_3, - 2 \zeta_3, \dots \}
\, , \\
\vev{\vev{\vev{\Gamma^{(2)}_{K-2} \Gamma^{(1)}_{K-2} }}}
&
=
\{ - \ft{\pi^4}{18}, - \ft{4 \pi^4}{45}, - \ft{\pi^4}{15}, - \ft{\pi^4}{15}, - \ft{\pi^4}{15}, \dots \}
\, , \\
\vev{\vev{\vev{\tilde\Gamma^{(1)}_{K-2} \big( \Gamma^{(1)}_{K-2} \big)^2}}}
&
=
\{
- \ft{\pi^4}{180}, - \ft{\pi^4}{180}, 0, 0, 0, \dots
\}
\, , \\
\vev{\vev{\vev{\Gamma^{(2)}_{K-2} \big( \Gamma^{(1)}_{K-2} \big)^2}}}
&
=
\{
- \ft{2}{3} \zeta_3 + 6 \zeta_5, - \ft{2}{3} \zeta_3 + 7 \zeta_5, \zeta_5, 0, 0, \dots
\}
\, .
\end{align}

Substituting these results into Eqs.\ \re{c20}--\re{c10}, we immediately see that $q_{2,2}$ must vanish to preserve the proper
transcendental weight; the only alternative, $q_{2,2}\sim\pi$, is inadmissible. At this stage, the sole restriction on $q_{1,2}$ is
that it be proportional to $\pi^2$, which does not spoil the UT structure. The additional terms are then
\begin{align}
c_{20}^{[K,2]}
&= 
\dots +
\left\{ -g^8 \ft{1}{1440} \pi^4 q_{1,2}, -g^8 \ft{1}{1440} \pi^4 q_{1,2}, 0, 0, \dots \right\}
\, , \\
c_{01}^{[K,2]}
&= 
\dots +
\left\{ g^8 \ft{1}{3} \pi^2\zeta_3 q_{1,2}, g^8 \ft{2}{3} \pi^2\zeta_3 q_{1,2}, g^8 \ft{2}{3} \pi^2\zeta_3 q_{1,2}, g^8 \ft{2}{3} \pi^2\zeta_3 q_{1,2}, \dots \right\}
\, , \\
c_{10}^{[K,2]}
&= 
\dots +
\left\{ -g^6 \ft{1}{2}\zeta_3 q_{1,2}, -g^6 \zeta_3 q_{1,2}, -g^6 \zeta_3 q_{1,2}, -g^6 \zeta_3 q_{1,2}, \dots \right\}
\, .
\end{align}
A nonzero $q_{1,2}$ at three loops would, however, spoil the agreement already achieved with the data, so we set it to zero.
Thus, the non-minimal gauge contributes nothing beyond the minimal UT gauge: the UT structure is rigid, and the UT gauge is robust
for this correlator. As we now show, this ceases to be true for more complicated correlators.

\section[Twist-K--twist-K null correlator]{Twist-\textbf{\textit K}--twist-\textbf{\textit K} null correlator}
\label{GKK3Lsection}

We finally turn to the correlators $\cG_{[K,K]}$. Because of their greater gauge freedom, we restrict the calculation to three
loops. We use the crossing solutions of Sect.\ \ref{3LGKKsection}, supplemented by a set of non-minimal gauge parameters.

We begin with the `superleading' logarithms. Substituting the cumulants from Appendix \ref{CumulantsApp}, we obtain
\begin{align}
c_{30}^{[K,K]} = c_{03}^{[K,K]}
&=
\left\{
{-3g^6 \zeta_3, \, K=3
\atop
- \ft{8}{3}g^6\zeta_3, \, K>3}
\right.
\, ,
\\
c_{20}^{[K,K]} = c_{02}^{[K,K]}
&=
\left\{
{\ft{1}{2} \pi^2 g^4 - \ft{16}{45} \pi^4g^6,
\, K=3
\atop
\ft{1}{2} \pi^2 g^4 -\ft{2}{5} \pi^4 g^6,
\, \, K>3}
\right.
\, .
\end{align}
These coefficients are gauge-independent.

For the double-logarithmic coefficient $c_{11}^{[K,K]}$, consider, for example, $K=3$:
\begin{align}
c_{11}^{[3,3]}
=
-
2g^2
+
\ft{2}{3} \pi^2 g^4
+g^6
\big[
2\pi^2
-
\ft{52}{45} \pi^4
-
4
\left(
1 - \ft{1}{12} \pi^2
\right)
\textcolor{red}{r_{0,2}}
\big]
\, .
\end{align}
The gauge parameter $r_{0,2}$ remains free. Uniform transcendentality requires
\begin{align}
r_{0,2}
&=
\frac{\pi^2}{2}
\, .
\end{align}
This immediately yields a UT coefficient $c_{11}^{[K,K]}$ for every $K$, in agreement with the data:
\begin{align}
c_{11}^{[K,K]}
=
- 2 g^2
+
\ft{2}{3} \pi^2 g^4
-
\ft{89}{90} \pi^4g^6
\, .
\end{align}
Incidentally, with this value of $r_{0,2}$, Eq.\ \re{GKKconsistencyW1} is satisfied provided the gauge parameter $\lambda$
is set to $-\ft18$.

The $q_{2,2}$ shift in Eq.\ \re{NonMinimalGT} multiplies a weight-four covariance. Uniform transcendentality would therefore require
$q_{2,2}$ to carry weight one, i.e., to be proportional to $\pi$, which we exclude. Hence $q_{2,2}=0$ also for this correlator.

Finally, we come to the coefficients $c_{10}^{[K,K]}=c_{01}^{[K,K]}$. They contain three gauge parameters in addition to the
previously fixed $r_{0,2}$. Two of them, $r_{1,2}$ and $r_{2,2}$, are minimal: with an appropriate choice, the coefficients are
UT for every $K$. Leaving the remaining parameter $q_{1,2}$ unfixed and choosing
\begin{align}
r_{1,2}
=
\frac{3}{2}\zeta_3
\, , \qquad
r_{2,2}
=
\frac{\pi^2}{32}
\, ,
\end{align}
we obtain, for $K>3$,
\begin{align}
c_{10}^{[K,K]}
=
-
6 g^4\zeta_3
+
g^6
\big(
\ft{5}{3} \pi^2\zeta_3
-
\zeta_3 \textcolor{red}{q_{1,2}}
+
40 \zeta_5
\big)
\, .
\end{align}
Agreement with the data requires
\begin{align}
q_{1,2}
&=
-\frac{\pi^2}{3}
\, .
\end{align}
At $K=3$, however, the three-loop coefficient is incorrect:
\begin{align}
c_{10}^{[3,3]}
&=
-
6 g^4 \zeta_3
+
g^6
\big(
\ft{143}{60} \pi^2\zeta_3
+20\zeta_5
\big)
\, .
\end{align}
This was anticipated: hopping interactions invalidate the flux-tube picture for the $\cG_{[3,3]}$ correlator starting at three loops.

The lesson is that crossing becomes less restrictive as the intermediate states in the partial-wave expansion grow more involved.
The resulting larger space of gauge parameters cannot be fixed unambiguously by a universal principle. We have used uniform
transcendentality for this purpose, but it does not appear to be sufficient on its own.

\section{Conclusions}

In this paper, we developed a framework based on crossing and flux-tube dynamics for studying four-point correlators of half-BPS
operators in the null limit. Using the conformal bootstrap to constrain such correlators is not new; our formulation makes precise,
however, just how much freedom crossing leaves behind. We proposed a set of resummed formulas for correlators carrying a single
R-symmetry representation in each OPE channel. Their drawback is a new weight factor that, under our present assumptions, can be fixed
only order by order in perturbation theory.

OPE channels populated by high-twist operators necessarily involve averages over quantum numbers that enumerate anomalous
dimensions within their allowed bands. In the large-spin limit, we recast these averages in terms of flux-tube excitations propagating
on top of the GKP background. The resulting self-consistent picture agrees with the available three-loop data for the correlators
considered here.

The most unsettling feature of our construction is the large space of gauge transformations that leave the crossing solutions intact
but alter the null correlators. We managed to fix this freedom perturbatively, using uniform transcendentality as a guide. Whether a
nonperturbative principle can enforce the same choice remains an open question.

Our analysis opens several avenues. Let us mention a few.

First, can the framework be generalized to correlators containing several R-symmetry representations in their partial-wave expansions?
The simplest testing ground is the correlator with sextuple $\{0,0,K_1-2,K_2-2,0,0\}$. Its $t$-channel receives only twist-two
operators, whereas its $s$-channel is dominated by twist $K_1+K_2-2$. Whether the various internal-symmetry irreps can be encoded
by a rational function of hole rapidities, in analogy with the pentagon OPE for null Wilson loops
\cite{Basso:2015uxa,Belitsky:2016vyq}, remains to be seen.

Second, the same framework extends to the infinite-weight limit of polarized correlators and to the emergence of the
octagon. Its double-logarithmic exponent is governed by the octagon anomalous dimension, known exactly as a function of the
't Hooft coupling \cite{Belitsky:2019fan}, whereas the double-logarithmic behavior studied here begins with the cusp anomalous
dimension and is followed by a series of cumulants. In a forthcoming paper \cite{Belitsky:HoleGas}, we show that these two
pictures are reconciled by describing the octagon as a gas of holes on top of the GKP background: as the bridges grow, the
cumulants of the hole ensemble resum, and the octagon anomalous dimension emerges as a collective property of the hole gas.
The measures and form factors determined here provide the perturbative input for that description, and its exact form
should, in turn, constrain the creation and absorption form factors beyond the orders computed in this paper.

Finally, can the flux-tube picture advocated here be bootstrapped to all orders in the 't Hooft coupling? Do the creation and
annihilation form factors obey some version of the Watson and mirror equations, and can those equations be solved? We leave these
questions for the future.

\begin{acknowledgments}

I would like to thank G.P. Korchemsky for his interest in this work and for reading the manuscript.

\end{acknowledgments}

\appendix

\section{Integral basis}
\label{app:basis}

In this appendix, we present explicit definitions of all off-shell position-space integrals entering the perturbative expansion through
three loops. We work in Euclidean signature. We rescale these integrals relative to the conventions of Ref.\ \cite{Chicherin:2015edu}
to match the 't Hooft-coupling convention standard in integrability, so that every $L$-loop integral below includes the factor
\begin{equation}
I^{(L)}_{\tt here} = 4^L I^{(L)}_{\tt 1512.02926}
\, .
\end{equation}
We also relabel the space-time points according to our correlator conventions:
\begin{align*}
G_{2,K,K,2}
&= \vev{\cO_2(x_1)\cO_K(x_2)\cO_K(x_3)\cO_2(x_4)},
\\
G_{2,K,2K-2,K}
&= \vev{\cO_2(x_1)\cO_K(x_2)\cO_{2K-2}(x_3)\cO_K(x_4)}.
\end{align*}

The ladder integrals at one, two, and three loops---denoted by $g$, $h$, and $L$, respectively---together with the factorized box--ladder integral, are
\begin{align}
 g_{1234}
 &
 =
 -\frac1{\pi^2}\int
 \frac{\dd^4x_5}{x_{15}^2x_{25}^2x_{35}^2x_{45}^2}
 \, , \\
 h_{12;34}
 &
 =
 +
 \frac{x_{34}^2}{\pi^4}\int 
 \frac{\dd^4x_5\,\dd^4x_6}{(x_{15}^2x_{35}^2x_{45}^2)x_{56}^2 (x_{26}^2x_{36}^2x_{46}^2)}
\, , \\
L_{12;34}
&
=
-
\frac{x_{34}^4}{\pi^6} \int
\frac{\dd^4x_5\,\dd^4x_6\,\dd^4x_7} {(x_{15}^2x_{35}^2x_{45}^2)x_{56}^2 (x_{36}^2x_{46}^2)x_{67}^2 (x_{27}^2x_{37}^2x_{47}^2)}
\, ,\\
(g\!\times\!h)_{12;34}
& =
 -\frac{x_{12}^2x_{34}^4}{\pi^6}
\int 
\frac{\dd^4x_5\,\dd^4x_6\,\dd^4x_7}{(x_{15}^2x_{25}^2x_{35}^2x_{45}^2) (x_{16}^2x_{36}^2x_{46}^2)x_{67}^2 (x_{27}^2x_{37}^2x_{47}^2)} 
\nonumber\\
&
=
x_{12}^2x_{34}^2 \, g_{1234} h_{12;34}.
\end{align}
Here the last integral is simply the factorized box--ladder integral. 

It is convenient to absorb the ubiquitous factor $x_{13}^2 x_{24}^2$ into the definitions of all these integrals, including the factorized topology:
\begin{equation}
\label{HattedIntegrals}
 \widehat I_{ij;kl} \equiv x_{13}^2 x_{24}^2 I_{ij;kl}
 \, .
\end{equation}
Recall the Davydychev--Usyukina ladder functions $\Phi^{(L)}(u,v)$, defined by
\begin{align}\notag
& \Phi^{(L)}(u,v) = -{1\over z-\bar z} f^{(L)}\lr{{z\over z-1},{\bar z\over \bar z -1}}
\\
& f^{(L)}(z,\bar z) = \sum_{r=0}^L {(-1)^r (2L-r)!\over r!(L-r)!L!} \ln^r(z\bar z) \big[{\rm Li}_{2L-r}(z)-{\rm Li}_{2L-r}(\bar z) \big]
\, ,
\end{align}
where $u=z\bar z$ and $v=(1-z)(1-\bar z)$. The direct-channel integrals can then be written as
\begin{align}
\widehat{g}_{1234}
&= - \Phi^{(1)}(u,v),&
\widehat{h}_{13;24}
&= + \Phi^{(2)}(u,v),\nonumber\\
\widehat{L}_{13;24}
&= - \Phi^{(3)}(u,v),&
\widehat{(g\!\times\!h)}_{13;24}
&= - \Phi^{(1)}(u,v)\Phi^{(2)}(u,v).
\label{eq:DU-direct}
\end{align}
The cross-channel orientations are
\begin{align}
\widehat{h}_{12;34}
&=
+\frac{1}{u}\Phi^{(2)}\!\left(\frac1u,\frac vu\right),&
\widehat{h}_{14;23}
&=
+\frac{1}{v}\Phi^{(2)}\!\left(\frac uv,\frac1v\right),\notag\\
\widehat{L}_{12;34}
&=
-\frac{1}{u}\Phi^{(3)}\!\left(\frac1u,\frac vu\right),&
\widehat{L}_{14;23}
&=
-\frac{1}{v}\Phi^{(3)}\!\left(\frac uv,\frac1v\right),\notag\\
\widehat{(g\!\times\!h)}_{12;34}
&= 
- \Phi^{(1)}(u,v) \Phi^{(2)}\!\left(\frac1u,\frac vu\right),&
\widehat{(g\!\times\!h)}_{14;23}
&=
- \Phi^{(1)}(u,v) \Phi^{(2)}\!\left(\frac uv,\frac1v\right)
\, ,
\label{eq:DU-crossed}
\end{align}
such that
\begin{equation}
\frac{1}{u} \widehat{(g\!\times\!h)}_{12;34} =\widehat g_{1234}\widehat h_{12;34}
\, ,\qquad
\widehat{(g\!\times\!h)}_{13;24} =\widehat g_{1234}\widehat h_{13;24}
\, ,\qquad
\frac{1}{v}\widehat{(g\!\times\!h)}_{14;23} = \widehat g_{1234}\widehat h_{14;23}
\, .
\label{eq:factorized-hat-relation}
\end{equation}
The easy and hard integrals are
\begin{align}
E_{12;34}
&=\frac{x_{23}^2x_{24}^2}{(-\pi^2)^3}\int
\frac{\dd^4x_5\,\dd^4x_6\,\dd^4x_7\,x_{16}^2}{(x_{15}^2x_{25}^2x_{35}^2)x_{56}^2 (x_{26}^2x_{36}^2x_{46}^2)x_{67}^2 (x_{17}^2x_{27}^2x_{47}^2)}
\, ,\\
H_{12;34}
&
=\frac{x_{41}^2x_{23}^2x_{34}^2}{(-\pi^2)^3}\int
\frac{\dd^4x_5\,\dd^4x_6\,\dd^4x_7\,x_{57}^2}{(x_{15}^2x_{25}^2x_{35}^2x_{45}^2)x_{56}^2(x_{36}^2x_{46}^2)x_{67}^2(x_{17}^2x_{27}^2x_{37}^2x_{47}^2)}
\, ,
\end{align}
respectively. Their hatted versions are defined as in Eq.\ \re{HattedIntegrals}, and all other configurations follow by permuting the external labels. 
The complete basis needed through three loops for the correlators discussed in the body of the paper is
\begin{align}
\mathcal I_1={}&\{g_{1234}\},\nonumber\\
\mathcal I_2={}&\{h_{12;34},h_{13;24},h_{14;23},[g_{1234}]^2\},\notag\\
\mathcal I_3={}&\{(g\!\times\!h)_{12;34},(g\!\times\!h)_{13;24}, (g\!\times\!h)_{14;23},
L_{12;34},L_{13;24},L_{14;23},\nonumber\\
&\qquad\qquad\qquad E_{12;34},E_{13;24},E_{14;23}, H_{12;34},H_{13;24},H_{14;23}\}
\, .
\end{align}
While the ladder-type integrals have been known for ages, the easy and hard integrals were evaluated rather more recently \cite{Drummond:2013nda}, 
using the Euclidean OPE constraints of Ref.\ \cite{Eden:2012rr}.

\section{Null limits of basis integrals}
\label{app:null-integrals}

Having established the definitions, we now list the null limits in which both cross ratios tend to zero at the same rate.
These expressions were obtained\footnote{We would like to thank G.P. Korchemsky for sharing his unpublished notes 
\cite{Kor2018}.} by applying the \texttt{HPL} package \cite{Maitre:2005uu} to the results of Ref.\ \cite{Drummond:2013nda}. 
We discard log-free and power-suppressed terms, denote the resulting equalities by $\logeq$, and write each answer as a terminating series in powers of the logarithms
\begin{align*}
\Lu=\log u, \quad \Lv=\log v .
\end{align*}
We find
\begin{align}
\widehat g_{1234}
\logeq{}&
-\Lu\Lv
\, , \\
\widehat h_{13;24}
\logeq{}&
\frac14\Lu^2\Lv^2+\frac{\pi^2}{3}\Lu\Lv +\frac{\pi^2}{12}(\Lu^2+\Lv^2)
\, , \\
\widehat h_{12;34}
\logeq{}&
-3\zeta_3\Lv+\frac{\pi^2}{12}\Lv^2
\, ,\\
\widehat h_{14;23}
\logeq{}&
-3\zeta_3\Lu+\frac{\pi^2}{12}\Lu^2
\, , \\
\widehat L_{13;24}
\logeq{}&
-\frac{\Lu^3\Lv^3}{36}
-\frac{\pi^2}{36}(\Lu^3\Lv+\Lu\Lv^3) 
-\frac{\pi^2}{12}\Lu^2\Lv^2
-\frac{7\pi^4}{60}\Lu\Lv
-\frac{7\pi^4}{180}(\Lu^2+\Lv^2)
\, ,\\
\widehat L_{12;34}
\logeq{}&
\frac{\zeta_3}{6}\Lv^3+10\zeta_5\Lv-\frac{\pi^4}{45}\Lv^2
\, ,\\
\widehat L_{14;23}
\logeq{}&
\frac{\zeta_3}{6}\Lu^3+10\zeta_5\Lu-\frac{\pi^4}{45}\Lu^2
\, , \\
\widehat{(g\!\times\!h)}_{13;24}
\logeq{}&
-\frac14\Lu^3\Lv^3
-\frac{\pi^2}{12}(\Lu^3\Lv+\Lu\Lv^3)
-\frac{5\pi^2}{12}\Lu^2\Lv^2
-\frac{41\pi^4}{180}\Lu\Lv
-\frac{\pi^4}{36}(\Lu^2+\Lv^2)
\, ,\\
\frac1u\widehat{(g\!\times\!h)}_{12;34}\logeq{}&
-3\zeta_3\Lu\Lv^2-\pi^2\zeta_3\Lv
+\frac{\pi^2}{12}\Lu\Lv^3
+\frac{\pi^4}{15}\Lu\Lv+\frac{\pi^4}{36}\Lv^2
\, ,\\
\frac1v\widehat{(g\!\times\!h)}_{14;23}
\logeq{}&
-3\zeta_3\Lv\Lu^2-\pi^2\zeta_3\Lu
+\frac{\pi^2}{12}\Lv\Lu^3
+\frac{\pi^4}{15}\Lu\Lv+\frac{\pi^4}{36}\Lu^2
\, .
\end{align}
The easy integrals read
\begin{align}
\widehat E_{13;24}
\logeq{}&
\frac{\pi^2\zeta_3}{3}(\Lu+\Lv)-\frac{\pi^4}{36}\Lu\Lv
\, , \\ 
\widehat E_{14;23}
\logeq{}&
\frac{2\pi^2\zeta_3}{3}\Lv-\frac{\pi^4}{90}\Lv^2
-\frac{2\zeta_3}{3}\Lv^3 
\nonumber\\
&
+\Lu\left(10\zeta_5+\frac{\pi^2\zeta_3}{3}
-\frac{\pi^4}{36}\Lv+\frac{3\zeta_3}{2}\Lv^2
-\frac{\pi^2}{18}\Lv^3\right),
\label{eq:null-E14}\\
\widehat E_{12;34}(\Lu,\Lv)\logeq{}&
\widehat E_{14;23}(\Lv,\Lu)
\, .
\label{eq:null-E12}
\end{align}
Finally, introducing the auxiliary function
\begin{align}
\mathcal H_{12}^{\log}(\Lu,\Lv)
&
=
\left(-20\zeta_5+\frac{4\pi^2\zeta_3}{3}\right)\Lv
+\frac{\pi^4}{45}\Lv^2-\zeta_3\Lv^3 
\nonumber\\
&+\Lu\left(-40\zeta_5-\frac{4\pi^4}{15}\Lv
+6\zeta_3\Lv^2-\frac{\pi^2}{9}\Lv^3\right)
\, ,
\end{align}
we can write the null limits of the hard integrals as
\begin{align}
\frac{1}{v} \widehat{H}_{12;34}
\logeq{}&
\mathcal{H}_{12}^{\log}(\Lu,\Lv),
\label{eq:null-H12}\\
\widehat{H}_{14;23}
\logeq{}&
\mathcal{H}_{12}^{\log}(\Lv,\Lu),
\label{eq:null-H14}\\
\widehat{H}_{13;24}
\logeq{}&
0
\, .
\label{eq:null-H13}
\end{align}

\section{Bessel moments}
\label{BesselMomApp}

The first six Bessel moments of the logarithms, corresponding to $p=1,\ldots,6$, are
\begin{alignat}{1}
m_K (p)
=
\big\{ &1,
\nonumber\\[2pt]
&\ft{1}{2}\psi^{(0)}(K-1),
\nonumber\\[2pt]
&\ft{\pi^2}{24}
 +\ft{1}{4}\bigl[\psi^{(0)}(K-1)\bigr]^2
 +\ft{1}{4}\psi^{(1)}(K-1),
\nonumber\\[2pt]
&\ft{\pi^2}{16}\psi^{(0)}(K-1)
 +\ft{1}{8}\bigl[\psi^{(0)}(K-1)\bigr]^3
 \nonumber\\
&\qquad
 +\ft{3}{8}\psi^{(0)}(K-1)\psi^{(1)}(K-1)
 +\ft{1}{8}\psi^{(2)}(K-1)
 -\ft{\zeta_3}{4},
\nonumber\\[2pt]
&\ft{3\pi^4}{320}
 +\ft{\pi^2}{16}\bigl[\psi^{(0)}(K-1)\bigr]^2
 +\ft{1}{16}\bigl[\psi^{(0)}(K-1)\bigr]^4
 +\ft{\pi^2}{16}\psi^{(1)}(K-1)
 \nonumber\\
&\qquad
 +\ft{3}{8}\bigl[\psi^{(0)}(K-1)\bigr]^2\psi^{(1)}(K-1)
 +\ft{3}{16}\bigl[\psi^{(1)}(K-1)\bigr]^2
 \nonumber\\
&\qquad
 +\ft{1}{4}\psi^{(0)}(K-1)\psi^{(2)}(K-1)
 \nonumber\\
&\qquad
 +\ft{1}{16}\psi^{(3)}(K-1)
 -\ft{1}{2}\psi^{(0)}(K-1)\zeta_3,
\nonumber\\[2pt]
&\ft{3\pi^4}{128}\psi^{(0)}(K-1)
 +\ft{5\pi^2}{96}\bigl[\psi^{(0)}(K-1)\bigr]^3
 \nonumber\\
&\qquad
 +\ft{1}{32}\bigl[\psi^{(0)}(K-1)\bigr]^5
 \nonumber\\
&\qquad
 +\ft{5\pi^2}{32}\psi^{(0)}(K-1)\psi^{(1)}(K-1)
 +\ft{5}{16}\bigl[\psi^{(0)}(K-1)\bigr]^3\psi^{(1)}(K-1)
 \nonumber\\
&\qquad
 +\ft{15}{32}\psi^{(0)}(K-1)
   \bigl[\psi^{(1)}(K-1)\bigr]^2
 +\ft{5\pi^2}{96}\psi^{(2)}(K-1)
 \nonumber\\
&\qquad
 +\ft{5}{16}\bigl[\psi^{(0)}(K-1)\bigr]^2
   \psi^{(2)}(K-1)
 +\ft{5}{16}\psi^{(1)}(K-1)\psi^{(2)}(K-1)
 \nonumber\\
&\qquad
 +\ft{5}{32}\psi^{(0)}(K-1)\psi^{(3)}(K-1)
 +\ft{1}{32}\psi^{(4)}(K-1)
 -\ft{5\pi^2}{48}\zeta_3
 \nonumber\\
&\qquad
 -\ft{5}{8}\bigl[\psi^{(0)}(K-1)\bigr]^2\zeta_3
 -\ft{5}{8}\psi^{(1)}(K-1)\zeta_3
 -\ft{3}{4}\zeta_5
 \big\}\,.
\end{alignat}
Here we have dropped all terms involving $\gamma_{\rm E}$ and $\log 2$, since they cancel from the combinations of moments entering the
perturbative coefficients of interest. They are merely a pesky nuisance of the intermediate calculation.

\section[Twist-K--twist-2 correlator: three and four loops]{Twist-\textbf{\textit K}--twist-2 correlator: three and four loops}
\label{2Kcorr34loopsApp}

In this appendix, we present explicit solutions of the crossing equations at three and four loops. They are far too long to quote
in the main text.

\subsection{Three loops}

At three loops, the $t$-channel expansion coefficients and the averages of their $s$-channel counterparts are
\begin{align}
C_t^{(3, 3)}
&= -\ft{1}{6} \vev{\big(\Gamma_{K-2}^{(1)}\big)^3} - \vev{\big(\Gamma_{K-2}^{(1)}\big)^2} \Gamma_{\rm cusp}^{(1)} m_K (2) 
- 
2 \vev{\Gamma_{K-2}^{(1)}} \big(\Gamma_{\rm cusp}^{(1)}\big)^2 m_K (3) - \ft{4}{3} \big(\Gamma_{\rm cusp}^{(1)}\big)^3 m_K (4)
\, , \\
C_t^{(3, 2)}
&= 
\ft{1}{2} \vev{C_s^{(1, 0)} \big(\Gamma_{K-2}^{(1)}\big)^2} 
+ 
\ft{1}{2} \vev{C_s^{(1, 1)} \big(\Gamma_{K-2}^{(1)}\big)^2} m_K (2) 
+ 
2 \vev{C_s^{(1, 0)} \Gamma_{K-2}^{(1)}} \Gamma_{\rm cusp}^{(1)} m_K (2) 
\nonumber\\
&
+ 
\vev{\Gamma_{K-2}^{(1)} \Gamma_{K-2}^{(2)}} 
+ 
\vev{\Gamma_{K-2}^{(1)}} \Gamma_{\rm virt}^{(2)} 
+ 
\ft{1}{2} \vev{\big(\Gamma_{K-2}^{(1)}\big)^3} m_2 (2) 
+ 
2 \vev{\Gamma_{K-2}^{(2)}} \Gamma_{\rm cusp}^{(1)} m_K (2) 
\nonumber\\
&
+ 
2 \Gamma_{\rm virt}^{(2)} \Gamma_{\rm cusp}^{(1)} m_K (2) 
+ 
2 \vev{\Gamma_{K-2}^{(1)}} \Gamma_{\rm cusp}^{(2)} m_K (2) 
+ 
3 \vev{\big(\Gamma_{K-2}^{(1)}\big)^2} \Gamma_{\rm cusp}^{(1)} m_2 (2) m_K (2) 
\nonumber\\
&
+ 
4 \Gamma_{\rm cusp}^{(1)} \Gamma_{\rm cusp}^{(2)} m_K (3) 
+ 
2 \vev{\Gamma_{K-2}^{(1)}} \big(\Gamma_{\rm cusp}^{(1)}\big)^2 m_2 (2) m_K (3)
\, , \\
C_t^{(3, 1)} 
&= 
- \vev{C_s^{(2, 0)} \Gamma_{K-2}^{(1)}} 
- \vev{C_s^{(1, 0)} \Gamma_{K-2}^{(2)}} 
- \vev{C_s^{(1, 0)} \big(\Gamma_{K-2}^{(1)}\big)^2} m_2 (2) 
\nonumber\\
&
- \vev{C_s^{(2, 1)} \Gamma_{K-2}^{(1)}} m_K (2) 
- \vev{C_s^{(1, 1)} \Gamma_{K-2}^{(2)}} m_K (2) 
- \vev{C_s^{(1, 1)} \big(\Gamma_{K-2}^{(1)}\big)^2} m_2 (2) m_K (2) 
\nonumber\\
&
- 4 \vev{C_s^{(1, 0)} \Gamma_{K-2}^{(1)}} \Gamma_{\rm cusp}^{(1)} m_2 (2) m_K (2) 
- \vev{C_s^{(2, 2)} \Gamma_{K-2}^{(1)}} m_K (3) 
\nonumber\\
&
- \vev{\Gamma_{K-2}^{(3)}} - \Gamma_{\rm virt}^{(3)} 
- 2 \vev{\Gamma_{K-2}^{(1)} \Gamma_{K-2}^{(2)}} m_2 (2) 
- 2 \vev{\Gamma_{K-2}^{(1)}} \Gamma_{\rm virt}^{(2)} m_2 (2) 
- \vev{\big(\Gamma_{K-2}^{(1)}\big)^3} \big(m_2 (2)\big)^2 
\nonumber\\
&
+ \ft{1}{2} \vev{\big(\Gamma_{K-2}^{(1)}\big)^3} m_2 (3) 
- 2 \Gamma_{\rm cusp}^{(3)} m_K (2) 
- 4 \vev{\Gamma_{K-2}^{(2)}} \Gamma_{\rm cusp}^{(1)} m_2 (2) m_K (2) 
\nonumber\\
&
- 4 \vev{\Gamma_{K-2}^{(1)}} \Gamma_{\rm cusp}^{(2)} m_2 (2) m_K (2) 
- 6 \vev{\big(\Gamma_{K-2}^{(1)}\big)^2} \Gamma_{\rm cusp}^{(1)} \big(m_2 (2)\big)^2 m_K (2) 
\nonumber\\
&
+ 3 \vev{\big(\Gamma_{K-2}^{(1)}\big)^2} \Gamma_{\rm cusp}^{(1)} m_2 (3) m_K (2) 
- 2 \Gamma_{\rm virt}^{(2)} \Gamma_{\rm cusp}^{(1)} \big(m_K (2)\big)^2 
+ 2 \Gamma_{\rm virt}^{(2)} \Gamma_{\rm cusp}^{(1)} m_K (3) 
\nonumber\\
&
+ 2 \vev{\Gamma_{K-2}^{(1)}} \big(\Gamma_{\rm cusp}^{(1)}\big)^2 m_2 (3) m_K (3)
\, , \\
\vev{C_s^{(3, 3)}} 
&= 
-\ft{4}{3} \big(\Gamma_{\rm cusp}^{(1)}\big)^3 m_2 (4)
\, , \\
\vev{C_s^{(3, 2)}} 
&= 
2 \Gamma_{\rm virt}^{(2)} \Gamma_{\rm cusp}^{(1)} m_2 (2) 
+ 
4 \Gamma_{\rm cusp}^{(1)} \Gamma_{\rm cusp}^{(2)} m_2 (3) 
+ 
2 \vev{\Gamma_{K-2}^{(1)}} \big(\Gamma_{\rm cusp}^{(1)}\big)^2 m_2 (2) m_2 (3) 
\nonumber\\
&
- 
2 \vev{\Gamma_{K-2}^{(1)}} \big(\Gamma_{\rm cusp}^{(1)}\big)^2 m_2 (4)
\, , \\
\vev{C_s^{(3, 1)}} 
&= 
-2 \vev{C_s^{(1, 0)} \Gamma_{K-2}^{(1)}} \Gamma_{\rm cusp}^{(1)} \big(m_2 (2)\big)^2 
+ 
2 \vev{C_s^{(1, 0)} \Gamma_{K-2}^{(1)}} \Gamma_{\rm cusp}^{(1)} m_2 (3) 
- 
\Gamma_{\rm virt}^{(3)} 
\nonumber\\
&
- 
2 \Gamma_{\rm cusp}^{(3)} m_2 (2) 
- 
2 \vev{\Gamma_{K-2}^{(2)}} \Gamma_{\rm cusp}^{(1)} \big(m_2 (2)\big)^2 
- 
2 \vev{\Gamma_{K-2}^{(1)}} \Gamma_{\rm cusp}^{(2)} \big(m_2 (2)\big)^2 
\nonumber\\
&
- 
2 \vev{\big(\Gamma_{K-2}^{(1)}\big)^2} \Gamma_{\rm cusp}^{(1)} \big(m_2 (2)\big)^3 
+ 
2 \vev{\Gamma_{K-2}^{(2)}} \Gamma_{\rm cusp}^{(1)} m_2 (3) 
+ 
2 \Gamma_{\rm virt}^{(2)} \Gamma_{\rm cusp}^{(1)} m_2 (3) 
\nonumber\\
&
+ 
2 \vev{\Gamma_{K-2}^{(1)}} \Gamma_{\rm cusp}^{(2)} m_2 (3) 
+ 
3 \vev{\big(\Gamma_{K-2}^{(1)}\big)^2} \Gamma_{\rm cusp}^{(1)} m_2 (2) m_2 (3) 
- 
\vev{\big(\Gamma_{K-2}^{(1)}\big)^2} \Gamma_{\rm cusp}^{(1)} m_2 (4)
\nonumber\\
&
- 
2 \Gamma_{\rm virt}^{(2)} \Gamma_{\rm cusp}^{(1)} m_2 (2) m_K (2)
\, . 
\end{align}
There are also relations among averages involving lower-order coefficients and the high-twist anomalous dimension:
\begin{align}
\vev{C_s^{(1, 1)} \big(\Gamma_{K-2}^{(1)}\big)^2} 
&= 
-2 \vev{\big(\Gamma_{K-2}^{(1)}\big)^2} \Gamma_{\rm cusp}^{(1)} m_2 (2)
\, , \\
\vev{C_s^{(2, 2)} \Gamma_{K-2}^{(1)}} 
&= 
2 \vev{\Gamma_{K-2}^{(1)}} \big(\Gamma_{\rm cusp}^{(1)}\big)^2 m_2 (3)
\, , \\
\vev{C_s^{(2, 1)} \Gamma_{K-2}^{(1)}} 
&= 
-\vev{C_s^{(1, 1)} \Gamma_{K-2}^{(2)}} 
- 
2 \vev{C_s^{(1, 0)} \Gamma_{K-2}^{(1)}} \Gamma_{\rm cusp}^{(1)} m_2 (2) 
- 
\vev{\Gamma_{K-2}^{(1)}} \Gamma_{\rm virt}^{(2)} 
\nonumber\\
&
- 
2 \vev{\Gamma_{K-2}^{(2)}} \Gamma_{\rm cusp}^{(1)} m_2 (2) 
- 
2 \vev{\Gamma_{K-2}^{(1)}} \Gamma_{\rm cusp}^{(2)} m_2 (2) 
- 
2 \vev{\big(\Gamma_{K-2}^{(1)}\big)^2} \Gamma_{\rm cusp}^{(1)} \big(m_2 (2)\big)^2 
\nonumber\\
&
+ 
2 \vev{\big(\Gamma_{K-2}^{(1)}\big)^2} \Gamma_{\rm cusp}^{(1)} m_2 (3)
\, . 
\end{align}
Finally, crossing constrains only the difference of the log-free coefficients:
\begin{align}
C_t^{(3,0)}
&
-\vev{C_s^{(3,0)}}
=
\vev{C_s^{(2,0)}\Gamma_{K-2}^{(1)}}m_2(2)
+
\vev{C_s^{(1,0)}\Gamma_{K-2}^{(2)}}m_2(2)
+
\vev{\Gamma_{K-2}^{(3)}}m_2(2)
\nonumber\\
&\quad
+
\vev{C_s^{(1,0)}}\Gamma_{\rm virt}^{(2)}m_2(2)
+
\Gamma_{\rm virt}^{(3)}m_2(2)
+
\vev{C_s^{(1,0)}\big(\Gamma_{K-2}^{(1)}\big)^2}m_2^2(2)
\nonumber\\
&\quad
+
2\vev{\Gamma_{K-2}^{(1)}\Gamma_{K-2}^{(2)}}m_2^2(2)
+
2\vev{\Gamma_{K-2}^{(1)}}\Gamma_{\rm virt}^{(2)}m_2^2(2)
+
\vev{\big(\Gamma_{K-2}^{(1)}\big)^3}m_2^3(2)
\nonumber\\
&\quad
-
\ft{1}{2}\vev{C_s^{(1,0)}\big(\Gamma_{K-2}^{(1)}\big)^2}m_2(3)
-
\vev{\Gamma_{K-2}^{(1)}\Gamma_{K-2}^{(2)}}m_2(3)
-
\vev{\Gamma_{K-2}^{(1)}}\Gamma_{\rm virt}^{(2)}m_2(3)
\nonumber\\
&\quad
-
\vev{\big(\Gamma_{K-2}^{(1)}\big)^3}m_2(2)m_2(3)
+
\ft{1}{6}\vev{\big(\Gamma_{K-2}^{(1)}\big)^3}m_2(4)
-
\vev{C_s^{(1,0)}}\Gamma_{\rm virt}^{(2)}m_K(2)
-
\Gamma_{\rm virt}^{(3)}m_K(2)
\nonumber\\
&\quad
+
\vev{C_s^{(2,1)}\Gamma_{K-2}^{(1)}}m_2(2)m_K(2)
+
\vev{C_s^{(1,1)}\Gamma_{K-2}^{(2)}}m_2(2)m_K(2)
\nonumber\\
&\quad
+
2\vev{C_s^{(1,0)}\Gamma_{K-2}^{(1)}}\Gamma_{\rm cusp}^{(1)}m_2^2(2)m_K(2)
+
2\vev{\Gamma_{K-2}^{(2)}}\Gamma_{\rm cusp}^{(1)}m_2^2(2)m_K(2)
\nonumber\\
&\quad
+
2\vev{\Gamma_{K-2}^{(1)}}\Gamma_{\rm cusp}^{(2)}m_2^2(2)m_K(2)
+
4\vev{\big(\Gamma_{K-2}^{(1)}\big)^2}\Gamma_{\rm cusp}^{(1)}m_2^3(2)m_K(2)
\nonumber\\
&\quad
-
\ft{1}{2}\vev{C_s^{(1,1)}\big(\Gamma_{K-2}^{(1)}\big)^2}m_2(3)m_K(2)
-
3\vev{\big(\Gamma_{K-2}^{(1)}\big)^2}\Gamma_{\rm cusp}^{(1)}m_2(2)m_2(3)m_K(2)
\nonumber\\
&\quad
+
\vev{C_s^{(2,2)}\Gamma_{K-2}^{(1)}}m_2(2)m_K(3)
-
2\vev{\Gamma_{K-2}^{(1)}}\big(\Gamma_{\rm cusp}^{(1)}\big)^2m_2(2)m_2(3)m_K(3)
\nonumber\\
&\quad
+
\vev{C_s^{(1,1)}\big(\Gamma_{K-2}^{(1)}\big)^2}m_2^2(2)m_K(2)
\, .
\end{align}

\subsection{Four loops}

At four loops, the pattern is the same, although the expressions become more cumbersome. Since they are used in the main text,
we display them here for completeness:
\begin{align}
C_t^{(4, 4)}
&= 
\ft{1}{24} \vev{\big(\Gamma_{K-2}^{(1)}\big)^4} 
+ 
\ft{1}{3} \vev{\big(\Gamma_{K-2}^{(1)}\big)^3} \Gamma_{\rm cusp}^{(1)} m_K (2) 
+ 
\vev{\big(\Gamma_{K-2}^{(1)}\big)^2} \big(\Gamma_{\rm cusp}^{(1)}\big)^2 m_K (3) 
\nonumber\\
&
+ 
\ft{4}{3} \vev{\Gamma_{K-2}^{(1)}} \big(\Gamma_{\rm cusp}^{(1)}\big)^3 m_K (4) 
+ 
\ft{2}{3} \big(\Gamma_{\rm cusp}^{(1)}\big)^4 m_K (5)
\, , \\
C_t^{(4, 3)}
&= 
-\ft{1}{6} \vev{C_s^{(1, 0)}\big(\Gamma_{K-2}^{(1)}\big)^3} 
- 
\ft{1}{6} \vev{C_s^{(1, 1)}\big(\Gamma_{K-2}^{(1)}\big)^3} m_K (2) 
- 
\vev{C_s^{(1, 0)}\big(\Gamma_{K-2}^{(1)}\big)^2} \Gamma_{\rm cusp}^{(1)} m_K (2) 
\nonumber\\
&
- 
2 \vev{C_s^{(1, 0)}\Gamma_{K-2}^{(1)}} \big(\Gamma_{\rm cusp}^{(1)}\big)^2 m_K (3) 
- 
\ft{1}{2} \vev{\big(\Gamma_{K-2}^{(1)}\big)^2 \Gamma_{K-2}^{(2)}} 
- 
\ft{1}{2} \vev{\big(\Gamma_{K-2}^{(1)}\big)^2} \Gamma_{\rm virt}^{(2)} 
\nonumber\\
&
- 
\ft{1}{6} \vev{\big(\Gamma_{K-2}^{(1)}\big)^4} m_2 (2) 
- 
2 \vev{\Gamma_{K-2}^{(1)} \Gamma_{K-2}^{(2)}} \Gamma_{\rm cusp}^{(1)} m_K (2) 
- 
2 \vev{\Gamma_{K-2}^{(1)}} \Gamma_{\rm virt}^{(2)} \Gamma_{\rm cusp}^{(1)} m_K (2) 
\nonumber\\
&
- 
\vev{\big(\Gamma_{K-2}^{(1)}\big)^2} \Gamma_{\rm cusp}^{(2)} m_K (2) 
- 
\ft{4}{3} \vev{\big(\Gamma_{K-2}^{(1)}\big)^3} \Gamma_{\rm cusp}^{(1)} m_2 (2) m_K (2) 
- 
2 \vev{\Gamma_{K-2}^{(2)}} \big(\Gamma_{\rm cusp}^{(1)}\big)^2 m_K (3) 
\nonumber\\
&
- 
2 \Gamma_{\rm virt}^{(2)} \big(\Gamma_{\rm cusp}^{(1)}\big)^2 m_K (3) 
- 
4 \vev{\Gamma_{K-2}^{(1)}} \Gamma_{\rm cusp}^{(1)} \Gamma_{\rm cusp}^{(2)} m_K (3) 
- 
2 \vev{\big(\Gamma_{K-2}^{(1)}\big)^2} \big(\Gamma_{\rm cusp}^{(1)}\big)^2 m_2 (2) m_K (3) 
\nonumber\\
&
- 
4 \big(\Gamma_{\rm cusp}^{(1)}\big)^2 \Gamma_{\rm cusp}^{(2)} m_K (4) 
- 
\ft{4}{3} \vev{\Gamma_{K-2}^{(1)}} \big(\Gamma_{\rm cusp}^{(1)}\big)^3 m_2 (2) m_K (4)
\, , \\
C_t^{(4, 2)}
&= 
\ft{1}{2} \vev{C_s^{(2, 0)}\big(\Gamma_{K-2}^{(1)}\big)^2} 
+ 
\vev{C_s^{(1, 0)}\Gamma_{K-2}^{(1)} \Gamma_{K-2}^{(2)}} 
+ 
\vev{C_s^{(1, 0)}\Gamma_{K-2}^{(1)}} \Gamma_{\rm virt}^{(2)} 
\nonumber\\
&
+ 
\ft{1}{2} \vev{C_s^{(1, 0)}\big(\Gamma_{K-2}^{(1)}\big)^3} m_2 (2) 
+ 
\ft{1}{2} \vev{C_s^{(2, 1)}\big(\Gamma_{K-2}^{(1)}\big)^2} m_K (2) 
\nonumber\\
&
+ 
\vev{C_s^{(1, 1)}\Gamma_{K-2}^{(1)} \Gamma_{K-2}^{(2)}} m_K (2) 
+ 
2 \vev{C_s^{(2, 0)}\Gamma_{K-2}^{(1)}} \Gamma_{\rm cusp}^{(1)} m_K (2) 
\nonumber\\
&
+ 
2 \vev{C_s^{(1, 0)}\Gamma_{K-2}^{(2)}} \Gamma_{\rm cusp}^{(1)} m_K (2) 
+ 
2 \vev{C_s^{(1, 0)}\Gamma_{K-2}^{(1)}} \Gamma_{\rm cusp}^{(2)} m_K (2) 
\nonumber\\
&
+ 
\ft{1}{2} \vev{C_s^{(1, 1)}\big(\Gamma_{K-2}^{(1)}\big)^3} m_2 (2) m_K (2) 
+ 
3 \vev{C_s^{(1, 0)}\big(\Gamma_{K-2}^{(1)}\big)^2} \Gamma_{\rm cusp}^{(1)} m_2 (2) m_K (2) 
\nonumber\\
&
+ 
\ft{1}{2} \vev{C_s^{(2, 2)}\big(\Gamma_{K-2}^{(1)}\big)^2} m_K (3) 
+ 
2 \vev{C_s^{(1, 0)}\Gamma_{K-2}^{(1)}} \big(\Gamma_{\rm cusp}^{(1)}\big)^2 m_2 (2) m_K (3) 
\nonumber\\
&
+ 
\ft{1}{2} \vev{\big(\Gamma_{K-2}^{(2)}\big)^2} 
+ 
\vev{\Gamma_{K-2}^{(1)} \Gamma_{K-2}^{(3)}} 
+ 
\vev{\Gamma_{K-2}^{(2)}} \Gamma_{\rm virt}^{(2)} 
+ 
\ft{1}{2} \big(\Gamma_{\rm virt}^{(2)}\big)^2 
+ 
\vev{\Gamma_{K-2}^{(1)}} \Gamma_{\rm virt}^{(3)} 
\nonumber\\
&
+ 
\ft{3}{2} \vev{\big(\Gamma_{K-2}^{(1)}\big)^2 \Gamma_{K-2}^{(2)}} m_2 (2) 
+ 
\ft{3}{2} \vev{\big(\Gamma_{K-2}^{(1)}\big)^2} \Gamma_{\rm virt}^{(2)} m_2 (2) 
+ 
\ft{1}{2} \vev{\big(\Gamma_{K-2}^{(1)}\big)^4} \big(m_2 (2)\big)^2 
\nonumber\\
&
- 
\ft{1}{4} \vev{\big(\Gamma_{K-2}^{(1)}\big)^4} m_2 (3) 
+ 
2 \vev{\Gamma_{K-2}^{(3)}} \Gamma_{\rm cusp}^{(1)} m_K (2) 
+ 
2 \Gamma_{\rm virt}^{(3)} \Gamma_{\rm cusp}^{(1)} m_K (2) 
\nonumber\\
&
+ 
2 \vev{\Gamma_{K-2}^{(2)}} \Gamma_{\rm cusp}^{(2)} m_K (2) 
+ 
2 \Gamma_{\rm virt}^{(2)} \Gamma_{\rm cusp}^{(2)} m_K (2) 
+ 
2 \vev{\Gamma_{K-2}^{(1)}} \Gamma_{\rm cusp}^{(3)} m_K (2) 
\nonumber\\
&
+ 
6 \vev{\Gamma_{K-2}^{(1)} \Gamma_{K-2}^{(2)}} \Gamma_{\rm cusp}^{(1)} m_2 (2) m_K (2) 
+ 
4 \vev{\Gamma_{K-2}^{(1)}} \Gamma_{\rm virt}^{(2)} \Gamma_{\rm cusp}^{(1)} m_2 (2) m_K (2) 
\nonumber\\
&
+ 
3 \vev{\big(\Gamma_{K-2}^{(1)}\big)^2} \Gamma_{\rm cusp}^{(2)} m_2 (2) m_K (2) 
+ 
4 \vev{\big(\Gamma_{K-2}^{(1)}\big)^3} \Gamma_{\rm cusp}^{(1)} \big(m_2 (2)\big)^2 m_K (2) 
\nonumber\\
&
- 
2 \vev{\big(\Gamma_{K-2}^{(1)}\big)^3} \Gamma_{\rm cusp}^{(1)} m_2 (3) m_K (2) 
- 
2 \vev{\Gamma_{K-2}^{(1)}} \Gamma_{\rm virt}^{(2)} \Gamma_{\rm cusp}^{(1)} m_K (3) 
+ 
2 \big(\Gamma_{\rm cusp}^{(2)}\big)^2 m_K (3) 
\nonumber\\
&
+ 
4 \Gamma_{\rm cusp}^{(1)} \Gamma_{\rm cusp}^{(3)} m_K (3) 
+ 
2 \vev{\Gamma_{K-2}^{(2)}} \big(\Gamma_{\rm cusp}^{(1)}\big)^2 m_2 (2) m_K (3) 
\nonumber\\
&
+ 
4 \vev{\Gamma_{K-2}^{(1)}} \Gamma_{\rm cusp}^{(1)} \Gamma_{\rm cusp}^{(2)} m_2 (2) m_K (3) 
+ 
2 \vev{\big(\Gamma_{K-2}^{(1)}\big)^2} \big(\Gamma_{\rm cusp}^{(1)}\big)^2 \big(m_2 (2)\big)^2 m_K (3) 
\nonumber\\
&
- 
2 \vev{\big(\Gamma_{K-2}^{(1)}\big)^2} \big(\Gamma_{\rm cusp}^{(1)}\big)^2 m_2 (3) m_K (3) 
+ 
2 \Gamma_{\rm virt}^{(2)} \big(\Gamma_{\rm cusp}^{(1)}\big)^2 m_K (2) m_K (3) 
\nonumber\\
&
- 
2 \Gamma_{\rm virt}^{(2)} \big(\Gamma_{\rm cusp}^{(1)}\big)^2 m_K (4)
\, , \\
C_t^{(4, 1)}
&= 
-
\vev{C_s^{(3, 0)} \Gamma_{K-2}^{(1)}} 
- 
\vev{C_s^{(2, 0)} \Gamma_{K-2}^{(2)}} 
- 
\vev{C_s^{(1, 0)} \Gamma_{K-2}^{(3)}} 
- 
\vev{C_s^{(2, 0)} \big(\Gamma_{K-2}^{(1)}\big)^2} m_2 (2) 
\nonumber\\
&
- 
2 \vev{C_s^{(1, 0)} \Gamma_{K-2}^{(1)} \Gamma_{K-2}^{(2)}} m_2 (2) 
- 
2 \vev{C_s^{(1, 0)} \Gamma_{K-2}^{(1)}} \Gamma_{\rm virt}^{(2)} m_2 (2) 
\nonumber\\
&
- 
\vev{C_s^{(1, 0)} \big(\Gamma_{K-2}^{(1)}\big)^3} \big(m_2 (2)\big)^2 
+ 
\ft{1}{2} \vev{C_s^{(1, 0)} \big(\Gamma_{K-2}^{(1)}\big)^3} m_2 (3) 
- 
\vev{C_s^{(3, 1)} \Gamma_{K-2}^{(1)}} m_K (2) 
\nonumber\\
&
- 
\vev{C_s^{(2, 1)} \Gamma_{K-2}^{(2)}} m_K (2) 
- 
\vev{C_s^{(1, 1)} \Gamma_{K-2}^{(3)}} m_K (2) 
- 
2 C_t^{(3, 0)} \Gamma_{\rm cusp}^{(1)} m_K (2) 
\nonumber\\
&
- 
\vev{C_s^{(2, 1)} \big(\Gamma_{K-2}^{(1)}\big)^2} m_2 (2) m_K (2) 
- 
2 \vev{C_s^{(1, 1)} \Gamma_{K-2}^{(1)} \Gamma_{K-2}^{(2)}} m_2 (2) m_K (2) 
\nonumber\\
&
- 
2 \vev{C_s^{(2, 0)} \Gamma_{K-2}^{(1)}} \Gamma_{\rm cusp}^{(1)} m_2 (2) m_K (2) 
- 
2 \vev{C_s^{(1, 0)} \Gamma_{K-2}^{(2)}} \Gamma_{\rm cusp}^{(1)} m_2 (2) m_K (2) 
\nonumber\\
&
- 
4 \vev{C_s^{(1, 0)} \Gamma_{K-2}^{(1)}} \Gamma_{\rm cusp}^{(2)} m_2 (2) m_K (2) 
- 
\vev{C_s^{(1, 1)} \big(\Gamma_{K-2}^{(1)}\big)^3} \big(m_2 (2)\big)^2 m_K (2) 
\nonumber\\
&
- 
4 \vev{C_s^{(1, 0)} \big(\Gamma_{K-2}^{(1)}\big)^2} \Gamma_{\rm cusp}^{(1)} \big(m_2 (2)\big)^2 m_K (2) 
+ 
\ft{1}{2} \vev{C_s^{(1, 1)} \big(\Gamma_{K-2}^{(1)}\big)^3} m_2 (3) m_K (2) 
\nonumber\\
&
+ 
2 \vev{C_s^{(1, 0)} \big(\Gamma_{K-2}^{(1)}\big)^2} \Gamma_{\rm cusp}^{(1)} m_2 (3) m_K (2) 
- 
\vev{C_s^{(3, 2)} \Gamma_{K-2}^{(1)}} m_K (3)
\nonumber\\
&
- 
\vev{C_s^{(2, 2)} \Gamma_{K-2}^{(2)}} m_K (3) 
- 
\vev{C_s^{(2, 2)} \big(\Gamma_{K-2}^{(1)}\big)^2} m_2 (2) m_K (3) 
\nonumber\\
&
+ 
2 \vev{C_s^{(1, 0)} \Gamma_{K-2}^{(1)}} \big(\Gamma_{\rm cusp}^{(1)}\big)^2 m_2 (3) m_K (3) 
- 
\vev{C_s^{(3, 3)} \Gamma_{K-2}^{(1)}} m_K (4) - \vev{\Gamma_{K-2}^{(4)}} 
- 
\Gamma_{\rm virt}^{(4)} 
\nonumber\\
&
- \vev{\big(\Gamma_{K-2}^{(2)}\big)^2} m_2 (2) 
- 
2 \vev{\Gamma_{K-2}^{(1)} \Gamma_{K-2}^{(3)}} m_2 (2) 
- 
2 \vev{\Gamma_{K-2}^{(2)}} \Gamma_{\rm virt}^{(2)} m_2 (2) 
\nonumber\\
&
- 
2 \vev{\Gamma_{K-2}^{(1)}} \Gamma_{\rm virt}^{(3)} m_2 (2) 
\nonumber\\
&
- 
3 \vev{\big(\Gamma_{K-2}^{(1)}\big)^2 \Gamma_{K-2}^{(2)}} \big(m_2 (2)\big)^2 
- 
3 \vev{\big(\Gamma_{K-2}^{(1)}\big)^2} \Gamma_{\rm virt}^{(2)} \big(m_2 (2)\big)^2 
- 
\vev{\big(\Gamma_{K-2}^{(1)}\big)^4} \big(m_2 (2)\big)^3 
\nonumber\\
&
+ 
\ft{3}{2} \vev{\big(\Gamma_{K-2}^{(1)}\big)^2 \Gamma_{K-2}^{(2)}} m_2 (3) 
+ 
\ft{3}{2} \vev{\big(\Gamma_{K-2}^{(1)}\big)^2} \Gamma_{\rm virt}^{(2)} m_2 (3) 
+ 
\vev{\big(\Gamma_{K-2}^{(1)}\big)^4} m_2 (2) m_2 (3) 
\nonumber\\
&
- 
\ft{1}{6} \vev{\big(\Gamma_{K-2}^{(1)}\big)^4} m_2 (4) 
- 
2 \Gamma_{\rm cusp}^{(4)} m_K (2) 
- 
2 \vev{\Gamma_{K-2}^{(3)}} \Gamma_{\rm cusp}^{(1)} m_2 (2) m_K (2) 
\nonumber\\
&
- 
4 \vev{\Gamma_{K-2}^{(2)}} \Gamma_{\rm cusp}^{(2)} m_2 (2) m_K (2) 
- 
4 \vev{\Gamma_{K-2}^{(1)}} \Gamma_{\rm cusp}^{(3)} m_2 (2) m_K (2) 
\nonumber\\
&
- 
8 \vev{\Gamma_{K-2}^{(1)} \Gamma_{K-2}^{(2)}} \Gamma_{\rm cusp}^{(1)} \big(m_2 (2)\big)^2 m_K (2) 
- 
2 \vev{\Gamma_{K-2}^{(1)}} \Gamma_{\rm virt}^{(2)} \Gamma_{\rm cusp}^{(1)} \big(m_2 (2)\big)^2 m_K (2) 
\nonumber\\
&
- 
6 \vev{\big(\Gamma_{K-2}^{(1)}\big)^2} \Gamma_{\rm cusp}^{(2)} \big(m_2 (2)\big)^2 m_K (2) 
- 
6 \vev{\big(\Gamma_{K-2}^{(1)}\big)^3} \Gamma_{\rm cusp}^{(1)} \big(m_2 (2)\big)^3 m_K (2) 
\nonumber\\
&
+ 
4 \vev{\Gamma_{K-2}^{(1)} \Gamma_{K-2}^{(2)}} \Gamma_{\rm cusp}^{(1)} m_2 (3) m_K (2) 
+ 
2 \vev{\Gamma_{K-2}^{(1)}} \Gamma_{\rm virt}^{(2)} \Gamma_{\rm cusp}^{(1)} m_2 (3) m_K (2) 
\nonumber\\
&
+ 
3 \vev{\big(\Gamma_{K-2}^{(1)}\big)^2} \Gamma_{\rm cusp}^{(2)} m_2 (3) m_K (2) 
+ 
6 \vev{\big(\Gamma_{K-2}^{(1)}\big)^3} \Gamma_{\rm cusp}^{(1)} m_2 (2) m_2 (3) m_K (2) 
\nonumber\\
&
- 
\vev{\big(\Gamma_{K-2}^{(1)}\big)^3} \Gamma_{\rm cusp}^{(1)} m_2 (4) m_K (2) 
- 
2 \Gamma_{\rm virt}^{(3)} \Gamma_{\rm cusp}^{(1)} \big(m_K (2)\big)^2 
- 
2 \Gamma_{\rm virt}^{(2)} \Gamma_{\rm cusp}^{(2)} \big(m_K (2)\big)^2 
\nonumber\\
&
- 
2 \vev{\Gamma_{K-2}^{(1)}} \Gamma_{\rm virt}^{(2)} \Gamma_{\rm cusp}^{(1)} m_2 (2) \big(m_K (2)\big)^2 
+ 
2 \Gamma_{\rm virt}^{(3)} \Gamma_{\rm cusp}^{(1)} m_K (3) 
+ 
2 \Gamma_{\rm virt}^{(2)} \Gamma_{\rm cusp}^{(2)} m_K (3) 
\nonumber\\
&
+ 
4 \vev{\Gamma_{K-2}^{(1)}} \Gamma_{\rm virt}^{(2)} \Gamma_{\rm cusp}^{(1)} m_2 (2) m_K (3) 
+ 
2 \vev{\Gamma_{K-2}^{(2)}} \big(\Gamma_{\rm cusp}^{(1)}\big)^2 m_2 (3) m_K (3) 
\nonumber\\
&
+ 
4 \vev{\Gamma_{K-2}^{(1)}} \Gamma_{\rm cusp}^{(1)} \Gamma_{\rm cusp}^{(2)} m_2 (3) m_K (3) 
+ 
4 \vev{\big(\Gamma_{K-2}^{(1)}\big)^2} \big(\Gamma_{\rm cusp}^{(1)}\big)^2 m_2 (2) m_2 (3) m_K (3) 
\nonumber\\
&
- 
2 \vev{\big(\Gamma_{K-2}^{(1)}\big)^2} \big(\Gamma_{\rm cusp}^{(1)}\big)^2 m_2 (4) m_K (3) 
- 
\ft{4}{3} \vev{\Gamma_{K-2}^{(1)}} \big(\Gamma_{\rm cusp}^{(1)}\big)^3 m_2 (4) m_K (4)
\, , \\
\vev{C_s^{(4, 4)}} 
&= \ft{2}{3} \big(\Gamma_{\rm cusp}^{(1)}\big)^4 m_2 (5)
\, , \\
\vev{C_s^{(4, 3)}} &= 
-2 \Gamma_{\rm virt}^{(2)} \big(\Gamma_{\rm cusp}^{(1)}\big)^2 m_2 (3) 
- 
4 \big(\Gamma_{\rm cusp}^{(1)}\big)^2 \Gamma_{\rm cusp}^{(2)} m_2 (4) 
- 
\ft{4}{3} \vev{\Gamma_{K-2}^{(1)}} \big(\Gamma_{\rm cusp}^{(1)}\big)^3 m_2 (2) m_2 (4) 
\nonumber\\
&
+ 
\ft{4}{3} \vev{\Gamma_{K-2}^{(1)}} \big(\Gamma_{\rm cusp}^{(1)}\big)^3 m_2 (5)
\, , \\
\vev{C_s^{(4, 2)}} 
&= 
2 \vev{C_s^{(1, 0)} \Gamma_{K-2}^{(1)}} \big(\Gamma_{\rm cusp}^{(1)}\big)^2 m_2 (2) m_2 (3) 
- 
2 \vev{C_s^{(1, 0)} \Gamma_{K-2}^{(1)}} \big(\Gamma_{\rm cusp}^{(1)}\big)^2 m_2 (4) 
+ 
\ft{1}{2} \big(\Gamma_{\rm virt}^{(2)}\big)^2 
\nonumber\\
&
+ 
2 \Gamma_{\rm virt}^{(3)} \Gamma_{\rm cusp}^{(1)} m_2 (2) 
+ 
2 \Gamma_{\rm virt}^{(2)} \Gamma_{\rm cusp}^{(2)} m_2 (2) 
+ 
2 \vev{\Gamma_{K-2}^{(1)}} \Gamma_{\rm virt}^{(2)} \Gamma_{\rm cusp}^{(1)} \big(m_2 (2)\big)^2 
\nonumber\\
&
- 
2 \vev{\Gamma_{K-2}^{(1)}} \Gamma_{\rm virt}^{(2)} \Gamma_{\rm cusp}^{(1)} m_2 (3) 
+ 
2 \big(\Gamma_{\rm cusp}^{(2)}\big)^2 m_2 (3) 
+ 
4 \Gamma_{\rm cusp}^{(1)} \Gamma_{\rm cusp}^{(3)} m_2 (3) 
\nonumber\\
&
+ 
2 \vev{\Gamma_{K-2}^{(2)}} \big(\Gamma_{\rm cusp}^{(1)}\big)^2 m_2 (2) m_2 (3) 
+ 
4 \vev{\Gamma_{K-2}^{(1)}} \Gamma_{\rm cusp}^{(1)} \Gamma_{\rm cusp}^{(2)} m_2 (2) m_2 (3) 
\nonumber\\
&
+ 
2 \vev{\big(\Gamma_{K-2}^{(1)}\big)^2} \big(\Gamma_{\rm cusp}^{(1)}\big)^2 \big(m_2 (2)\big)^2 m_2 (3) 
- 
\vev{\big(\Gamma_{K-2}^{(1)}\big)^2} \big(\Gamma_{\rm cusp}^{(1)}\big)^2 \big(m_2 (3)\big)^2 
\nonumber\\
&
-
2 \vev{\Gamma_{K-2}^{(2)}} \big(\Gamma_{\rm cusp}^{(1)}\big)^2 m_2 (4) 
- 
2 \Gamma_{\rm virt}^{(2)} \big(\Gamma_{\rm cusp}^{(1)}\big)^2 m_2 (4) 
- 
4 \vev{\Gamma_{K-2}^{(1)}} \Gamma_{\rm cusp}^{(1)} \Gamma_{\rm cusp}^{(2)} m_2 (4) 
\nonumber\\
&
- 
2 \vev{\big(\Gamma_{K-2}^{(1)}\big)^2} \big(\Gamma_{\rm cusp}^{(1)}\big)^2 m_2 (2) m_2 (4) 
+ 
\vev{\big(\Gamma_{K-2}^{(1)}\big)^2} \big(\Gamma_{\rm cusp}^{(1)}\big)^2 m_2 (5) 
\nonumber\\
&
+ 
2 \Gamma_{\rm virt}^{(2)} \big(\Gamma_{\rm cusp}^{(1)}\big)^2 m_2 (3) m_K (2)
\, , \\
\vev{C_s^{(4, 1)}} 
&= 
-
2 C_t^{(3, 0)} \Gamma_{\rm cusp}^{(1)} m_2 (2) 
- 
2 \vev{C_s^{(1, 0)} \Gamma_{K-2}^{(1)}} \Gamma_{\rm cusp}^{(2)} \big(m_2 (2)\big)^2 
\nonumber\\
&
+ 
2 \vev{C_s^{(2, 0)} \Gamma_{K-2}^{(1)}} \Gamma_{\rm cusp}^{(1)} m_2 (3) 
+ 
2 \vev{C_s^{(1, 0)} \Gamma_{K-2}^{(2)}} \Gamma_{\rm cusp}^{(1)} m_2 (3) 
\nonumber\\
&
+ 
2 \vev{C_s^{(1, 0)} \Gamma_{K-2}^{(1)}} \Gamma_{\rm cusp}^{(2)} m_2 (3) 
+ 
2 \vev{C_s^{(1, 0)} \big(\Gamma_{K-2}^{(1)}\big)^2} \Gamma_{\rm cusp}^{(1)} m_2 (2) m_2 (3) 
\nonumber\\
&
- 
\vev{C_s^{(1, 0)} \big(\Gamma_{K-2}^{(1)}\big)^2} \Gamma_{\rm cusp}^{(1)} m_2 (4) 
- 
\Gamma_{\rm virt}^{(4)} 
+ 
\big(\Gamma_{\rm virt}^{(2)}\big)^2 m_2 (2) 
- 
2 \Gamma_{\rm cusp}^{(4)} m_2 (2) 
\nonumber\\
&
- 
2 \vev{\Gamma_{K-2}^{(2)}} \Gamma_{\rm cusp}^{(2)} \big(m_2 (2)\big)^2 
- 
2 \vev{\Gamma_{K-2}^{(1)}} \Gamma_{\rm cusp}^{(3)} \big(m_2 (2)\big)^2 
- 
2 \vev{\big(\Gamma_{K-2}^{(1)}\big)^2} \Gamma_{\rm cusp}^{(2)} \big(m_2 (2)\big)^3 
\nonumber\\
&
+ 
2 \vev{\Gamma_{K-2}^{(3)}} \Gamma_{\rm cusp}^{(1)} m_2 (3) 
+ 
2 \Gamma_{\rm virt}^{(3)} \Gamma_{\rm cusp}^{(1)} m_2 (3) 
+ 
2 \vev{\Gamma_{K-2}^{(2)}} \Gamma_{\rm cusp}^{(2)} m_2 (3) 
+ 
2 \Gamma_{\rm virt}^{(2)} \Gamma_{\rm cusp}^{(2)} m_2 (3) 
\nonumber\\
&
+ 
2 \vev{\Gamma_{K-2}^{(1)}} \Gamma_{\rm cusp}^{(3)} m_2 (3) 
+ 
4 \vev{\Gamma_{K-2}^{(1)} \Gamma_{K-2}^{(2)}} \Gamma_{\rm cusp}^{(1)} m_2 (2) m_2 (3) 
\nonumber\\
&
+ 
4 \vev{\Gamma_{K-2}^{(1)}} \Gamma_{\rm virt}^{(2)} \Gamma_{\rm cusp}^{(1)} m_2 (2) m_2 (3) 
+ 
3 \vev{\big(\Gamma_{K-2}^{(1)}\big)^2} \Gamma_{\rm cusp}^{(2)} m_2 (2) m_2 (3) 
\nonumber\\
&
+ 
2 \vev{\big(\Gamma_{K-2}^{(1)}\big)^3} \Gamma_{\rm cusp}^{(1)} \big(m_2 (2)\big)^2 m_2 (3) 
- 
\vev{\big(\Gamma_{K-2}^{(1)}\big)^3} \Gamma_{\rm cusp}^{(1)} \big(m_2 (3)\big)^2 
\nonumber\\
&
- 
2 \vev{\Gamma_{K-2}^{(1)} \Gamma_{K-2}^{(2)}} \Gamma_{\rm cusp}^{(1)} m_2 (4) 
- 
2 \vev{\Gamma_{K-2}^{(1)}} \Gamma_{\rm virt}^{(2)} \Gamma_{\rm cusp}^{(1)} m_2 (4) 
- 
\vev{\big(\Gamma_{K-2}^{(1)}\big)^2} \Gamma_{\rm cusp}^{(2)} m_2 (4) 
\nonumber\\
&
- 
\vev{\big(\Gamma_{K-2}^{(1)}\big)^3} \Gamma_{\rm cusp}^{(1)} m_2 (2) m_2 (4) 
+ 
\ft{1}{3} \vev{\big(\Gamma_{K-2}^{(1)}\big)^3} \Gamma_{\rm cusp}^{(1)} m_2 (5) 
- 
\big(\Gamma_{\rm virt}^{(2)}\big)^2 m_K (2) 
\nonumber\\
&
- 
2 \Gamma_{\rm virt}^{(3)} \Gamma_{\rm cusp}^{(1)} m_2 (2) m_K (2) 
- 
2 \Gamma_{\rm virt}^{(2)} \Gamma_{\rm cusp}^{(2)} m_2 (2) m_K (2) 
\nonumber\\
&
- 
2 \vev{\Gamma_{K-2}^{(1)}} \Gamma_{\rm virt}^{(2)} \Gamma_{\rm cusp}^{(1)} \big(m_2 (2)\big)^2 m_K (2)
\, . 
\end{align}
Finally, the solution for averages involving both powers of the anomalous dimensions and expansion coefficients is
\begin{align}
\vev{C_s^{(1, 1)} \big(\Gamma_{K-2}^{(1)}\big)^3} 
&= 
-2 \vev{\big(\Gamma_{K-2}^{(1)}\big)^3} \Gamma_{\rm cusp}^{(1)} m_2 (2)
\, , \\
\vev{C_s^{(2, 1)} \big(\Gamma_{K-2}^{(1)}\big)^2} 
&= 
-
2 \vev{C_s^{(1, 1)} \Gamma_{K-2}^{(1)} \Gamma_{K-2}^{(2)}} 
- 
2 \vev{C_s^{(1, 0)} \big(\Gamma_{K-2}^{(1)}\big)^2} \Gamma_{\rm cusp}^{(1)} m_2 (2) 
\nonumber\\
&
- 
2 \vev{C_s^{(2, 2)} \big(\Gamma_{K-2}^{(1)}\big)^2} m_K (2) 
- 
\vev{\big(\Gamma_{K-2}^{(1)}\big)^2} \Gamma_{\rm virt}^{(2)} 
- 
4 \vev{\Gamma_{K-2}^{(1)} \Gamma_{K-2}^{(2)}} \Gamma_{\rm cusp}^{(1)} m_2 (2) 
\nonumber\\
&
- 
2 \vev{\big(\Gamma_{K-2}^{(1)}\big)^2} \Gamma_{\rm cusp}^{(2)} m_2 (2) 
- 
2 \vev{\big(\Gamma_{K-2}^{(1)}\big)^3} \Gamma_{\rm cusp}^{(1)} \big(m_2 (2)\big)^2 
\nonumber\\
&
+ 
2 \vev{\big(\Gamma_{K-2}^{(1)}\big)^3} \Gamma_{\rm cusp}^{(1)} m_2 (3) 
+ 
4 \vev{\big(\Gamma_{K-2}^{(1)}\big)^2} \big(\Gamma_{\rm cusp}^{(1)}\big)^2 m_2 (3) m_K (2)
\, , \\
\vev{C_s^{(3, 3)} \Gamma_{K-2}^{(1)}} 
&= 
-\ft{4}{3} \vev{\Gamma_{K-2}^{(1)}} \big(\Gamma_{\rm cusp}^{(1)}\big)^3 m_2 (4)
\, , \\
\vev{C_s^{(3, 2)} \Gamma_{K-2}^{(1)}} 
&
= 
-
\vev{C_s^{(2, 2)} \Gamma_{K-2}^{(2)}} 
+ 
2 \vev{C_s^{(1, 0)} \Gamma_{K-2}^{(1)}} \big(\Gamma_{\rm cusp}^{(1)}\big)^2 m_2 (3) 
+ 
2 \vev{\Gamma_{K-2}^{(1)}} \Gamma_{\rm virt}^{(2)} \Gamma_{\rm cusp}^{(1)} m_2 (2) 
\nonumber\\
&
+ 
2 \vev{\Gamma_{K-2}^{(2)}} \big(\Gamma_{\rm cusp}^{(1)}\big)^2 m_2 (3) 
+ 
4 \vev{\Gamma_{K-2}^{(1)}} \Gamma_{\rm cusp}^{(1)} \Gamma_{\rm cusp}^{(2)} m_2 (3) 
\nonumber\\
&
+ 
2 \vev{\big(\Gamma_{K-2}^{(1)}\big)^2} \big(\Gamma_{\rm cusp}^{(1)}\big)^2 m_2 (2) m_2 (3) 
- 
2 \vev{\big(\Gamma_{K-2}^{(1)}\big)^2} \big(\Gamma_{\rm cusp}^{(1)}\big)^2 m_2 (4)
\, , \\
\vev{C_s^{(3, 1)} \Gamma_{K-2}^{(1)}} 
&= 
-
\vev{C_s^{(2, 1)} \Gamma_{K-2}^{(2)}} 
- 
\vev{C_s^{(1, 1)} \Gamma_{K-2}^{(3)}} 
- 
\vev{C_s^{(1, 0)} \Gamma_{K-2}^{(1)}} \Gamma_{\rm virt}^{(2)} 
\nonumber\\
&
- 
2 \vev{C_s^{(2, 0)} \Gamma_{K-2}^{(1)}} \Gamma_{\rm cusp}^{(1)} m_2 (2) 
- 
2 \vev{C_s^{(1, 0)} \Gamma_{K-2}^{(2)}} \Gamma_{\rm cusp}^{(1)} m_2 (2) 
\nonumber\\
&
- 
2 \vev{C_s^{(1, 0)} \Gamma_{K-2}^{(1)}} \Gamma_{\rm cusp}^{(2)} m_2 (2) 
- 
2 \vev{C_s^{(1, 0)} \big(\Gamma_{K-2}^{(1)}\big)^2} \Gamma_{\rm cusp}^{(1)} \big(m_2 (2)\big)^2 
\nonumber\\
&
+ 
2 \vev{C_s^{(1, 0)} \big(\Gamma_{K-2}^{(1)}\big)^2} \Gamma_{\rm cusp}^{(1)} m_2 (3) 
- 
\vev{\Gamma_{K-2}^{(2)}} \Gamma_{\rm virt}^{(2)} 
\nonumber\\
&
- 
\vev{\Gamma_{K-2}^{(1)}} \Gamma_{\rm virt}^{(3)} 
- 
2 \vev{\Gamma_{K-2}^{(3)}} \Gamma_{\rm cusp}^{(1)} m_2 (2) 
- 
2 \vev{\Gamma_{K-2}^{(2)}} \Gamma_{\rm cusp}^{(2)} m_2 (2) 
\nonumber\\
&
- 
2 \vev{\Gamma_{K-2}^{(1)}} \Gamma_{\rm cusp}^{(3)} m_2 (2) 
- 
4 \vev{\Gamma_{K-2}^{(1)} \Gamma_{K-2}^{(2)}} \Gamma_{\rm cusp}^{(1)} \big(m_2 (2)\big)^2 
\nonumber\\
&
- 
2 \vev{\Gamma_{K-2}^{(1)}} \Gamma_{\rm virt}^{(2)} \Gamma_{\rm cusp}^{(1)} \big(m_2 (2)\big)^2 
- 
2 \vev{\big(\Gamma_{K-2}^{(1)}\big)^2} \Gamma_{\rm cusp}^{(2)} \big(m_2 (2)\big)^2 
\nonumber\\
&
- 
2 \vev{\big(\Gamma_{K-2}^{(1)}\big)^3} \Gamma_{\rm cusp}^{(1)} \big(m_2 (2)\big)^3 
+ 
4 \vev{\Gamma_{K-2}^{(1)} \Gamma_{K-2}^{(2)}} \Gamma_{\rm cusp}^{(1)} m_2 (3) 
\nonumber\\
&
+ 
2 \vev{\Gamma_{K-2}^{(1)}} \Gamma_{\rm virt}^{(2)} \Gamma_{\rm cusp}^{(1)} m_2 (3) 
+ 
2 \vev{\big(\Gamma_{K-2}^{(1)}\big)^2} \Gamma_{\rm cusp}^{(2)} m_2 (3) 
\nonumber\\
&
+ 
3 \vev{\big(\Gamma_{K-2}^{(1)}\big)^3} \Gamma_{\rm cusp}^{(1)} m_2 (2) m_2 (3) 
- 
\vev{\big(\Gamma_{K-2}^{(1)}\big)^3} \Gamma_{\rm cusp}^{(1)} m_2 (4)
\, , \\
\vev{C_s^{(2, 2)} \big(\Gamma_{K-2}^{(1)}\big)^2} 
&= 
2 \vev{\big(\Gamma_{K-2}^{(1)}\big)^2} \big(\Gamma_{\rm cusp}^{(1)}\big)^2 m_2 (3)
\, .
\end{align}
We do not display the solution for $C_t^{(4,0)}-\vev{C_s^{(4,0)}}$, as it will not be needed below.

\section[Twist-K--twist-K correlator: three loops]{Twist-\textbf{\textit K}--twist-\textbf{\textit K} correlator: three loops}
\label{KKcorr3loopsApp}

At three loops, we find ten equations, six of which determine the averages of the three-loop expansion coefficients
$\vev{C_{s,t}^{(3,n)}}$ for $n=1,2,3$:
\begin{align}
\vev{C_t^{(3,3)}}
&=
-
\ft{1}{6}\vev{\big(\Gamma_{K-2}^{(1)}\big)^3}
-
\vev{\big(\Gamma_{K-2}^{(1)}\big)^2}\Gamma_{\rm cusp}^{(1)}m_K(2)
\nonumber\\
&\quad
-
2\vev{\Gamma_{K-2}^{(1)}}\big(\Gamma_{\rm cusp}^{(1)}\big)^2m_K(3)
-
\ft{4}{3}\big(\Gamma_{\rm cusp}^{(1)}\big)^3m_K(4)\, ,
\\
\vev{C_t^{(3,2)}}
&=
\ft{1}{2}\vev{C_s^{(1,0)}\big(\Gamma_{K-2}^{(1)}\big)^2}
+
\vev{\Gamma_{K-2}^{(1)}\Gamma_{K-2}^{(2)}}
+
\vev{\Gamma_{K-2}^{(1)}}\Gamma_{\rm virt}^{(2)}
+
\ft{1}{2}\vev{C_s^{(1,1)}\big(\Gamma_{K-2}^{(1)}\big)^2}m_K(2)
\nonumber\\
&\quad
+
\ft{1}{2}\vev{\big(\Gamma_{K-2}^{(1)}\big)^3}m_K(2)
+
2\vev{C_s^{(1,0)}\Gamma_{K-2}^{(1)}}\Gamma_{\rm cusp}^{(1)}m_K(2)
+
2\vev{\Gamma_{K-2}^{(2)}}\Gamma_{\rm cusp}^{(1)}m_K(2)
\nonumber\\
&\quad
+
2\Gamma_{\rm virt}^{(2)}\Gamma_{\rm cusp}^{(1)}m_K(2)
+
2\vev{\Gamma_{K-2}^{(1)}}\Gamma_{\rm cusp}^{(2)}m_K(2)
+
3\vev{\big(\Gamma_{K-2}^{(1)}\big)^2}\Gamma_{\rm cusp}^{(1)}m_K^2(2)
\nonumber\\
&
+
2\vev{C_s^{(1,1)}\Gamma_{K-2}^{(1)}}\Gamma_{\rm cusp}^{(1)}m_K(3)
+
4\Gamma_{\rm cusp}^{(1)}\Gamma_{\rm cusp}^{(2)}m_K(3)
+
8\vev{\Gamma_{K-2}^{(1)}}\big(\Gamma_{\rm cusp}^{(1)}\big)^2m_K(2)m_K(3)
\nonumber\\
&
-
2\vev{\Gamma_{K-2}^{(1)}}\big(\Gamma_{\rm cusp}^{(1)}\big)^2m_K(4)\, ,
\\
\vev{C_t^{(3,1)}}
&=
-
\vev{C_s^{(2,0)}\Gamma_{K-2}^{(1)}}
-
\vev{C_s^{(1,0)}\Gamma_{K-2}^{(2)}}
-
\vev{\Gamma_{K-2}^{(3)}}
-
\Gamma_{\rm virt}^{(3)}
-
\vev{C_s^{(2,1)}\Gamma_{K-2}^{(1)}}m_K(2)
\nonumber\\
&
-
\vev{C_s^{(1,0)} \big(\Gamma_{K-2}^{(1)}\big)^2}m_K(2)
-
\vev{C_s^{(1,1)}\Gamma_{K-2}^{(2)}}m_K(2)
-
2\vev{\Gamma_{K-2}^{(1)}\Gamma_{K-2}^{(2)}}m_K(2)
\nonumber\\
&
-
2\vev{\Gamma_{K-2}^{(1)}}\Gamma_{\rm virt}^{(2)}m_K(2)
-
2\Gamma_{\rm cusp}^{(3)}m_K(2)
-
\vev{C_s^{(1,1)}\big(\Gamma_{K-2}^{(1)}\big)^2}m_K^2(2)
-
\vev{\big(\Gamma_{K-2}^{(1)}\big)^3}m_K^2(2)
\nonumber\\
&
-
4\vev{C_s^{(1,0)}\Gamma_{K-2}^{(1)}}\Gamma_{\rm cusp}^{(1)}m_K^2(2)
-
4\vev{\Gamma_{K-2}^{(2)}}\Gamma_{\rm cusp}^{(1)}m_K^2(2)
-
2\vev{C_t^{(1,0)} \bar{\Gamma}_{K-2}^{(1)}}\Gamma_{\rm cusp}^{(1)}m_K^2(2)
\nonumber\\
&
-
2\vev{\bar{\Gamma}_{K-2}^{(2)}}\Gamma_{\rm cusp}^{(1)}m_K^2(2)
-
2\Gamma_{\rm virt}^{(2)}\Gamma_{\rm cusp}^{(1)}m_K^2(2)
-
6
\vev{\Gamma_{K-2}^{(1)}}\Gamma_{\rm cusp}^{(2)}m_K^2(2)
\nonumber\\
&
-
6\vev{\big(\Gamma_{K-2}^{(1)}\big)^2}\Gamma_{\rm cusp}^{(1)}m_K^3(2)
-
2\vev{C_s^{(1,1)}\Gamma_{K-2}^{(1)}}\Gamma_{\rm cusp}^{(1)}m_K^3(2)
-
2\vev{\big(\bar{\Gamma}_{K-2}^{(1)}\big)^2}\Gamma_{\rm cusp}^{(1)}m_K^3(2)
\nonumber\\
&
-
4\vev{\Gamma_{K-2}^{(1)}}\big(\Gamma_{\rm cusp}^{(1)}\big)^2m_K^4(2)
-
\vev{C_s^{(2,2)}\Gamma_{K-2}^{(1)}}m_K(3)
+
\ft{1}{2}\vev{\big(\Gamma_{K-2}^{(1)}\big)^3}m_K(3)
\nonumber\\
&
+
2\vev{C_t^{(1,0)}\bar{\Gamma}_{K-2}^{(1)}}\Gamma_{\rm cusp}^{(1)}m_K(3)
+
2\vev{\bar{\Gamma}_{K-2}^{(2)}}\Gamma_{\rm cusp}^{(1)}m_K(3)
+
2\Gamma_{\rm virt}^{(2)}\Gamma_{\rm cusp}^{(1)}m_K(3)
\nonumber\\
&
+
2\vev{\Gamma_{K-2}^{(1)}}\Gamma_{\rm cusp}^{(2)}m_K(3)
+
3\vev{\big(\Gamma_{K-2}^{(1)}\big)^2}\Gamma_{\rm cusp}^{(1)}m_K(2)m_K(3)
\nonumber\\
&
-
2\vev{C_s^{(1,1)}\Gamma_{K-2}^{(1)}}\Gamma_{\rm cusp}^{(1)}m_K(2)m_K(3)
+
3\vev{\big(\bar{\Gamma}_{K-2}^{(1)}\big)^2}\Gamma_{\rm cusp}^{(1)}m_K(2)m_K(3)
\nonumber\\
&
-
4\vev{\Gamma_{K-2}^{(1)}}\big(\Gamma_{\rm cusp}^{(1)}\big)^2m_K^2(2)m_K(3)
+
2\vev{\Gamma_{K-2}^{(1)}}\big(\Gamma_{\rm cusp}^{(1)}\big)^2m_K^2(3)
\nonumber\\
&
-
\vev{\big(\bar{\Gamma}_{K-2}^{(1)}\big)^2}\Gamma_{\rm cusp}^{(1)}m_K(4)\, ,
\\
\vev{C_s^{(3,3)}}
&=
-
\ft{1}{6}\vev{\big(\bar{\Gamma}_{K-2}^{(1)}\big)^3}
-
\vev{\big(\bar{\Gamma}_{K-2}^{(1)}\big)^2}\Gamma_{\rm cusp}^{(1)}m_K(2)
\nonumber\\
&\quad
-
2\vev{\bar{\Gamma}_{K-2}^{(1)}}\big(\Gamma_{\rm cusp}^{(1)}\big)^2m_K(3)
-
\ft{4}{3}\big(\Gamma_{\rm cusp}^{(1)}\big)^3m_K(4)\, ,
\\
\vev{C_s^{(3,2)}}
&=
\ft{1}{2}\vev{C_t^{(1,0)}\big(\bar{\Gamma}_{K-2}^{(1)}\big)^2}
+
\vev{\bar{\Gamma}_{K-2}^{(1)}\bar{\Gamma}_{K-2}^{(2)}}
+
\vev{\bar{\Gamma}_{K-2}^{(1)}}\Gamma_{\rm virt}^{(2)}
+
\ft{1}{2}\vev{C_t^{(1,1)}\big(\bar{\Gamma}_{K-2}^{(1)}\big)^2}m_K(2)
\nonumber\\
&
+
\ft{1}{2}\vev{\big(\bar{\Gamma}_{K-2}^{(1)}\big)^3}m_K(2)
+
2\vev{C_t^{(1,0)}\bar{\Gamma}_{K-2}^{(1)}}\Gamma_{\rm cusp}^{(1)}m_K(2)
+
2\vev{\bar{\Gamma}_{K-2}^{(2)}}\Gamma_{\rm cusp}^{(1)}m_K(2)
\nonumber\\
&
+
2\Gamma_{\rm virt}^{(2)}\Gamma_{\rm cusp}^{(1)}m_K(2)
+
2\vev{\bar{\Gamma}_{K-2}^{(1)}}\Gamma_{\rm cusp}^{(2)}m_K(2)
+
3\vev{\big(\bar{\Gamma}_{K-2}^{(1)}\big)^2}\Gamma_{\rm cusp}^{(1)}m_K^2(2)
\nonumber\\
&
+
2\vev{C_t^{(1,1)}\bar{\Gamma}_{K-2}^{(1)}}\Gamma_{\rm cusp}^{(1)}m_K(3)
+
4\Gamma_{\rm cusp}^{(1)}\Gamma_{\rm cusp}^{(2)}m_K(3)
+
8\vev{\bar{\Gamma}_{K-2}^{(1)}}\big(\Gamma_{\rm cusp}^{(1)}\big)^2m_K(2)m_K(3)
\nonumber\\
&
-
2\vev{\bar{\Gamma}_{K-2}^{(1)}}\big(\Gamma_{\rm cusp}^{(1)}\big)^2m_K(4)\, ,
\\
\vev{C_s^{(3,1)}}
&=
-
\vev{C_t^{(2,0)}\bar{\Gamma}_{K-2}^{(1)}}
-
\vev{C_t^{(1,0)}\bar{\Gamma}_{K-2}^{(2)}}
-
\vev{\bar{\Gamma}_{K-2}^{(3)}}
-
\Gamma_{\rm virt}^{(3)}
-
\vev{C_t^{(2,1)}\bar{\Gamma}_{K-2}^{(1)}}m_K(2)
\nonumber\\
&
-
\vev{C_t^{(1,0)}\big(\bar{\Gamma}_{K-2}^{(1)}\big)^2}m_K(2)
-
\vev{C_t^{(1,1)}\bar{\Gamma}_{K-2}^{(2)}}m_K(2)
-
2\vev{\bar{\Gamma}_{K-2}^{(1)}\bar{\Gamma}_{K-2}^{(2)}}m_K(2)
\nonumber\\
&
-
2\vev{\bar{\Gamma}_{K-2}^{(1)}}\Gamma_{\rm virt}^{(2)}m_K(2)
-
2\Gamma_{\rm cusp}^{(3)}m_K(2)
-
\vev{C_t^{(1,1)}\big(\bar{\Gamma}_{K-2}^{(1)}\big)^2}m_K^2(2)
\nonumber\\
&
-
\vev{\big(\bar{\Gamma}_{K-2}^{(1)}\big)^3}m_K^2(2)
-
2\vev{C_s^{(1,0)}\Gamma_{K-2}^{(1)}}\Gamma_{\rm cusp}^{(1)}m_K^2(2)
-
2\vev{\Gamma_{K-2}^{(2)}}\Gamma_{\rm cusp}^{(1)}m_K^2(2)
\nonumber\\
&
-
4\vev{C_t^{(1,0)}\bar{\Gamma}_{K-2}^{(1)}}\Gamma_{\rm cusp}^{(1)}m_K^2(2)
-
4\vev{\bar{\Gamma}_{K-2}^{(2)}}\Gamma_{\rm cusp}^{(1)}m_K^2(2)
-
2\Gamma_{\rm virt}^{(2)}\Gamma_{\rm cusp}^{(1)}m_K^2(2)
\nonumber\\
&
-
6\vev{\bar{\Gamma}_{K-2}^{(1)}}\Gamma_{\rm cusp}^{(2)}m_K^2(2)
-
2\vev{\big(\Gamma_{K-2}^{(1)}\big)^2}\Gamma_{\rm cusp}^{(1)}m_K^3(2)
-
2\vev{C_t^{(1,1)}\bar{\Gamma}_{K-2}^{(1)}}\Gamma_{\rm cusp}^{(1)}m_K^3(2)
\nonumber\\
&
-
6\vev{\big(\bar{\Gamma}_{K-2}^{(1)}\big)^2}\Gamma_{\rm cusp}^{(1)}m_K^3(2)
-
4\vev{\bar{\Gamma}_{K-2}^{(1)}}\big(\Gamma_{\rm cusp}^{(1)}\big)^2m_K^4(2)
-
\vev{C_t^{(2,2)}\bar{\Gamma}_{K-2}^{(1)}}m_K(3)
\nonumber\\
&
+
\ft{1}{2}\vev{\big(\bar{\Gamma}_{K-2}^{(1)}\big)^3}m_K(3)
+
2\vev{C_s^{(1,0)}\Gamma_{K-2}^{(1)}}\Gamma_{\rm cusp}^{(1)}m_K(3)
+
2\vev{\Gamma_{K-2}^{(2)}}\Gamma_{\rm cusp}^{(1)}m_K(3)
\nonumber\\
&
+
2\Gamma_{\rm virt}^{(2)}\Gamma_{\rm cusp}^{(1)}m_K(3)
+
2\vev{\bar{\Gamma}_{K-2}^{(1)}}\Gamma_{\rm cusp}^{(2)}m_K(3)
+
3\vev{\big(\Gamma_{K-2}^{(1)}\big)^2}\Gamma_{\rm cusp}^{(1)}m_K(2)m_K(3)
\nonumber\\
&
-
2\vev{C_t^{(1,1)}\bar{\Gamma}_{K-2}^{(1)}}\Gamma_{\rm cusp}^{(1)}m_K(2)m_K(3)
+
3\vev{\big(\bar{\Gamma}_{K-2}^{(1)}\big)^2}\Gamma_{\rm cusp}^{(1)}m_K(2)m_K(3)
\nonumber\\
&
-
4\vev{\bar{\Gamma}_{K-2}^{(1)}}\big(\Gamma_{\rm cusp}^{(1)}\big)^2m_K^2(2)m_K(3)
+
2\vev{\bar{\Gamma}_{K-2}^{(1)}}\big(\Gamma_{\rm cusp}^{(1)}\big)^2m_K^2(3)
-
\vev{\big(\Gamma_{K-2}^{(1)}\big)^2}\Gamma_{\rm cusp}^{(1)}m_K(4)\, .
\end{align}
It also fixes averages involving lower-loop coefficients, namely $\vev{C_s^{(2,2)} \Gamma_{K-2}^{(1)}}$ 
and $\vev{C_t^{(2,2)} \bar{\Gamma}_{K-2}^{(1)}}$:
\begin{align}
\vev{C_t^{(2,2)}\bar{\Gamma}_{K-2}^{(1)}}
&=
-\ft{1}{2}\vev{C_s^{(1,1)}\big(\Gamma_{K-2}^{(1)}\big)^2}
-
\vev{\big(\Gamma_{K-2}^{(1)}\big)^2}\Gamma_{\rm cusp}^{(1)}m_K(2)
-
2\vev{C_t^{(1,1)}\bar{\Gamma}_{K-2}^{(1)}}\Gamma_{\rm cusp}^{(1)}m_K(2)
\nonumber\\
&\quad
-
4\vev{\bar{\Gamma}_{K-2}^{(1)}}\big(\Gamma_{\rm cusp}^{(1)}\big)^2m_K^2(2)
+
2\vev{\bar{\Gamma}_{K-2}^{(1)}}\big(\Gamma_{\rm cusp}^{(1)}\big)^2m_K(3)
\, ,
\\
\vev{C_s^{(2,2)}\Gamma_{K-2}^{(1)}}
&=
-\ft{1}{2}\vev{C_t^{(1,1)}\big(\bar{\Gamma}_{K-2}^{(1)}\big)^2}
-
2\vev{C_s^{(1,1)}\Gamma_{K-2}^{(1)}}\Gamma_{\rm cusp}^{(1)}m_K(2)
-
\vev{\big(\bar{\Gamma}_{K-2}^{(1)}\big)^2}\Gamma_{\rm cusp}^{(1)}m_K(2)
\nonumber\\
&\quad
-
4\vev{\Gamma_{K-2}^{(1)}}\big(\Gamma_{\rm cusp}^{(1)}\big)^2m_K^2(2)
+
2\vev{\Gamma_{K-2}^{(1)}}\big(\Gamma_{\rm cusp}^{(1)}\big)^2m_K(3)
\, .
\end{align}
However, it does not fix $\vev{C_s^{(2,1)} \Gamma_{K-2}^{(1)}}$ and $\vev{C_t^{(2,1)} \bar{\Gamma}_{K-2}^{(1)}}$ separately.
Instead, it fixes their difference,
\begin{align}
\vev{C_s^{(2,1)}\Gamma_{K-2}^{(1)}}
&
-
\vev{C_t^{(2,1)}\bar{\Gamma}_{K-2}^{(1)}}
=
-\vev{C_s^{(1,1)}\Gamma_{K-2}^{(2)}}
+
\vev{C_t^{(1,1)}\bar{\Gamma}_{K-2}^{(2)}}
-
\vev{C_s^{(1,1)}\big(\Gamma_{K-2}^{(1)}\big)^2}m_K(2)
\nonumber\\
&
+
\vev{C_t^{(1,1)}\big(\bar{\Gamma}_{K-2}^{(1)}\big)^2}m_K(2)
-
2\vev{C_s^{(1,0)}\Gamma_{K-2}^{(1)}}\Gamma_{\rm cusp}^{(1)}m_K(2)
-
2\vev{\Gamma_{K-2}^{(2)}}\Gamma_{\rm cusp}^{(1)}m_K(2)
\nonumber\\
&
+
2\vev{C_t^{(1,0)}\bar{\Gamma}_{K-2}^{(1)}}\Gamma_{\rm cusp}^{(1)}m_K(2)
+
2\vev{\bar{\Gamma}_{K-2}^{(2)}}\Gamma_{\rm cusp}^{(1)}m_K(2)
-
4\vev{\big(\Gamma_{K-2}^{(1)}\big)^2}\Gamma_{\rm cusp}^{(1)}m_K^2(2)
\nonumber\\
&
+
4\vev{\big(\bar{\Gamma}_{K-2}^{(1)}\big)^2}\Gamma_{\rm cusp}^{(1)}m_K^2(2)
+
2\vev{\big(\Gamma_{K-2}^{(1)}\big)^2}\Gamma_{\rm cusp}^{(1)}m_K(3)
-
2\vev{\big(\bar{\Gamma}_{K-2}^{(1)}\big)^2}\Gamma_{\rm cusp}^{(1)}m_K(3)
\, ,
\end{align}
as well as the difference of $C_{s,t}^{(3,0)}$:
\begin{align}
\vev{C_t^{(3,0)}}
&
-
\vev{C_s^{(3,0)}}
=
\vev{C_s^{(2,0)}\Gamma^{(1)}_{K-2}}m_K(2)
+
\vev{C_s^{(1,0)}\Gamma^{(2)}_{K-2}}m_K(2)
+
\vev{\Gamma^{(3)}_{K-2}}m_K(2)
\nonumber\\
&
-
\vev{C_t^{(2,0)}\bar{\Gamma}^{(1)}_{K-2}}m_K(2)
-
\vev{C_t^{(1,0)}\bar{\Gamma}^{(2)}_{K-2}}m_K(2)
-
\vev{\bar{\Gamma}^{(3)}_{K-2}}m_K(2)
\nonumber\\
&
+
\vev{C_s^{(2,1)}\Gamma^{(1)}_{K-2}}m_K^2(2)
+
\vev{C_s^{(1,0)}\big(\Gamma^{(1)}_{K-2}\big)^2}m_K^2(2)
+
\vev{C_s^{(1,1)}\Gamma^{(2)}_{K-2}}m_K^2(2)
\nonumber\\
&
+
2\vev{\Gamma^{(1)}_{K-2}\Gamma^{(2)}_{K-2}}m_K^2(2)
-
\vev{C_t^{(2,1)}\bar{\Gamma}^{(1)}_{K-2}}m_K^2(2)
-
\vev{C_t^{(1,0)}\big(\bar{\Gamma}^{(1)}_{K-2}\big)^2}m_K^2(2)
\nonumber\\
&
-
\vev{C_t^{(1,1)}\bar{\Gamma}^{(2)}_{K-2}}m_K^2(2)
-
2\vev{\bar{\Gamma}^{(1)}_{K-2}\bar{\Gamma}^{(2)}_{K-2}}m_K^2(2)
+
2\vev{\Gamma^{(1)}_{K-2}}\Gamma_{\rm virt}^{(2)}m_K^2(2)
\nonumber\\
&
-
2\vev{\bar{\Gamma}^{(1)}_{K-2}}\Gamma_{\rm virt}^{(2)}m_K^2(2)
+
\vev{C_s^{(1,1)}\big(\Gamma^{(1)}_{K-2}\big)^2}m_K^3(2)
+
\vev{\big(\Gamma^{(1)}_{K-2}\big)^3}m_K^3(2)
\nonumber\\
&
-
\vev{C_t^{(1,1)}\big(\bar{\Gamma}^{(1)}_{K-2}\big)^2}m_K^3(2)
-
\vev{\big(\bar{\Gamma}^{(1)}_{K-2}\big)^3}m_K^3(2)
+
2\vev{C_s^{(1,0)}\Gamma^{(1)}_{K-2}}\Gamma_{\rm cusp}^{(1)}m_K^3(2)
\nonumber\\
&
+
2\vev{\Gamma^{(2)}_{K-2}}\Gamma_{\rm cusp}^{(1)}m_K^3(2)
-
2\vev{C_t^{(1,0)}\bar{\Gamma}^{(1)}_{K-2}}\Gamma_{\rm cusp}^{(1)}m_K^3(2)
-
2\vev{\bar{\Gamma}^{(2)}_{K-2}}\Gamma_{\rm cusp}^{(1)}m_K^3(2)
\nonumber\\
&
+
2\vev{\Gamma^{(1)}_{K-2}}\Gamma_{\rm cusp}^{(2)}m_K^3(2)
-
2\vev{\bar{\Gamma}^{(1)}_{K-2}}\Gamma_{\rm cusp}^{(2)}m_K^3(2)
+
4\vev{\big(\Gamma^{(1)}_{K-2}\big)^2}\Gamma_{\rm cusp}^{(1)}m_K^4(2)
\nonumber\\
&
-
4\vev{\big(\bar{\Gamma}^{(1)}_{K-2}\big)^2}\Gamma_{\rm cusp}^{(1)}m_K^4(2)
-
\ft{1}{2}\vev{C_s^{(1,0)}\big(\Gamma^{(1)}_{K-2}\big)^2}m_K(3)
-
\vev{\Gamma^{(1)}_{K-2}\Gamma^{(2)}_{K-2}}m_K(3)
\nonumber\\
&
+
\ft{1}{2}\vev{C_t^{(1,0)}\big(\bar{\Gamma}^{(1)}_{K-2}\big)^2}m_K(3)
+
\vev{\bar{\Gamma}^{(1)}_{K-2}\bar{\Gamma}^{(2)}_{K-2}}m_K(3)
-
\vev{\Gamma^{(1)}_{K-2}}\Gamma_{\rm virt}^{(2)}m_K(3)
\nonumber\\
&
+
\vev{\bar{\Gamma}^{(1)}_{K-2}}\Gamma_{\rm virt}^{(2)}m_K(3)
+
\vev{C_s^{(2,2)}\Gamma^{(1)}_{K-2}}m_K(2)m_K(3)
-
\ft{1}{2}\vev{C_s^{(1,1)}\big(\Gamma^{(1)}_{K-2}\big)^2}m_K(2)m_K(3)
\nonumber\\
&
-
\vev{\big(\Gamma^{(1)}_{K-2}\big)^3}m_K(2)m_K(3)
-
\vev{C_t^{(2,2)}\bar{\Gamma}^{(1)}_{K-2}}m_K(2)m_K(3)
\nonumber\\
&
+
\ft{1}{2}\vev{C_t^{(1,1)}\big(\bar{\Gamma}^{(1)}_{K-2}\big)^2}m_K(2)m_K(3)
+
\vev{\big(\bar{\Gamma}^{(1)}_{K-2}\big)^3}m_K(2)m_K(3)
\nonumber\\
&
-
3\vev{\big(\Gamma^{(1)}_{K-2}\big)^2}\Gamma_{\rm cusp}^{(1)}m_K^2(2)m_K(3)
+
3\vev{\big(\bar{\Gamma}^{(1)}_{K-2}\big)^2}\Gamma_{\rm cusp}^{(1)}m_K^2(2)m_K(3)
\nonumber\\
&
-
2\vev{\Gamma^{(1)}_{K-2}}\big(\Gamma_{\rm cusp}^{(1)}\big)^2m_K(2)m_K^2(3)
+
2\vev{\bar{\Gamma}^{(1)}_{K-2}}\big(\Gamma_{\rm cusp}^{(1)}\big)^2m_K(2)m_K^2(3)
\nonumber\\
&
+
\ft{1}{6}\vev{\big(\Gamma^{(1)}_{K-2}\big)^3}m_K(4)
-
\ft{1}{6}\vev{\big(\bar{\Gamma}^{(1)}_{K-2}\big)^3}m_K(4)
\, .
\end{align}
If we further impose $s\leftrightarrow t$ symmetry---appropriate because identical towers are exchanged in the two channels---crossing
leaves the corresponding mixed averages with anomalous dimensions entirely unfixed. All we can conclude is that
\begin{align}
\label{SymEqsThreeLoopApp}
\vev{C_s^{(2,1)}\Gamma_{K-2}^{(1)}}
&=
\vev{C_t^{(2,1)}\bar{\Gamma}_{K-2}^{(1)}}
\, ,  \\
\vev{C_s^{(3,0)}}
&=
\vev{C_t^{(3,0)}}
\, .
\end{align}
This is precisely the intrinsic freedom in the OPE coefficients left by the crossing equation.

\section{Flux-tube cumulants}
\label{CumulantsApp}

In this appendix, we provide a compendium of cumulants through twist eight. The first cumulants are known through three loops,
whereas cumulants of orders two through four are given through four loops. In the first six formulas below, $\Gamma_{\rm cusp}$
is to be expanded consistently and the result truncated after $g^6$. The four-loop coefficients were determined from 34-digit numerical evaluations of the Born ensemble dressed by the
measures and form factors of Sect.~\ref{TwoLoopMeasureFFsection}, and identified in the basis
$\{\pi^6,\zeta_3^2\}$. The four-loop part of $\vev{\vev{\Gamma_N}}_2$ is then free of $\zeta_3^2$ once $\Gamma_{\rm cusp}$ is
expanded, and equals $\frac{568}{945}\pi^6g^8$ for every $N\ge3$. They are
\begin{align}
\vev{\vev{\Gamma_1}}_1
&=
-\Gamma_{\rm cusp}H_1^{(1)}
+
3g^2\Gamma_{\rm cusp}\zeta_3
\nonumber\\
&\quad
+
g^4\Gamma_{\rm cusp}
\left(
\frac{\pi^2}{15}\zeta_3
-
10\zeta_5
\right)
+
g^6
\left(
\frac{4\pi^2}{3}\zeta_3
-
80\zeta_5
\right)
\, ,
\\
\vev{\vev{\Gamma_2}}_1
&=
-\Gamma_{\rm cusp}H_2^{(1)}
-
12g^4\zeta_3
+
6g^2\Gamma_{\rm cusp}\zeta_3
\nonumber\\
&\quad
+
g^4\Gamma_{\rm cusp}
\left(
-\frac{\pi^2}{3}\zeta_3
-
20\zeta_5
\right)
+
\frac{16}{3}g^6\pi^2\zeta_3
\, ,
\\
\vev{\vev{\Gamma_3}}_1
&=
-\Gamma_{\rm cusp}H_3^{(1)}
-
20g^4\zeta_3
+
8g^2\Gamma_{\rm cusp}\zeta_3
\nonumber\\
&\quad
+
g^4\Gamma_{\rm cusp}
\left(
-\frac{\pi^2}{3}\zeta_3
-
25\zeta_5
\right)
+
g^6
\left(
8\pi^2\zeta_3
+
20\zeta_5
\right)
\, ,
\\
\vev{\vev{\Gamma_4}}_1
&=
-\Gamma_{\rm cusp}H_4^{(1)}
-
28g^4\zeta_3
+
10g^2\Gamma_{\rm cusp}\zeta_3
\nonumber\\
&\quad
+
g^4\Gamma_{\rm cusp}
\left(
-\frac{\pi^2}{3}\zeta_3
-
30\zeta_5
\right)
+
g^6
\left(
\frac{32\pi^2}{3}\zeta_3
+
40\zeta_5
\right)
\, ,
\\
\vev{\vev{\Gamma_5}}_1
&=
-\Gamma_{\rm cusp}H_5^{(1)}
+
12g^2\Gamma_{\rm cusp}\zeta_3
-
36g^4\zeta_3
\nonumber\\
&\quad
+
g^4\Gamma_{\rm cusp}
\left(
-\frac{\pi^2}{3}\zeta_3
-
36\zeta_5
\right)
+
g^6
\left(
\frac{40\pi^2}{3}\zeta_3
+
64\zeta_5
\right)
\, ,
\\
\vev{\vev{\Gamma_6}}_1
&=
-\Gamma_{\rm cusp}H_6^{(1)}
+
14g^2\Gamma_{\rm cusp}\zeta_3
-
44g^4\zeta_3
\nonumber\\
&\quad
+
g^4\Gamma_{\rm cusp}
\left(
-\frac{\pi^2}{3}\zeta_3
-
42\zeta_5
\right)
+
g^6
\left(
16\pi^2\zeta_3
+
88\zeta_5
\right)
\, ,
\end{align}
and
\begin{align}
\vev{\vev{\Gamma_1}}_2
&=
\left(
H_1^{(2)}-\frac{\pi^2}{12}
\right)\Gamma_{\rm cusp}^2
-
\frac{1}{9}g^4\pi^4\Gamma_{\rm cusp}
+
\frac{1}{60}g^2\pi^4\Gamma_{\rm cusp}^2
\nonumber\\
&\quad
+
g^4\Gamma_{\rm cusp}^2
\left(
-\frac{31\pi^6}{6300}
+
\zeta_3^2
\right)
+
g^6\Gamma_{\rm cusp}
\left(
\frac{169\pi^6}{2835}
-
16\zeta_3^2
\right)
+
\frac{208}{945}g^8\pi^6
\, ,
\\
\vev{\vev{\Gamma_2}}_2
&=
\left(
H_2^{(2)}-\frac{\pi^2}{12}
\right)\Gamma_{\rm cusp}^2
-
\frac{8}{45}g^4\pi^4\Gamma_{\rm cusp}
+
\frac{1}{90}g^2\pi^4\Gamma_{\rm cusp}^2
\nonumber\\
&\quad
+
g^4\Gamma_{\rm cusp}^2
\left(
-\frac{23\pi^6}{4725}
+
2\zeta_3^2
\right)
+
g^6\Gamma_{\rm cusp}
\left(
\frac{124\pi^6}{945}
-
8\zeta_3^2
\right)
+
\frac{32}{105}g^8\pi^6
\, ,
\\
\vev{\vev{\Gamma_3}}_2
&=
\left(
H_3^{(2)}-\frac{\pi^2}{12}
\right)\Gamma_{\rm cusp}^2
-
\frac{2}{15}g^4\pi^4\Gamma_{\rm cusp}
+
g^4\Gamma_{\rm cusp}^2
\left(
\frac{\pi^6}{378}
+
3\zeta_3^2
\right)
\nonumber\\
&\quad
+
g^6\Gamma_{\rm cusp}
\left(
\frac{16\pi^6}{189}
-
12\zeta_3^2
\right)
+
\frac{208}{945}g^8\pi^6
\, ,
\\
\vev{\vev{\Gamma_4}}_2
&=
\left(
H_4^{(2)}-\frac{\pi^2}{12}
\right)\Gamma_{\rm cusp}^2
-
\frac{2}{15}g^4\pi^4\Gamma_{\rm cusp}
\nonumber\\
&\quad
+
g^4\Gamma_{\rm cusp}^2
\left(
\frac{\pi^6}{630}
+
4\zeta_3^2
\right)
+
g^6\Gamma_{\rm cusp}
\left(
\frac{74\pi^6}{945}
-
16\zeta_3^2
\right)
\nonumber\\
&\quad
+
\frac{248}{945}g^8\pi^6
\, ,
\\
\vev{\vev{\Gamma_5}}_2
&=
\left(
H_5^{(2)}-\frac{\pi^2}{12}
\right)\Gamma_{\rm cusp}^2
-
\frac{2}{15}g^4\pi^4\Gamma_{\rm cusp}
\nonumber\\
&\quad
+
g^4\Gamma_{\rm cusp}^2
4\zeta_3^2
+
g^6\Gamma_{\rm cusp}
\left(
\frac{82\pi^6}{945}
-
16\zeta_3^2
\right)
\nonumber\\
&\quad
+
\frac{16}{63}g^8\pi^6
\, ,
\\
\vev{\vev{\Gamma_6}}_2
&=
\left(
H_6^{(2)}-\frac{\pi^2}{12}
\right)\Gamma_{\rm cusp}^2
-
\frac{2}{15}g^4\pi^4\Gamma_{\rm cusp}
\nonumber\\
&\quad
+
g^4\Gamma_{\rm cusp}^2
4\zeta_3^2
+
g^6\Gamma_{\rm cusp}
\left(
\frac{82\pi^6}{945}
-
16\zeta_3^2
\right)
\nonumber\\
&\quad
+
\frac{16}{63}g^8\pi^6
\, ,
\end{align}
and 
\begin{align}
\vev{\vev{\Gamma_1}}_3
&=
\Gamma_{\rm cusp}^3
\left(
-2H_1^{(3)}
+
\frac{7}{4}\zeta_3
\right)
+
g^2\Gamma_{\rm cusp}^3
\left(
\frac{3\pi^2}{4}\zeta_3
-
\frac{33}{4}\zeta_5
\right)
+
g^4\Gamma_{\rm cusp}^2
\left(
-2\pi^2\zeta_3
+
18\zeta_5
\right)
\, ,
\\
\vev{\vev{\Gamma_2}}_3
&=
\Gamma_{\rm cusp}^3
\left(
-2H_2^{(3)}
+
2\zeta_3
\right)
+
g^2\Gamma_{\rm cusp}^3
\left(
\frac{\pi^2}{2}\zeta_3
-
\frac{13}{2}\zeta_5
\right)
+
g^4\Gamma_{\rm cusp}^2
\left(
-2\pi^2\zeta_3
+
21\zeta_5
\right)
\, ,
\\
\vev{\vev{\Gamma_3}}_3
&=
\Gamma_{\rm cusp}^3
\left(
-2H_3^{(3)}
+
2\zeta_3
\right)
-
\frac{3}{4}g^2\Gamma_{\rm cusp}^3\zeta_5
+
3g^4\Gamma_{\rm cusp}^2\zeta_5
\, ,
\\
\vev{\vev{\Gamma_4}}_3
&=
\Gamma_{\rm cusp}^3
\left(
-2H_4^{(3)}
+
2\zeta_3
\right)
\, ,
\\
\vev{\vev{\Gamma_5}}_3
&=
\Gamma_{\rm cusp}^3
\left(
-2H_5^{(3)}
+
2\zeta_3
\right)
\, ,
\\
\vev{\vev{\Gamma_6}}_3
&=
\Gamma_{\rm cusp}^3
\left(
-2H_6^{(3)}
+
2\zeta_3
\right)
\, ,
\end{align}
and
\begin{align}
\vev{\vev{\Gamma_1}}_4
&=
\Gamma_{\rm cusp}^4
\left(
6H_1^{(4)}
-
\frac{11\pi^4}{180}
\right)
\, ,
\\
\vev{\vev{\Gamma_2}}_4
&=
\Gamma_{\rm cusp}^4
\left(
6H_2^{(4)}
-
\frac{47\pi^4}{720}
\right)
\, ,
\\
\vev{\vev{\Gamma_3}}_4
&=
\Gamma_{\rm cusp}^4
\left(
6H_3^{(4)}
-
\frac{\pi^4}{15}
\right)
\, ,
\\
\vev{\vev{\Gamma_4}}_4
&=
\Gamma_{\rm cusp}^4
\left(
6H_4^{(4)}
-
\frac{\pi^4}{15}
\right)
\, ,
\\
\vev{\vev{\Gamma_5}}_4
&=
\Gamma_{\rm cusp}^4
\left(
6H_5^{(4)}
-
\frac{\pi^4}{15}
\right)
\, ,
\\
\vev{\vev{\Gamma_6}}_4
&=
\Gamma_{\rm cusp}^4
\left(
6H_6^{(4)}
-
\frac{\pi^4}{15}
\right)
\, .
\end{align}
Here $H_\ell^{(k)}=\sum_{j=1}^{\ell}j^{-k}$ denotes the generalized harmonic number of order $k$.

\end{document}